\documentclass{statsoc}

\usepackage[a4paper]{geometry}
\usepackage{graphicx}
\usepackage[textwidth=8em,textsize=small]{todonotes}

\usepackage{caption}

\usepackage{threeparttable}
\usepackage{tikz}
\usetikzlibrary{shapes.geometric, arrows}

\usepackage{latexsym, color}
\usepackage{amsmath}
\usepackage{amsfonts}
\usepackage{amssymb}
\usepackage{psfrag}
\usepackage{graphicx}
\usepackage{epsfig}
\usepackage{multirow}
\usepackage{multicol}
\usepackage{subfigure}
\usepackage{color}
\usepackage[ruled,vlined]{algorithm2e}
\usepackage{floatrow}
\usepackage{calc}
\usepackage{natbib}
\usepackage{soul}

\definecolor{eGreen}{rgb}{.057, .549,.065}

\newcommand{\balpha}{\mbox{\boldmath $\alpha$}}

\newcommand{\bmu}{\mbox{\boldmath $\mu$}}
\newcommand{\bphi}{\mbox{\boldmath $\phi$}}

\newcommand{\bnu}{\mbox{\boldmath $\nu$}}
\newcommand{\bSigma}{\mbox{\boldmath $\Sigma$}}

\newcommand{\bzeta}{\mbox{\boldmath $\zeta$}}
\newcommand{\bbeta}{\mbox{\boldmath $\beta$}}
\newcommand{\beeta}{\mbox{\boldmath $\eta$}}
\newcommand{\btheta}{\mbox{\boldmath $\theta$}}
\newcommand{\bTheta}{\mbox{\boldmath $\Theta$}}

\newcommand{\bkappa}{\mbox{\boldmath $\kappa$}}

\newcommand{\bDelta}{\mbox{\boldmath $\Delta$}}
\newcommand{\bdelta}{\mbox{\boldmath $\delta$}}

\newcommand{\bpsi}{\mbox{\boldmath $\psi$}}

\newcommand{\bsigma}{\mbox{\boldmath $\sigma$}}
\newcommand{\blambda}{\mbox{\boldmath $\lambda$}}

\newcommand{\bgamma}{\mbox{\boldmath $\gamma$}}

\newcommand{\bX}{\mathbf{X}}
\newcommand{\bx}{\mathbf{x}}
\newcommand{\bz}{\mathbf{z}}

\newcommand{\pkg}[1]{{\normalfont\fontseries{b}\selectfont #1}}
\let\proglang=\textsf

\tikzstyle{random} = [circle, radius=3cm,text centered, draw=black]
\tikzstyle{fixed} = [rectangle, minimum width=3em, minimum height=3em, text centered, draw=black]
\tikzstyle{arrow} = [thick,->,>=stealth]

\title[Unsupervised Bayesian classification]{Unsupervised Bayesian classification for models with scalar and functional covariates}

 \author[Garcia et. al.]{Nancy L. Garcia$^{1}$\thanks{{\it Corresponding author}: Department of Statistics, University of Campinas, S\~{a}o Paulo, Brazil. {e-mail: nancyg@unicamp.br}}, Mariana Rodrigues-Motta$^{1}$, Helio S. Migon$^{2}$,\\  Eva Petkova$^{3,4}$,  Thaddeus Tarpey$^{3}$, R. Todd Ogden$^{5}$,\\  Julio O. Giodano$^{6}$ and Martin Matias Perez$^{6}$ \\
{\footnotesize $^{1}$ Department of Statistics, University of Campinas, Brazil, \\
\footnotesize $^{2}$ Department of Statistics, Federal University of Rio de Janeiro, Brazil, \\
\footnotesize $^{3}$ Department of Population Health, Grossman School of Medicine, New York University, USA, \\ \footnotesize $^{4}$ Department of Child and Adolescent Psychiatry, Grossman School of Medicine, New York University, USA \\
\footnotesize $^{5}$ Department of Biostatistics, Columbia University, USA  \\
\footnotesize $^{6}$ College of Agriculture and Life Sciences, Cornell University, USA}}

\begin{document}
\maketitle

\begin{abstract}
We consider unsupervised classification by means of a latent
multinomial variable which categorizes a scalar response into one of L components of a mixture model. This process can be thought as a hierarchical model with first level modelling a scalar response according to a mixture of parametric distributions, the second level models the mixture probabilities by means of a generalised linear model with functional and scalar covariates. The traditional approach of treating functional covariates as vectors not only suffers from the curse of dimensionality since functional covariates can be measured at very small intervals leading to a highly parametrised model but also does not take into account the nature of the data. We use basis expansion to reduce the dimensionality and a Bayesian approach to estimate the parameters while providing predictions of the latent classification vector. By means of a simulation study we investigate the behaviour of our approach considering  normal mixture model and zero inflated mixture of Poisson distributions.  We also compare the performance of the classical Gibbs sampling approach with Variational Bayes Inference. \\
{\it Key words: latent vector, functional covariates, variable selection, unsupervised clustering, variational inference.}
\end{abstract}

\section{Introduction}

Mixture models are popular statistical tools for classification purposes in a broad range of applied fields. 
The Gaussian mixture model is by far the most used approach for model based cluster analysis \citep[e.g.,][]{day1969estimating,fraley2006mclust,mcnicholas2010model}. However, classification problems based on mixture models often require non-Gaussian mixture distributions. As an example, consider zero-inflated regression models \citep[e.g.,][]{lambert1992zero,ridout1998models}, whose distribution of the count outcome is a mixture of two components and the goal is to classify zero outcomes as coming from either a degenerate at zero distribution  or zeros generated by means of a count distribution, such as a Poisson or Negative Binomial distribution. Other examples are the zero-augmented models for semi-continuous data \citep[e.g.,][]{rodrigues2015mixed}. Such models fit data using a mixture of two components where one component models the zero by means of a degenerate at zero distribution and the other component models the positive outcome using a continuous positive distribution, as for example gamma or lognormal distributions. 

 Mixture probabilities often depend on scalar explanatory variables \citep[e.g.,][]{lambert1992zero,ridout1998models,hall2000zero, hall2004marginal}. However, many modern applications routinely have more complex covariates   in the form of vectors,  matrices, functions, images. The main question of interest is to examine how these complex covariates affect the response.  The prevailing approaches in these cases use either a parametric or a nonparametric approach and model the mean of the distribution as a function of the covariates;  see for example, \cite{cardot1999functional, james2002generalized, ramsay2005principal, ferraty2006nonparametric, ferraty2009additive, goldsmith2011penalized, mclean2014functional} among others. On the other hand, there are other applications where the interest is to study the effect of the covariates on the entire distribution of the response, for example quantile regression     \citep{koenker1978regression} \citep[e.g.,][and references therein]{park2019conditional}.

In this study, we have a different objective, which is to classify an outcome as accurately as possible using the information on the scalar and functional covariates as explanatory variables for the mixture probability. We model the mixture model in terms of a latent variable. The role of the latent variable is not only to divide a sample of subjects into subgroups according to some similarity measure, but also to provide practitioners interpretable clustering results. Many authors have studied the classification problem, see for example \cite{titterington1985statistical, everitt1981finite, mclachlan2004finite} and references therein.  Moreover, there are \proglang{R} packages \citep{ref_rlang} that perform inference for mixture models such as \pkg{mixtools} \citep{benaglia:hal-00384896} but these tools cannot implement  one or more functional covariates. First of all, it is necessary to reduce the dimensionality of the data. For example, for one of our illustrations, the dataset has 14 functional covariates, each one of which is observed at 45 points.  \cite{jiang2017latent} consider these functional covariates as $14 \times 45$  matrices and propose a variation of principal components analysis built upon a low rank Candecomp/Parafac decomposition applied to the rows and columns of the matrices. Although their method is powerful, it requires all functional covariates to be observed at the same points across observations. Also, it does not take advantage of the functional structure of the data and  does not consider how the covariates are ordered in the matrix. In this paper, similarly to \cite{ciarleglio2018constructing}, we propose a flexible classification procedure, more general than a Gaussian mixture, that can incorporate functional covariates, either through a linear or non-linear effect, that not only reduces the dimensionality of the problem  but also takes advantage of the functional nature of the covariates. Additionally, the method does not restrict the functions to be observed at common points. 

The remainder of this paper is organized as follows. Section \ref{sec:model} introduces the hierarchical mixture model with latent variable and regression of mixture probabilities as function of functional covariates and a Bayesian approach is presented in Section \ref{sec:bayesian}. Our method can be applied to regression with a response of continuous, semi-continuous and discrete nature, but in this article we illustrate the method considering a normal mixture model and zero inflated mixture of Poisson (ZIMP) model, as shown in Sections \ref{sec:mixnormal} and \ref{sec:zimp}, respectively. An extensive simulation study is given in Section \ref{sec:simulations}, with the primary goal of examining the performance of the normal mixture model and the ZIMP model by considering aspects of sample size and ability of the functional curves to predict the latent variables correctly. The secondary goal of the simulation study is to compare the estimation ability of the MCMC and Variational Bayes (VB) methods \citep{ormerod2012gaussian, hoffman2013stochastic, blei2017variational}, as well as their performance with respect to computational time-consuming. Finally, two applications with real data are presented in Section \ref{sec:realdata}, one for the normal mixture model and another for the ZIMP model.

\section{General model} \label{sec:model}

For each subject $i = 1, \ldots, n$, we observe: $y_i$ the scalar response, $\mathbf{z}_{i}$ a vector of scalar covariates  and ${\bX}_{ij}=\{t_{ijs}, X_{ij}(t_{ijs}), 1 \le s \le S_{ij}, 1 \le j \le J\}$, $J$ functional covariates observed at discrete points in closed domains $\tau_1, \ldots, \tau_J$. In general, these sets are closed intervals on the real line, and although we interpret $t_{ijs}$ as time in this study, our proposed model works for functional covariates observed at points in space or time-space. Each functional covariate $X_{ij}$ is observed at $S_{ij}$ points $t_{ijs}, 1 \le s \le S_{ij}$ which do need to be the same along the subjects. That is,  neither the domains in which we observe the functional covariates nor the observation points  need to be the same.  We model the distribution of $y_i$  hierarchically by means of a latent class model, postulating a mixture distribution for the observed response to classify subjects into $L$ classes.   We will assume an $L$-mixture latent class model with unobserved multinomial random variables indicating  class membership $\{\bgamma_i = (\gamma_{i1}, \ldots, \gamma_{iL}), i = 1, \ldots, n\}$ where  $\sum_{l=1}^{L} \gamma_{il} = 1$ and 
\begin{equation*}
p_{il} = P(\gamma_{il} = 1) \label{eq:pil}.
\end{equation*}
Probability of the random variable $\gamma_{il}$ is modelled as a function of the scalar and functional covariates by
\begin{equation}
g(p_{il}) = \mathbf{z}_{i}^{\top} \btheta_{l} + \sum_{j=1}^J \int_{\tau_j} F_{jl}(X_{ij}(t),t,\bphi_{jl}) dt \label{eq:link}
\end{equation}
where $g$ is a known link function (e.g., probit or logit), $\btheta_l$ is a vector of parameters that captures the linear additive effect of the scalar covariates, and $F_{jl}(\cdot,\cdot)$ is a bivariate smooth function related to the $j$th functional covariate in component $l$ which depends on the vector of parameters $\bphi_{jl}$. This is termed the functional generalized additive model by \cite{mclean2014functional}. Let $\mathbf{y}=(y_1,\ldots,y_n)^{\top}$, $\bz=(\bz_1^{\top}, \ldots, \bz_n^{\top})^{\top}$ and $\bX = \{\bX_{ij}, i=1, \ldots, n, j=1, \ldots, J\}$ be vectors and let $f_l(y_i;\blambda_{il})$, $i=1, \ldots, n$ be the pdf of the scalar response $y_i$ in class $l$, such as $\blambda_{il}$ is the vector of parameters in that class. Then the likelihood of the mixture model for $\mathbf{y}$ is given by

\begin{eqnarray*}
f(\mathbf{y}|\blambda_1,\ldots,\blambda_L,\btheta,\bphi,\mathbf{z},\bX) = \prod_{i=1}^n \left( \sum_{l=1}^{L} p_{il} f_l(y_i;\blambda_{il})\right),   \label{eq:like}
\end{eqnarray*}
 and the complete-data likelihood can be written as
 \begin{equation}
 f(\mathbf{y},\gamma|\blambda_1,\ldots,\blambda_L,\btheta,\bphi,\mathbf{z},\bX) = \prod_{i=1}^{n} \prod_{l=1}^{L} f_l(y_i;\blambda_{il})^{\gamma_{il}} p_{il}  \label{eq:complete_like} \mbox{,}
 \end{equation}
where the relationship between $p_{ij}$ with $\btheta,\bphi,\mathbf{z}$ and $\bX$ is given by\eqref{eq:link}.

In general, a nonparametric model is used to represent the effects of the covariates on $\blambda_{il}$ but not on the mixing probabilities $p_{il}$, see for example \citet{cardot1999functional}, \citet{james2002generalized}, \citet{ferraty2006nonparametric}, \citet{ramsay2007applied}, \cite{mclean2014functional} and references therein. In many applications, as will be shown in Section \ref{sec:realdata}, the interest lies in using the covariates solely to classify subjects into $L$ classes. Therefore it is necessary to relate the covariates to the mixing probabilities $p_{il}$ and not to the parameters $\blambda$. To demonstrate the strength of our method, in our simulations and applications, we will analyse  two cases explicitly. The first one is the {\it Mixture of Normal distributions} where $\lambda_{il} = (\mu_{l}, \sigma_l)$ and $f$ corresponds to  the normal density with mean $\mu_l$ and standard deviation $\sigma_l$. The second one is {\it Zero Inflated Mixture of Poisson distributions (ZIMP)}, considering  $f_{1} = \delta_{\{0\}}$ as the distribution for the class with ``Pure Zero'' and $ f_l(y_i;\lambda_{l})$ for $l = 1,2$ correspond to  Poisson distributions.

\subsection{Functional linear model}

Model~(\ref{eq:link}) can be restricted to be linear by specifying $F_{jl}(x(t),t) = w_{jl}(t) x(t)$, yielding the more common generalized linear functional model
\begin{equation}
g(p_{il}) = \mathbf{z}_{i}^\top \btheta_{l} + \sum_{j=1}^J \int_{\tau_j} w_{jl}(t) X_{ij}(t) dt, \label{linear_model}
\end{equation}
$l=1,\ldots,L$.
To fit model (\ref{linear_model}), we consider each weight function $w_{jl}(.)$ as a smooth function approximated by a function belonging to the finite-dimensional space spanned by  $B$-splines basis functions. This is not the only possibility, as other bases could be chosen such as Fourier expansion, wavelets, natural splines, etc.\  \citep{silverman2018density}. Also, we are going to  choose the number of knots and knots placement in an {\it ad-hoc} manner. Although knot determination and placement are important issues, they are not the objective of this work and will not be discussed here. 

Therefore, for a positive integer $K_j$ and a vector of $(K_j - 4)$ interior knots $\Upsilon_j \subset \tau_j$, we express the weight function as
\begin{eqnarray}
w_{jl}(t) &=& \sum_{k=1}^{K_j} \phi_{jlk} B^{(j)}_{k}(t), \label{eq:weight}
\end{eqnarray}
where $\{B^{(j)}_1, \ldots, B^{(j)}_{K_j}\}$ are cubic B-spline basis functions determined by $\Upsilon_j$.

Substituting (\ref{eq:weight}) into (\ref{linear_model}) yields
\begin{equation}
g(p_{il}) = \mathbf{z}_{i}^{\top} \btheta_{l} +  \sum_{j=1}^J \mathbf{R}_{ij}^\top \bphi_{jl}\label{eq:6}, 
\end{equation}
where, for each pair $(j,l)$, $\bphi_{jl} = (\phi_{jl1}, \ldots, \phi_{jlK_j})$ and  $\mathbf{R}_{ij}^{\top} = (R_{ij1}, \ldots, R_{ijK_j})$ are vectors of length $K_j$,  with $R_{ijk} = \int_{\tau_j}  B_k^{(j)} (t) X_{ij}(t) dt$. 

The linear case has the advantage of easy interpretability of the weight functions. If the weight $w_{jl}(t)$ is positive (negative) over the interval $(t_a,t_b)$, this means that the higher the value of $X_{ij}(t)$ in this interval the higher (lower) the probability of $\gamma_{il} = 1$, considering all other explanatory variables fixed. 

\subsection{Functional nonlinear model}

For the more general model~(\ref{eq:link}), we consider each function $F_{jl}(\cdot,\cdot)$ to be a smooth surface which can be well approximated by a family of tensor products of cubic $B$-splines (see for example, \cite{kim2018additive}). That is, for each function $F_{jl}$, there exist positive integers $K_{1j}$ and $K_{2j}$ and  vectors of $(K_{1j} - 4)$ interior knots $\Upsilon_{1j} \subset \tau_j$ and $(K_{2j} - 4)$ interior knots $\Upsilon_{2j} \subset \chi_{j}$, the image of $X_{ij}$, such that
\begin{equation}
F_{jl}(s,t) = \sum_{k_1=1}^{K_{1j}} \sum_{k_2=1}^{K_{2j}} \phi_{jl k_1 k_2} B^{(\Upsilon_{1j})}_{k_1}(t) B^{(\Upsilon_{2j})}_{k_2}(s), \label{eq:weight.1}
\end{equation}
where $\{B^{(\Upsilon_{1j})}_{1}, \ldots, B^{(\Upsilon_{1j})}_{K_{1j}}\}$ and $\{B^{(\Upsilon_{2j})}_{1}, \ldots, B^{(\Upsilon_{1j})}_{K_{2j}}\}$ are $B$-spline basis determined by $\Upsilon_{1j}$ and $\Upsilon_{2j}$, respectively. 

Substituting (\ref{eq:weight.1}) into (\ref{eq:link}) yields
\begin{eqnarray}
g(p_{il}) &=& \mathbf{z}_{i}^{\top} \btheta_{l} +  \sum_{j=1}^J \mathbf{R}_{ij}^{\top} \bphi_{jl} \label{eq:6.1}, 
\end{eqnarray}
where $\mathbf{R}_{ij}$ are vectors of dimension $(K_1 * K_2) \times 1$ with $\bphi_{jl}(k_1,k_2) =  \phi_{jl k_1 k_2}$ and \\ $\mathbf{R}_{ij}(k_1,k_2)= \int_{\chi_j} \int_{\tau_j}  B^{(X_{ij})}_{k_1}(X_{ij}(s)) B^{(T_{ij})}_{k_2}(t) \, dt \, ds$ properly stacked.\\

As we can see, from \eqref{eq:6} and \eqref{eq:6.1}, using basis expansion for both the linear and non-linear case, we end up with the same linear structure in terms of parameters $\btheta$ and $\bphi$.

\section{A Bayesian approach to the mixture model regression with functional covariates} \label{sec:bayesian}

Model (\ref{eq:like}) specifies the distribution of the response $y_i$ depending on which mixture component subject $i$ belongs. The mixture components are parameterized by the vector $\blambda^{\top}=(\blambda^{\top}_1,\ldots,\blambda^{\top}_L)$ whose components are related to the densities $f_{1},\ldots,f_{L}$, respectively, considering category $L$ as the baseline. Therefore, we will denote by $\Theta^{\top}=(\blambda^{\top},\bbeta^{\top})$ the vector of unknown parameters where $\bbeta^{\top} = (\bbeta^{\top}_1, \ldots, \bbeta^{\top}_{L-1})$ and
$\bbeta_{l}^{\top} = (\btheta^{\top}_l,\bphi^{\top}_{1l},\ldots,\bphi^{\top}_{J l})$
indicate the parameters for the regression coefficients of the model with $\mathbf{x}^{\top}_{i} := (\mathbf{z}^{\top}_{i}, \mathbf{R}^{\top}_{i1}, \ldots, \mathbf{R}^{\top}_{iJ})$ as covariates for subject $i$.

\subsection{Hierarchical structure specification and prior specification}

A formal Bayesian analysis of a mixture model usually leads to intractable calculations. Data augmentation is an efficient procedure for mixture models that leads to feasible computations using Gibbs sampling \citep{diebolt1994estimation}. 
The joint augmented posterior distribution is the product of (\ref{eq:complete_like}) and the prior distributions and has no closed form. Therefore, the Gibbs sampling algorithm is suitable to sample from the posterior distribution of $\bgamma$, $\blambda$ and $\bbeta$. 

The nature of the application under study dictates the form of $f_l(y_i;\blambda_l)$ which in turn provides knowledge about the nature of parameters in $\blambda_l$.
There is a very rich family of distributions $f_l(y_i;\blambda_l)$ that may characterize the mixture distribution of $y_i$. Instead of focusing on a specific distribution for mixture components, we focus on a general solution for posterior sampling of the latent variables $\bgamma$ and parameters in $\bTheta$, which are developed with a general $f_l(y_i;\blambda_l)$ without loss of generality. Therefore, prior distributions for $\blambda$ are problem specific. In particular, for each of the components of $\bbeta$, we will assume a Student-$t$ prior distribution with mean $0$, degrees-of-freedom parameter $df$, and scale $s$, with $df$ and $s$ providing minimal prior information to constrain the coefficients to lie in a reasonable range \citep[see Section 2 of][]{gelman2008weakly}. An advantage of the $t$ family is that fat-tailed distributions allow for flexible inference, since it includes both the Gaussian ($df=\infty$) and the Cauchy ($df=1$) distributions.

\subsection{Posterior computation of parameters} \label{sec:posteriors}

We sample from the posterior distribution using a Gibbs sampling scheme, and most of the full conditional posterior distribution of the latent variables $\bgamma$ and parameters in $\bTheta$ are given by standard methods. For the sake of completeness, we describe briefly the conditional posterior distributions for $\bgamma$ and $\bbeta$. 

\paragraph{Full conditional posterior distribution of $\bgamma_{i}$}

Let $\bgamma_{-i} = (\bgamma_{1},\ldots,\bgamma_{(i-1)},\bgamma_{(i+1)},\ldots,\bgamma_{n})$, i.e., the vector $(\bgamma_{1},\ldots,\bgamma_{n})$ leaving out the $i$th element. The full conditional posterior distribution of $\bgamma_{i}$ is given by
\begin{equation*}
P(\gamma_{il} = 1|\bTheta,\mathbf{y},\bgamma_{-i}) \propto f_l(y_i;\blambda_l) g^{-1}_{l}(\bx_i^{\top}\bbeta_{l}).
\end{equation*}

 For example, for the logit link function we have 
\begin{eqnarray*}
g^{-1}_{l}\left(\bx_i^{\top}\bbeta_{l}\right) &\propto& \exp\left(\bx_i^{\top}\bbeta_l\right)
\end{eqnarray*}
whereas for the probit link function we have
$$g^{-1}_{l}\left(\bx_i^{\top}\bbeta_{l}\right) \propto \Phi\left(\bx_i^{\top}\bbeta_l\right)$$
for $l=1, \ldots, L-1$ and
$$g^{-1}_{L}\left(\bx_i^{\top}\bbeta\right) = 1 - \sum_{l=1}^{L-1} g^{-1}_{l}\left(\bx_i^{\top}\bbeta_{l}\right).$$

\paragraph{Full conditional posterior distribution of $\bbeta_{l}$} 

The full conditional posterior distribution of $\bbeta_{l}$ cannot be computed explicitly except for when we are using the probit link function and thus  we can apply the simple latent-variable method of \citet{albert1993bayesian}. Other methods for calculating the full conditional have been proposed using data-augmentation or multiple layers of latent variables, see for example  \citet{holmes2006bayesian}, \citet{fruhwirth2010data}, \citet{gramacy2012simulation} and \citet{polson2013bayesian}. In our approach, we follow \cite{gelman2008weakly} by considering Student $t$ prior distribution for each component of $\bbeta_{l}$, in which the standard logistic regression algorithm proceeds by approximately linearizing the score function, solving using weighted least squares, and then iterating this process, each step evaluating
the derivatives at the latest estimate $\hat{p}_{ilk}$. As in the classical logistic regression, at iteration $k$, the algorithm determines pseudo-data $\psi_i$ given by
\begin{equation}
\psi_{il} = g(\hat{p}_{ilk}) + (\gamma_{il}-\hat{p}_{ilk})  g'(\hat{p}_{ilk}) , \,\, i=1, \ldots, n \label{zi}
\end{equation}
and weights 
\begin{equation}
\mathbf{W}_{lk}^{-1} = \mbox{diag} \left\{  \left( g'(\hat{p}_{ilk}) \right)^{2} V_{ilk} \right\} \label{wi}
\end{equation}
where $V_{ilk}$ is the variance function evaluated at $\hat{p}_{1lk}$ in iteration $k$ \citep{mccullagh1989generalized}. We then perform weighted least squares, regressing the working variable $\bpsi_{l}=(\psi_{1l},\ldots,\psi_{nl})$ on the design matrix ${\bf x} = (\mathbf{x}_{1},\ldots,\mathbf{x}_{n})^{\top}$ of dimension $n \times p_{l}$ with weights $\mathbf{W}_{lk}$ to give a new estimate of $\bbeta_{l}$, and the iteration proceeds until approximate convergence.

We add prior information to the classical logistic regression algorithm given in \eqref{zi} and \eqref{wi} by augmenting the approximate likelihood with the prior distribution $\bbeta_{l} \sim N(\bmu_{bl},\mathbf{\Sigma}_{\beta})$, with $\mathbf{\Sigma}_{\beta} = \sigma^{2}_{l}I$, $l=1,\ldots,L-1$.   Considering a normal distribution as an approximation to the generalized linear model likelihood, the full conditional posterior density is given by
\begin{eqnarray}
\log p(\bbeta_{l}|\bpsi) &\propto& \exp \left\{-\frac{1}{2} (\bpsi_{l}-\mathbf{x} \bbeta_{l})^{\top} \bSigma^{-1} (\bpsi_{l}-\mathbf{x} \bbeta_{l}) \right\} \nonumber \\
&\times& \exp \left\{-\frac{1}{2} (\bbeta_{l} - \bmu_{bl})^{\top} \bSigma_{\beta} (\bbeta_{l} - \bmu_{bl}) \right\} \label{fullbeta_p}
\end{eqnarray}
where $\bSigma = \mathbf{W}_{lk}^{-1}$, with $\mathbf{W}_{lk}^{-1}$ as in (\ref{wi}) and elements of $\bpsi_{l}$ given in (\ref{zi}). Rearranging terms in $(\ref{fullbeta_p})$, the full conditional posterior density of $\bbeta_{l}$ is given by a normal distribution with covariance matrix $\mathbf{V}_{\beta} = ({\bf x}^{\top} \bSigma^{-1} {\bf x}+\bSigma_{\beta}^{-1})^{-1}$ and mean $\mathbf{V}_{\beta} ({\bf x}^{\top} \bSigma^{-1} \bpsi_{l}+\bSigma_{\beta}^{-1} \bmu_{bl})$.

\subsection{A generic discussion of variational inference}

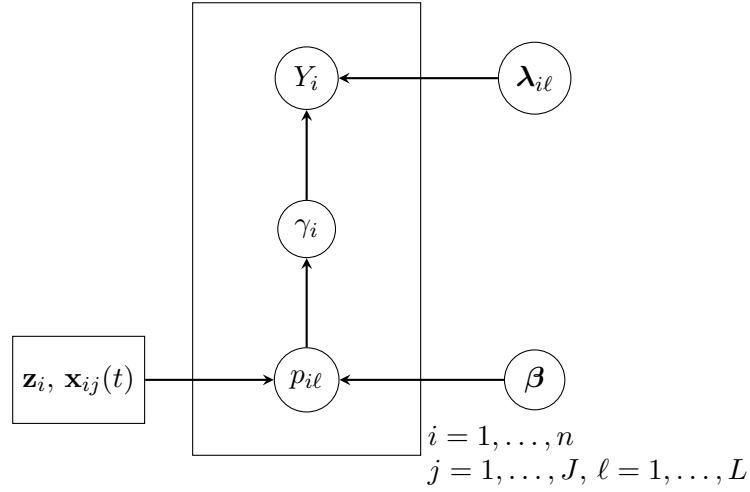
\begin{figure}[htp]
\begin{tikzpicture}[node distance=2cm]

\node (response) [random] {$Y_i$};
\node (gamma) [random, below of=response] {$\gamma_{i}$};
\node (pi) [random, below of=gamma] {$p_{i\ell}$};
\node (lambda) [random, right of=response, xshift=1cm] {$\blambda_{i\ell}$};
\node (beta) [random, right of=pi, xshift=1cm] {$\bbeta$};
\node (x) [fixed, left of=pi, xshift=-1cm] {$\mathbf{z}_i$, $\mathbf{x}_{ij}(t)$};
\draw [arrow] (gamma) -- (response);
\draw [arrow] (lambda) -- (response);
\draw [arrow] (pi) -- (gamma);
\draw [arrow] (x) -- (pi);
\draw [arrow] (beta) -- (pi);
\draw (-1.5,1) rectangle (1.5,-5);
\node[align=left] at (3.7,-5) {$i=1,\ldots,n$ \\ $j=1, \ldots,J$, $\ell = 1, \ldots, L$};
\end{tikzpicture}
\caption{Direct Acyclical Graph describing the general model. The circles represent random quantities and rectangles represent deterministic quantities. The variable $Y_i$
is the $i$th individual observation, $\bgamma_{i}$ is a latent variable indicating the class of the $i$th observation, $p_{i\ell} = g^{-1}(\mathbf{x}_i'\bbeta)$ is the probability of individual $i$ belongs to category $\ell$, where $\mathbf{x}_i$
is a vector of known regressors and $\bbeta$ the regression parameters.} \label{fig:dag}
\end{figure}

Modern data analysis often demands computation with complex models
and massive datasets. To scale the problem described in the introduction of this paper for large samples and to include more functional covariates, we must resort to approximate posterior inference. Variational Bayes inference (VI)  is a machine learning technique that facilitates  approximation of the posterior distribution in complex models using massive datasets \citep{blei2017variational, ormerod2012gaussian, hoffman2013stochastic}. VB inference provides the main alternative to the Markov Chain Monte Carlo (MCMC) algorithm \citep{robert_monte_2004, gamerman2006markov}.
To fix ideas, let us consider the model described by the DAG (direct acyclical graph) shown in Figure \ref{fig:dag}, where $\bbeta$ is a vector of  regression parameters, $\gamma_i$ are categorical latent variables and ${\bf y}$ the observations.

VB inference starts by introducing a variational family of distributions, indexed by some variational parameters $\bkappa$ and a criterion function to search for the member $q(\cdot|\kappa)$ of the family that best approximates the predictive distribution. The optimisation criterion is derived based on the log-marginal posterior distribution of the observed data, a usual model selection criterion, $\log(p(y|{\cal M}))$. Often this quantity evolves to where it requires the solution to an intractable integral. To avoid this tedious calculation, a lower bound quantity, called ELBO (Evidence Lower Bound) is easily evaluated as:
\begin{eqnarray}
\log(p(y|{\cal M}  )) & = & \log \left( \int \int p(y,\blambda,\bbeta,\bgamma) d\bgamma d\bbeta d\blambda \right) \nonumber \\
& = & \log \left( E_{q}\left[ \frac{p(y,\blambda,\bbeta,\bgamma)}{q(\blambda,\bbeta,\bgamma|\bkappa)} \right] \right) \nonumber  \\
& \ge & E_{q} \left[ \log \frac{p(y,\blambda,\bbeta,\bgamma)}{q(\blambda,\bbeta,\bgamma|\bkappa)} \right] \label{eq:elbo.1}
\end{eqnarray}
where $\bgamma$ and $(\blambda,\bbeta)$ represent local and global quantities/parameters, respectively. The inequality in \eqref{eq:elbo.1} is obtained by Jensen's inequality. It is natural to use this lower bound as a model selection criterion in place of the predictive distribution, avoiding cumbersome high dimensional integration. Therefore, the VI inference objective is to maximize ELBO, which is equivalent to minimizing the Kulback-Leibler  divergence up to an additive constant \citep{blei2017variational}. For the variational family of distributions, $q(\blambda,\bbeta,\bgamma|\bkappa)$, in this paper, we focus on the mean-field inference although many researchers have also studied more complex families \citep[e.g.,][among others]{hoffman2013stochastic, ranganath2016hierarchical}.

Based on the illustrative Figure \ref{fig:dag}, we propose the following partition of the joint distribution of local and global, denominated mean field family \citep{parisi1988statistical}, 
\begin{equation}
    \label{eq:meanfield}
q(\blambda,\bbeta,\bgamma|\bkappa) = q(\blambda|\bkappa) q(\bbeta| \blambda,\bkappa) \prod_{i=1}^{n} q(\gamma_{i} |\bbeta,\bkappa)   
\end{equation}
where $\bkappa$ comprises all the parameters of the variational family. To avoid a cumbersome notation we are using the same notation $q$ for the joint variational distribution of (\ref{eq:elbo.1}) and the conditional distributions in (\ref{eq:meanfield}).

The approximate conditional inference is viewed as an optimisation problem. Given the above setup, the mean field family and the ELBO criterion, one can find the optimal solution via the coordinate ascent variational inference (CAVI) algorithm (Bishop, 2006). Each factor of the mean-field variational density is optimised iteratively, while keeping the others fixed, climbing the ELBO to a local optimum.

Letting $\bnu = (\blambda,\bbeta,\bgamma)$, we need to compute $q^*(\bnu | \bkappa) = q^*(\blambda | \bkappa) q^*(\bbeta|\lambda,\bkappa) \prod_{i=1}^{n} q^*(\gamma_{i} |\bbeta,\bkappa)$ where
\begin{eqnarray}
q^*(\blambda|\bkappa) &\propto& \exp\left\{ E_{(\bbeta,\bgamma)} \log \, p(\blambda|\bbeta,\bgamma,\mathbf{y},\bkappa)  \right\}, \label{eq:vb1} \\
q^*(\bbeta|\bkappa) &\propto& \exp\left\{ E_{(\blambda,\bgamma)} \log \, p(\bbeta|\blambda,\bgamma,\mathbf{y},\bkappa)  \right\}, \mbox{and} \label{eq:vb2} \\
q^*(\gamma_i|\bkappa) &\propto& \exp \left\{ E_{(\bgamma(-\gamma_i),\blambda,\bbeta)} \log \, p(\gamma_i|\bgamma(-\gamma_i),\blambda,\bbeta,\mathbf{y})  \right\}.  \label{eq:vb3}
\end{eqnarray}

It is worth it pointing out that the form of the optimal densities involves the full conditional distributions, revealing a link with Gibbs sampling. However, the VB algorithm does not repeatedly simulate from the full conditional distributions as  is done by the Gibbs sampler.

Alternative ways to maximize the ELBO are discussed in \cite{hoffman2013stochastic} and \cite{ranganath2014black}. They propose to calculate the ELBO gradient and use one of many alternative gradient ascent algorithms. Much effort has been done to take care of more general settings and developed generic algorithms for conjugate exponential-family models \citep{attias1999inferring, xing2003generalized}, leading to the automated variational
inference, allowing users to write down a model and immediately use variational inference to approximate its posterior distribution \citep{bishop2006pattern}. 

\section{Normal mixture regression model with functional covariates} \label{sec:mixnormal}
In this section, inspired by the dataset to be analysed in Section \ref{sec:early}, we deal with the mixture model of two normal distributions with different means but the same variance. Let $Y_1,\ldots,Y_n$, be independent random variables with 
\begin{equation}
\label{eq:normal}
    p(y_i|\mu_0,\mu_1,\sigma^2,\gamma_i) = \gamma_i \phi(y_i; \mu_{1}, \sigma^2) + (1-\gamma_i) \phi(y_i; \mu_{0}, \sigma^2) \mbox{, $\mu_1 > \mu_0$}
\end{equation}
where $\phi(.;\mu, \sigma^2)$ is the normal density with parameters $\mu$ and $\sigma^{2}$, and $\gamma_1,\ldots,\gamma_n$ are binary latent random variables, and let
\begin{equation*}
p(\gamma_i|\bbeta) \,=\, p_i(\bbeta)^{\gamma_i} (1-p_i(\bbeta))^{1-\gamma_i}, \quad \gamma_i = 0 \mbox{ or } 1
\end{equation*}
where
\begin{equation}
    \label{eq:pi}
p_i(\bbeta) = g^{-1}\left(\mathbf{x}_{i}^{\top} \bbeta \right),   
\end{equation}
$g$ is a link function, and $\bbeta = (\btheta,\bphi)$ with $\btheta$ being a vector of parameters associated to scalar effects and $\bphi$ a vector representing the coefficients of a function written from a $B$-splines expansion given by \eqref{eq:6} or \eqref{eq:6.1}.

For the parameters $\mu_0, \mu_1,$ and  $\sigma^2$ in the model, we use diffuse priors: 
\begin{itemize}
\item $\mu_0 \sim N(0,\tau_0^{2})$, $\mu_1 \sim N(0,\tau_1^{2})$ 
\item $\sigma^2 \sim$ inverse gamma$(a_0, b_0)$.
\end{itemize}
For each coefficient $\btheta, \bphi_1, \ldots, \bphi_J$, we specify weakly informative $t$ family of prior distributions with mean 0, degrees-of-freedom parameter $\nu$, and scale $s$, with $\nu$ and $s$ providing minimal prior information to constrain the coefficients to lie in a reasonable range (see Section 2 of \cite{gelman2008weakly}). The Gaussian distribution is obtained when $\nu \rightarrow \infty$, whereas the Cauchy distribution corresponds to $\nu=1$.

The observed data likelihood for the hierarchical model is difficult to optimize directly because the unobserved vector $\bgamma=\{\gamma_{i}\}_{i=1}^{n}$. Denoting by $\bTheta = (\mu_0,\mu_1,\sigma^2,\bbeta)$, the complete likelihood is given by
\begin{eqnarray}
f(\mathbf{y},\bgamma|\bTheta) &=& \prod_{i=1}^{n}   \frac{1}{\sqrt{2 \pi \sigma^{2}}} \mbox{exp} \left\{ -\frac{(y_i-\mu_0(1-\gamma_i)-\mu_1 \gamma_i)^2}{2 \, \sigma^2} \right \} \nonumber \\ &\times& [p_i(\bbeta)]^{\gamma_i} \, [1-p_i(\bbeta)]^{(1-\gamma_i)}   \mbox{.} \label{eq:7} 
\end{eqnarray}

\subsection{Full conditional posterior distributions} \label{sec:n_posteriors}

The joint augmented posterior distribution is proportional to the product of the likelihood given by \eqref{eq:7} and prior distributions specified in the previous section and has no closed form. Therefore, we adapt the Gibbs sampling algorithm to sample from the full conditional posterior distribution of  $\bTheta$ and the latent variables $\bgamma$.

\subsubsection{Full Conditional posterior distribution of $\gamma_i$} \label{sec:fullgamma}
Let $\bgamma_{-i}$ be the vector $\bgamma = (\gamma_1,\ldots,\gamma_{i-1},\gamma_{i+1},\ldots,\gamma_n)$. The full conditional posterior distribution of $\gamma_i$ is given by
\begin{eqnarray*}
 P(\gamma_i=1|\bTheta,\mathbf{y},\bgamma_{-i})  
&=& \frac{\phi\left(\frac{y_i-\mu_1}{\sigma}\right) p_i(\bbeta)}{\phi\left(\frac{y_i-\mu_1}{\sigma}\right) p_i(\bbeta)+\phi\left(\frac{y_i-\mu_0}{\sigma}\right) \left(1-p_i(\bbeta)\right)}
\end{eqnarray*}
since $P(\gamma_i = x,\bTheta,\mathbf{y},\bgamma_{-i}) = \phi\left(\frac{y_i-\mu_1 x - \mu_0 (1-x)}{\sigma}\right) p_i(\bbeta)$ where $p_i(\bbeta)$ depends on the covariates through the regression term and it is given by \eqref{eq:pi}. 
\subsubsection{Full Conditional posterior distribution of $\mu_0$ and $\mu_1$} \label{sec:fulleta}
We update $\mu_0$ using a normal distribution with mean 
$$ \left(\frac{1}{\tau_{0}^{2}}+\frac{1}{\sigma^2}\sum_{i=1}^{n} (1-\gamma_i)\right)^{-1}  \sum_{i=1}^{n} y_i (1-\gamma_i)$$
and variance
$$\left(\frac{1}{\tau_{0}^{2}}+\frac{1}{\sigma^2}\sum_{i=1}^{n} (1-\gamma_i)\right)^{-1},$$
while $\mu_1$ is updated, conditionally on $\mu_0$, with a truncated normal distribution on $(\mu_0,\infty)$, with mean 
$$ \left(\frac{1}{\tau_{1}^{2}}+\frac{1}{\sigma^2}\sum_{i=1}^{n} \gamma_i \right)^{-1}  \sum_{i=1}^{n} y_i \gamma_i$$
and variance
$$\left(\frac{1}{\tau_{1}^{2}}+\frac{1}{\sigma^2}\sum_{i=1}^{n} \gamma_i\right)^{-1}.$$

\subsubsection{Full Conditional posterior distribution of $\sigma^2$} \label{sec:fullsigma}
We update $\sigma^{2}$ using an inverse-gamma distribution with parameters $$a_0 + \frac{n}{2} \quad \mbox{and} \quad  b_0+\sum_{i=1}^{n} (y_i \gamma_i - \mu_1)^2 + (y_i (1-\gamma_i) - \mu_0)^2.$$
\subsubsection{Full Conditional posterior distribution of $\bbeta= (\btheta, \bphi_1, \ldots, \bphi_J)$} \label{sec:fullthetaphi}
Here we implement the computation of the full conditional posterior distribution of $(\btheta, \bphi_1, \ldots, \bphi_J)$ simultaneously, following \cite{gelman2008weakly}. We sample from the full conditional distribution of $\bbeta=(\btheta, \bphi_1, \ldots, \bphi_J)$ by assuming a $t$ prior distribution to each parameter $\beta_{d}$ in $\btheta$ and in $\bphi_j$, $j=1,\ldots,J$. However, instead of using a $t$ distribution directly, \cite{gelman2008weakly} assume $\beta_{d} \sim N(\mu_{d},\sigma^{2}_{d})$ and $\sigma^{2}_{d} \sim \mbox{Inv-}\chi^{2}(\nu_j,s^{2})$. The parameters $\beta_{d}$'s are treated as missing data and performing the EM algorithm, we estimate $\sigma^{2}_{d}$'s. The algorithm proceeds by alternating one step of iteratively weighted least squares to calculate the expectation of the logarithm of the full conditional posterior distribution using the estimate $\hat{\beta}_{j}$ and one step of the EM algorithm to calculate $\hat{\sigma}^{2}_d$ by maximization. Once enough iterations have been performed to reach approximate convergence, we get an estimate for the vector parameter $\bbeta = (\btheta, \bphi_1, \ldots, \bphi_J)$. This step is performed inside the Gibbs algorithm to sample from the conditional posterior distribution of $(\btheta, \bphi_1, \ldots, \bphi_J)$. To perform the calculations, we use the {\bf bayesglm} function implemented in R by \cite{gelman2008weakly}. To use the function {\bf bayesglm}, we specify the link function $g$ as either the probit or logit function, and inform the degrees of freedom $\nu$ and scale parameter $s$ as appropriate to consider a Normal, $t$ or Cauchy prior. 
\subsection{Variational Bayes for normal mixed model} \label{section:mixnormal}

For the mixture of normal distributions model, the  augmented vector of unknown parameters is 
 $\bnu = (\blambda,\bbeta,\bgamma),$ where $\blambda = (\mu_0,\mu_1,\sigma^{2})$,  $\bbeta=(\btheta,\bphi)$ and $\bgamma=(\gamma_1,\ldots,\gamma_n)$. Denote the parameters of the variational distributions as 
 $$\bkappa = \left(m_0,s_0^2,m_1,s_1^2,\balpha,A_0,B_0,\bmu^*_{b}, \mathbf{V}_{\bbeta}\right),$$ 
 and define 
 $q^*(\bnu|\bkappa) = q^*(\mu_0|\bkappa) q^*(\mu_1|\bkappa) q^*(\sigma^{2}|\bkappa) q^*(\bbeta|\bkappa) \prod_{i=1}^n q^*(\gamma_i|\bkappa)$. According to Equations \eqref{eq:vb1}, \eqref{eq:vb2} and \eqref{eq:vb3}, we 
have to calculate the variationals of $\mu_0, \mu_1, \sigma^{2},\bbeta$, and $\gamma_1, \ldots, \gamma_n$.

In the next sections, to simplify the notation, we will omit the dependence on $\bkappa$ when writing the variational distributions $q^*$. The details of computations can be found in Appendix \ref{ap:vb_normal}. 

\subsubsection{Variational density $q^*(\bgamma_i)$}\label{sec:vb_gamma}

 If we consider $q^*(\mu_0)$ and $q^*(\mu_1)$ belonging to the family of independent distributions with means $m_0$ and $m_1$ and variances $s_0^2$ and $s_1^2$, respectively, we get that $\gamma_i$ is a Bernoulli random variable with variational parameters for $\alpha_{i}$ given by
 $$\alpha_i = \frac{\alpha_{i1}}{\alpha_{i0}+\alpha_{i1}}$$
 where

\begin{eqnarray*} 
\alpha_{i0} &=& \exp\left\{E_{q^*(\btheta,\bphi)} \log \left[1-p_i(\bbeta,\bphi)\right] - E_{q^*(\sigma^{2})} \left( \frac{1}{2\sigma^{2}} \right) [(y_i-m_0)^{2} + s^{2}_{0}] \right\}. \label{eq:alphai0}
\end{eqnarray*}
and
\begin{eqnarray*}
\alpha_{i1} &=& \exp\left\{E_{q^*(\btheta,\bphi)} \log \left[p_i(\bbeta,\bphi)\right] - E_{q^*(\sigma^{2})} \left( \frac{1}{2\sigma^{2}} \right) [(y_i-m_1)^{2} + s^{2}_{1}] \right\}. \label{eq:alphai1}
\end{eqnarray*}

\subsubsection{Variational densities $q^*(\mu_0)$ and $q^*(\mu_1)$}

For $k = 0,1$, the variational distribution of $\mu_k$ is Gaussian with mean $m_k$ and variance $s_k^2$ given by

\begin{equation*}
     m_0 =  s_0^2 E_{q^*(\sigma^{2})} [1/\sigma^2] \sum_{i=1}^n (1-\alpha_i)y_i, \quad   s_0^2 =  \frac{1}{E_{q^*(\sigma^{2})} [1/\sigma^2] \sum_{i=1}^n (1-\alpha_i) +  1/\tau_0^2}
\end{equation*}
and 
\begin{equation*}
   m_1 =  s_1^2 E_{q^*(\sigma^{2})} [1/\sigma^2] \sum_{i=1}^n \alpha_iy_i, \quad 
     s_1^2 =  \frac{1}{E_{\sigma^{2}} [1/\sigma^2] \sum_{i=1}^n \alpha_i +  1/\tau_1^2},
\end{equation*}
where $\tau_0^2$ and $\tau_1^2$ are the parameters from the prior distribution.

\subsubsection{Variational density $q^*(\sigma^{2})$}

 The variational density of $\sigma^{2}$, considering the likelihood and the prior distribution of $\sigma^{2} \sim IG(a_0,b_0)$, is given by an inverse gamma with parameters
$A_0 = a_0+n/2$ and 
$$    B_0 = b_0 + \sum_{i=1}^n  \frac{\alpha_i}{2} \left((y_i - m_1)^2 + s_1^2\right)  + \sum_{i=1}^n  \frac{(1-\alpha_i)}{2} \left((y_i - m_0)^2 + s_0^2\right).$$

\subsubsection{Variational density $q^*(\bbeta_{l})$} \label{sec:VBbeta}

{We consider the full conditional posterior density of $\bbeta_{l},\bsigma$ as Equation (6) of \cite{gelman2008weakly} and derive the variational density of $\bbeta_{l}$ and $\bsigma$  as
\begin{eqnarray}
q^*(\bbeta_{l},\bsigma) &\propto& -\frac{1}{2} E_{q^*(\bgamma)} \left\{   (\bpsi_{l}-{\bf x}^{\top} \bbeta_{l})^{\top} \bSigma_{\psi}^{-1} (\bpsi_{l}-{\bf x}^{\top} \bbeta_{l}) \right. \nonumber \\
&\,\,\,\,\,\,\,\,\,\,\,\,\,\,\,\,\,\,\,\,\, \times&  \left. -\frac{1}{2} (\bbeta_{l} - \bmu_{\beta})^{\top} \bSigma_{\beta}^{-1} (\bbeta_{l} - \bmu_{l}) + \sum_{j} \log(\sigma_j)-p(\sigma_j|\nu_j,s_j)\right\} \label{fullbeta}
\end{eqnarray}
where $\bmu_{l}$ and $\bSigma_{\beta}$ are parameters of the prior $t$ distribution of $\bbeta_{l}$} in $\btheta$ or $\bphi_j$, $j=1, \ldots, J$, {  $\bSigma_{\psi} =\mathbf{W}_{l}^{-1}$ as in (\ref{wi})}, {and elements of $\bpsi_{l}$} are {given in (\ref{zi}). The expectation in (\ref{fullbeta}) is taken with respect to $\psi_{il}$ and is derived from
\begin{eqnarray}
E_{q*(\gamma_i)}(\psi_{il}) &=& g(\hat{p}_{ilk}) + (\alpha_i-\hat{p}_{ilk}) g{'}(\hat{p}_{ilk}) \label{eq:exppsi}
\end{eqnarray}
with $\alpha_{i}= E_{q^*(\gamma_i)}(\gamma_i) $ given in Section \ref{sec:vb_gamma}.
There is no closed form for $q^*(\bbeta_{l},\bsigma)$ and therefore we can not compute the ELBO.} However, when $\bsigma$ is known, the prior distribution of $\bbeta_{l}$ becomes a normal distribution. In that case, the variational $q^*(\bbeta_{l},\bsigma)=q^*(\bbeta_{l})$ is given by
\begin{eqnarray}
q^*(\bbeta_{l}) &\propto& -\frac{1}{2} E_{q^*(\bgamma)} \left\{ \sum_{i=1}^{n}  (\psi_{il}-{\bf x}_{i}^{\top} \bbeta_{l})^{\top} \bSigma_{\psi}^{-1} (\psi_{il}-{\bf x}_{i}^{\top}\bbeta_{l}) \right. \nonumber \\
&\,\,\,\,\,\,\,\,\,\,\,\,\,\,\,\,\,\,\,\,\,\times&  \left. -\frac{1}{2} (\bbeta_{l} - \bmu_{\beta})^{\top} \bSigma_{\beta}^{-1} (\bbeta_{l} - \bmu_{\beta})\right\} \label{fullbeta2}
\end{eqnarray}
Rearranging terms in $(\ref{fullbeta2})$, $q^*(\bbeta_{l})$ is given by a normal distribution with covariance matrix $\mathbf{V}_{\beta}^{*} = ({\bf x}^{\top} \bSigma_{\psi}^{-1} {\bf x}+\bSigma_{\beta}^{-1})^{-1}$ and mean $\bmu^{*}_{\beta} = \mathbf{V}_{\beta}^{*} ({\bf x}^{\top} \bSigma_{\psi}^{-1} E_{q^{*}(\gamma)}(\bpsi_{l})+\bSigma_{\beta}^{-1} \bmu_{\beta})$, with elements in $E_{q^{*}(\gamma)}(\bpsi_{l})$ given by (\ref{eq:exppsi}).

\subsubsection{Calculating the ELBO}

The ELBO is given by
\begin{eqnarray*}
\mbox{ELBO}(\bkappa) &=&  
\sum_{i=1}^{n} E_{q^*}[\log p(y_i|\gamma_i,\mu_0,\mu_1,\sigma^{2}, \bbeta)] + \sum_{i=1}^{n} E_{q^*}[\log p(\gamma_i)] \nonumber \\
&&\;\;\;+  E_{q^*}[\log p(\mu_0)] + E_{q^*}[\log p(\mu_1)]+ E_{q^*}[\log p(\sigma^{2})] + E_{q^*}[\log p(\bbeta)] \nonumber \\
&&\;\;\;- \sum_{i=1}^{n} (E_{q^*}[\log q^*(\gamma_i)] -  E_{q^*}[\log q^*(\mu_0)] -  E_{q^*}[\log q^*(\mu_1)]\nonumber \\ 
&&\;\;\;- E_{q^*}[\log q^*(\sigma^{2})] - E_{q^*}[\log q^*(\bbeta)] \nonumber \\
& = & E_0 + E_1 + E_2 + E_3 + E_4 + E_5  - F_1 - F_2 - F_3 - F_4 - F_5
\end{eqnarray*}
with the expectation is taken with respect to $q^*(\mu_0,\mu_1,\sigma^{2},\bgamma,\bbeta|\bkappa)$, that is $E_{q^*} := E_{q^*(\mu_0,\mu_1,\sigma^{2},\bgamma,\bbeta|\bkappa)}$. Recall that $q^*(\mu_0,\mu_1,\sigma^{2},\bgamma,\bbeta) = q^*(\mu_0)q^*(\mu_1)q^*(\sigma^{2})q^*(\bgamma)q^*(\bbeta)$. Let $\Psi(\cdot)$ denote the digamma function. Explicit computations are given in Appendix \ref{ap:elbo_normal}. \\

Therefore, to compute the ELBO, we need the following pieces:

\begin{align*}
E_0  = & 
 -\frac{n}{2} \log 2\pi -\frac{1}{2} \frac{A_0}{B_0} \sum_{i=1}^{n} \{\alpha_i [(y_i-m_1)^{2} + s_{1}^{2}] + (1-\alpha_i) [(y_i-m_0)^{2} + s_{0}^{2}]  \} \\ 
 & \quad + \frac{n}{2} \left(\log(B_0) - \Psi(A_0)\right), \\
    E_1 =& \sum_{i=1}^n \left[\alpha_i \int \log g^{-1}(\mathbf{x}_{i} \bbeta) q^*(\bbeta) \, d\bbeta] + (1-\alpha_i) \int \log [1-g^{-1}(\mathbf{x}_{i} \bbeta)] q^*(\bbeta) \, d\bbeta \right], \nonumber \\
E_2   = &  -\frac{1}{2} \log (2\pi  \tau^{2}_{0})  -\frac{1}{2\tau^{2}_{0}} (m_{0}^{2} + s^{2}_{0}), \nonumber \\
E_3   = &  -\frac{1}{2} \log (2\pi  \tau^{2}_{1})  -\frac{1}{2\tau^{2}_{1}} (m_{1}^{2} + s^{2}_{1}), \nonumber \\
E_4  = & a_0 \log b_0 - \log(\Gamma(a_0)) + (a_0 + 1)  \log\left(B_0 - \Psi(a_0 + n/2) \right)  - b_0 A_0 B_0^{-1}, \mbox{ and }\nonumber \\
E_5  = &  - \frac{R}{2} \log 2 \pi - \frac{1}{2} \log |\Sigma_{\bbeta}|  - \frac{1}{2} tr(\Sigma_{\bbeta}^{-1} \Sigma_{q^*(\bbeta)}) - \frac{1}{2} [(\bmu^* - \bmu_{\bbeta})^{\top} \Sigma_{\bbeta}^{-1} (\bmu^* - \bmu_{\bbeta})] \nonumber 
\end{align*}
where $R$ is the dimension of the $\bbeta$ vector and  $q^*(\bbeta)$ given by \eqref{fullbeta2}. The high dimensional integral in $E_1$ can be computed efficiently transforming it into a one-dimensional integral as described in Appendix \ref{ap:fast}. \\

On the other hand,

\begin{align*}
F_1  = & \sum_{i=1}^n \left(\alpha_i \log \alpha_i + (1-\alpha_i) \log (1 - \alpha_i)\right), \\
F_2  = &  -\frac{1}{2} \log 2\pi - \frac{1}{2}- \frac{1}{2} \log s^2_{0}\left(1/2s^{2}_{0}\right), \nonumber \\ 
F_3  = &  -\frac{1}{2} \log 2\pi - \frac{1}{2}- \frac{1}{2} \log s^2_{1}(1/2s^{2}_{1}), \nonumber \\ 
F_4  = & A_0 \log B_0 - \log \Gamma(A_0) + (A_0 + 1) \left(\log(B_0) - \Psi(A_0)\right)  - A_0, \mbox{ and} \\
F_5  =  
& - \frac{R}{2} \log 2 \pi - \frac{1}{2} \log |\Sigma_{q^*(\bbeta)}|  - \frac{R}{2}.
\end{align*}

\section{Zero Inflated mixture of Poisson regression model} \label{sec:zimp}
 
In this section we will analyse the case where the observed sample is given by  $Y_1, Y_2, \ldots, Y_n$, independent non-negative integer-valued random variables, such that
\begin{eqnarray}
P(Y_i=y|\lambda_1,\lambda_2,\bgamma_i) \,=\, [I(y_i=0)]^{\gamma_{i0}} \left[ \frac{1}{y_i!} e^{-\lambda_1} \lambda_{1}^{y_i}\right]^{\gamma_{i1}} \left[ \frac{1}{y_i!} e^{-\lambda_2} \lambda_{2}^{y_i}\right]^{\gamma_{i2}} \label{eq:zimp}
\end{eqnarray}
for $y_i=0, 1, \ldots$ where $\bgamma_i = (\gamma_{i0},\gamma_{i1},\gamma_{i2})$ are latent multinomial random variables  Multinomial$(1,p_{i0},p_{i1},p_{i2})$ with 
$$p_{i0} = \frac{1}{1 + \exp({\bf x}_i^{\top} \bbeta_1) + \exp({\bf x}_i^{\top} \bbeta_2)}, \quad p_{il} =  \frac{\exp({\bf x}_i^{\top} \bbeta_l)}{1 + \exp({\bf x}_i^{\top} \bbeta_1) + \exp({\bf x}_i^{\top} \bbeta_2)}, \,\,\, {l=1, 2}.$$ 
Therefore, the vector of unknowns is $\bnu = (\lambda_1,\lambda_2,\bbeta_1,\bbeta_2,\bgamma_1,\bgamma_2) = (\bTheta,\bgamma).$ 

For the model parameters we propose using the following priors: 
\begin{itemize}
\item $\lambda_1 \sim \mbox{Gamma}(a_1,b_1)$, $\lambda_2 \sim \mbox{Gamma}(a_2,b_2)$;
\item $\bbeta_{l}=$($\btheta_{l}, \bphi_{1l}, \ldots, \bphi_{Jl})$ will have weakly informative $t$ family of prior distributions as stated in Section \ref{sec:fullthetaphi}.
\end{itemize}

The observed data likelihood for the hierarchical model is difficult to optimize directly because the unobserved vector $\bgamma=\{\gamma_{il}, i=1, \ldots, n, l=0,1,2\}$. Therefore, we consider the complete likelihood given by
\begin{eqnarray}
f(\mathbf{y},\bgamma|\bTheta) &=& \prod_{i=1}^{n}  \left[ p_{i0} I(y_i=0) \right]^{\gamma_{i0}}  \left[\frac{1}{y_i!} p_{i1} e^{-\lambda_{1}} \lambda_{1}^{y_i}\right]^{\gamma_{i1}} \left[\frac{1}{y_i!} p_{i2} e^{\lambda_{2}} \lambda_{2}^{y_i}\right]^{\gamma_{i2}}  \mbox{.} 
 \label{eq:7.1} 
\end{eqnarray}

The joint augmented posterior distribution is the product of the likelihood and priors specified above and has no closed form. We adapt the Gibbs sampling algorithm to sample from the posterior distribution of  $\bTheta$ and the latent variables $\bgamma$. 

\subsection{Full conditional posterior distributions} \label{sec:zip_posteriors}
The posterior distribution of parameters is obtained based on a Gibbs sampling scheme. For that, we calculate the full conditional posterior distribution of parameters in $\bTheta$, similarly to described in Section \ref{sec:n_posteriors}.

\subsubsection{Full Conditional posterior distribution of $\gamma_i$} 
Let $\bgamma_{-i0}$, $\bgamma_{-i1}$ and $\bgamma_{-i2}$ be the vector $\bgamma_{l} = (\bgamma_{1l},\ldots,\bgamma_{nl})$ without observation $\bgamma_{il}$, $l=0,1,2$, respectively. The full conditional posterior distribution of $\bgamma_i$ for $l=1,2$ is given by
\begin{eqnarray}
P(\gamma_{il}=1|\bTheta,\mathbf{y},\bgamma_{-il}) &=& \hspace{-2mm}
 \frac{e^{-\lambda_{l}} \lambda_{l}^{y_i} \exp({\bf x}_i^{\top} \bbeta_l)}{I(y_i=0) + e^{-\lambda_{1}} \lambda_{1}^{y_i} \exp({\bf x}_i^{\top} \bbeta_1) + e^{-\lambda_{2}} \lambda_{2}^{y_i} \exp({\bf x}_i^{\top} \bbeta_2) }, 
\end{eqnarray}
for $l=1,2$ and
\begin{eqnarray}
P(\gamma_{i0}=1|\bTheta,\mathbf{y},\bgamma_{-i0}) &=&  \hspace{-2mm}
 \frac{I(y_i=0)}{I(y_i=0) + e^{-\lambda_{1}} \lambda_{1}^{y_i} \exp({\bf x}_i^{\top} \bbeta_1) + e^{-\lambda_{2}} \lambda_{2}^{y_i} \exp({\bf x}_i^{\top} \bbeta_2) }.
\end{eqnarray}

 \subsubsection{Full Conditional posterior distribution of $\lambda_{1}$ and $\lambda_{2}$}

We update $\lambda_{l}, l=1,2$ using a gamma distribution with parameters
$$ a_l + \sum_{i=1}^n y_i \gamma_{il} \quad \mbox{and} \quad  b_l + \sum_{i=1}^n \gamma_{il}, \; l=1,2.$$

\subsubsection{Full Conditional posterior distribution of $\bbeta_{l}= (\btheta_{l}, \bphi_{1l}, \ldots, \bphi_{Jl}), l=1,2$} 

These computations are exactly the same as the ones described in Section \ref{sec:fullthetaphi}.

\subsection{Variational Bayes of the ZIMP model}  \label{section:zimp}

Analogously to the normal case, we define the variational densities as $$q^*(\bnu|\bkappa) = q^*(\lambda_1|\bkappa) q^*(\lambda_2|\bkappa) q(\bbeta_1|\bkappa) q(\bbeta_2|\bkappa) q(\gamma|\bkappa)$$
where $\bkappa = (\balpha,\psi_1,\zeta_1,\psi_2,\zeta_2,\bmu^*_{\bbeta_1}, \bmu^*_{\bbeta_2}, \mathbf{V}_{\bbeta_1}, \mathbf{V}_{\bbeta_2})$ is the vector of variational parameters. For all the cases, the variational densities $q^*(\bbeta_1|\bkappa)$ and $q^*(\bbeta_2|\bkappa)$  will have exactly the same computations as in the normal case, see Section \ref{sec:VBbeta}.  The  vector of unknowns is 
$\bnu = (\lambda_1,\lambda_2,\bbeta_1, \bbeta_2, \bgamma) = (\bTheta,\bgamma),$
where $\bgamma=(\bgamma_1,\ldots,\bgamma_n)$.

Again, to simplify the notation, we will omit the dependence on $\bkappa$ when writing the variational distributions $q^*$. The details of computations can be found in Appendix \ref{ap:vb_zip}.

\subsubsection{Variational density $q^*(\bgamma_i)$}\label{sec:vb_gamma_p1}

 If we consider $q^*(\lambda_1)$ and $q^*(\lambda_2)$ belonging to the gamma family of  distributions with parameters $(\psi_1,\zeta_1)$ and $(\psi_2,\zeta_2)$ respectively,  we get $\bgamma_i$ is a multinomial random variable Multinomial$(1,\alpha_{i0},\alpha_{i1},\alpha_{i2})$ with
 $$\alpha_{il} = \frac{\rho_{il}}{\sum_{j=0}^2 \rho_{ij}}, \quad l=0,1,2$$
 where
\begin{eqnarray*}
    \rho_{i0} &=& I(y_i=0),  \\
    \rho_{i1} &=& \exp\left(-\frac{\psi1}{\zeta1} +  y_i (-\log(\zeta1) + \Psi(\psi1)) +  E_{q^*(\bbeta_1)}[{\bf x}_{i}^{\top} \bbeta_1 \right),  \quad \mbox{and} \\
    \rho_{i2} &=& \exp\left(-\frac{\psi1}{\zeta1} +  y_i (-\log(\zeta2) + \Psi(\psi2)) +  E_{q^*(\bbeta_2)}[{\bf x}_{i}^{\top} \bbeta_2 \right). 
\end{eqnarray*}
\vspace{-3cm}
\subsubsection{Variational density $q^*(\lambda_1)$ and $q^*(\lambda_2)$}

The variational distribution of  $\lambda_l, \;l=1,2$
is gamma density with parameters 
\begin{equation}
    \label{eq:zip_psi_zeta_1}
    \psi_1:= a_1 + \left(\sum_{i=1}^n  \alpha_{i1} y_i \right) \quad \mbox{and} \quad  \zeta_1:= b_1 + \sum_{i=1}^n  \alpha_{i1}
\end{equation}
and
\begin{equation}
    \label{eq:zip_psi_zeta_2}
    \psi_2:= a_2 + \left(\sum_{i=1}^n \alpha_{i2} y_i \right) \quad \mbox{and} \quad  \zeta_2:= b_2 + \sum_{i=1}^n    \alpha_{i2},
\end{equation}
respectively.

\subsection{Calculating the ELBO}

The ELBO is given by
\begin{eqnarray}
\mbox{ELBO}(\bkappa) &=&  
\sum_{i=1}^{n} E_{q^*}[\log p(y_i|\gamma_i,\lambda_{1},\lambda_{2})] + \sum_{i=1}^{n}  E_{q^*}[\log p(\bgamma_i)] \nonumber \\
&& \;\; + \, E_{q^*}[\log p(\lambda_{1})]  +  E_{q^*}[\log p(\lambda_{2})]  + E_{q^*}[\log p(\bbeta_1)]  +  E_{q^*}[\log p(\bbeta_2)]\nonumber \\
&& \;\; - \, \sum_{i=1}^{n} (E_{q^*}[\log q^*(\bgamma_i)] - E_{q^*}[\log q^*(\lambda_{1})] -  E_{q^*}[\log q^*(\lambda_{2})] \nonumber \\
&& \;\;-  E_{q^*}[\log q^*(\bbeta_1)] -  E_{q^*}[\log q^*(\bbeta_2)] \nonumber \\
&=& E_0+ E_1 + E_2 + E_3 + E_4 + E_5- F_1 - F_2 - F_3 -F_4 - F_5,
\end{eqnarray}
with the expectation taken with respect to $q^*(\lambda_1,\lambda_2,\bgamma,\bbeta|\bkappa)$, that is \\$E_{q^*} := E_{q^*(\lambda_1,\lambda_2,\bgamma,\bbeta|\bkappa)}$. Therefore, as shown in Appendix \ref{ap:elbo_zip} we have 
\begin{eqnarray*}
E_0 &=& \sum_{i=1}^{n}  \alpha_{i0}I(y_i=0) + \left\{ (-\log y_i!)  + \alpha_{i1} \left(  -  \frac{\psi_{1}}{\zeta_{1}}  + y_i \left( -\log(\zeta_{1}) + \Psi(\psi_{1})\right) \right) \right.\nonumber \\
& & \quad + \left. \alpha_{i2} \left( -\frac{ \psi_{2}}{\zeta_{2}}  + y_i \left(- \log(\zeta_{2}) + \Psi(\psi_{2}) \right) \right) \right\} \nonumber \\
E_1 &=&  \sum_{i=1}^n  \alpha_{i0} I(y_i=0) + \alpha_{i1} {\bf x'}_{i} \mu^*_{\bbeta_1} + \alpha_{i2} {\bf x'}_{i} \mu^*_{\bbeta_2} \nonumber \\
&& \quad - \int \log[1 + \exp({\bf x'}_{i} \bbeta_1) + \exp({\bf x'}_{i} \bbeta_2))] q^*(\bbeta_1) q^*(\bbeta_2) \, d\bbeta_1 d\bbeta_2 \nonumber \\
E_2 &=&  -  \log(\Gamma(a_1)) +  a_1 \log(b_1) + (a_1 -1)  (-\log(\zeta_{1}) +  \Psi(\psi_{1})) -b_1 \frac{\psi_{1}}{\zeta_{1}}  \nonumber \\
E_3 &=&  -  \log(\Gamma(a_2)) +  a_2 \log(b_2) + (a_2 -1)  (-\log(\zeta_{2}) +  \Psi(\psi_{2})) -b_2 \frac{\psi_{2}}{\zeta_{2}}, \nonumber \\
E_4 &=& - \frac{R}{2} \log 2 \pi - \frac{1}{2} \log |\Sigma_{\bbeta_1}|  - (1/2) [(\mu^{*}_{\bbeta_1} - \bmu_{\bbeta_1})^{\top} \Sigma_{\bbeta_1}^{-1} (\mu^{*}_{\bbeta_1} - \bmu_{\bbeta_1})] \nonumber \\
E_5 &=& - \frac{R}{2} \log 2 \pi - \frac{1}{2} \log |\Sigma_{\bbeta_2}|  - (1/2) [(\mu^{*}_{\bbeta_2} - \bmu_{\bbeta_2})^{\top} \Sigma_{\bbeta_2}^{-1} (\mu^{*}_{\bbeta_2} - \bmu_{\bbeta_2})]
\end{eqnarray*}
where $\psi_{1}, \zeta_{1}, \psi_{2}$ and $\zeta_{2}$ are given by \eqref{eq:zip_psi_zeta_1} and \eqref{eq:zip_psi_zeta_2}
and $\Psi$ is the digamma function and $R$ is the dimension of the $\bbeta_1$ (and $\bbeta_2$) vector.  The high dimensional integral in $E_1$ can be computed efficiently transforming it into a two-dimensional integral as described in Appendix \ref{ap:fast2}.

On the other hand,
\begin{eqnarray*}
F_1 &=& \sum_{i=1}^n \alpha_{i0} \log \alpha_{i0} I(y_i=0) + \alpha_{i1} \log \alpha_{i1} + \alpha_{i2} \log  \alpha_{i2}, \nonumber \\
F_2 &=&  -\log \Gamma(\psi_{1}) + \psi_{1} \log(\zeta_{1}) + (\psi_{1} - 1)(-\log(\zeta_{1}) +\Psi(\psi_{1}))  - \psi_{1}/\zeta_1 , \nonumber \\
F_3 &=&   - \log\Gamma(\psi_{2}) + \psi_{2} \log(\zeta_{2}) + (\psi_{2} - 1)(-\log(\zeta_{2}) +\Psi(\psi_{2}))  - \psi_{2}/\zeta_2 , \nonumber \\
F_4 &=& - \frac{R}{2} \log 2 \pi - \frac{1}{2} \log |\Sigma_{q^*(\bbeta_1)}|  - \frac{R}{2}, \nonumber \\
F_5 &=& - \frac{R}{2} \log 2 \pi - \frac{1}{2} \log |\Sigma_{q^*(\bbeta_2)}|  - \frac{R}{2}.
\end{eqnarray*}

\section{Simulations}\label{sec:simulations}

In this section, the primary goal is to examine the performance of the proposed model considering aspects of sample size and discrimination ability of the functional curves by considering the normal mixture model (\ref{eq:normal}) and the ZIMP model (\ref{eq:zimp}), and compare those aspects under the MCMC and VB estimation methods. We consider two simulation studies, given in Sections \ref{sec:sce1} and  \ref{sec:sce2}, respectively, each of which consists of three steps. In the first step, for each subject $i$, we generated the functional covariates $X_{i1}$  with domain $T$ and $X_{i2}$  with domain in $S$ respectively, and using linear weights $w_1(t), w_2(t), w_1^{'}(t), w_2^{'}(t)$, along with a logit model with $\log(p_{il}/p_{i1})$, we calculated
\begin{eqnarray}
p_{i1} =  \frac{\mbox{exp} \left(\sum_{t \in T} w_1(t) X_{i,1}(t) + \sum_{s \in S} w_2(s) X_{i,2}(s) \right)}{1 + \mbox{exp} \left(\sum_{t \in T} w_1(t) X_{i,1}(t) + \sum_{s \in S} w_2(s) X_{i,2}(s) \right)} \label{eq:p12}
\end{eqnarray}
and 
\begin{eqnarray}
p_{i2} = \frac{\mbox{exp} \left(\sum_{t \in T} w'_1(t) X_{i,1}(t) + \sum_{s \in S} w'_2(s) X_{i,2}(s) \right)}{1 + \mbox{exp} \left(\sum_{t \in T} w'_1(t) X_{i,1}(t) + \sum_{s \in S} w'_2(s) X_{i,2}(s) \right)} \mbox{.} \label{eq:p13}
\end{eqnarray}
Further, we sampled independent $\gamma_{i} \sim Ber(p_{i1})$, $p_{i1} = 1-p_{i2}$ to generate the response variable $Y_i$ from a normal mixture model as
\begin{eqnarray}
  Y_i &=& 9 \gamma_{i1} + \epsilon _i. \label{simdatanorm}
\end{eqnarray}

To sample from the ZIMP model (\ref{eq:zimp}) we sampled independent $\bgamma_{i} \sim \mbox{multinomial}(p_{i0},p_{i1},p_{i2})$, with $p_{i0} = 1-p_{i1}-p_{i2}$ and, given the generated $\bgamma_{i}$, we generated the response variable $Z_i$ as
\begin{eqnarray}
  Z_i =  \delta(0) \gamma_{i0} + \mbox{Poisson}(\lambda_1) \gamma_{i1} + \mbox{Poisson}(\lambda_2) \gamma_{i2} \label{simdatazimp}
\end{eqnarray}
where $\delta(0)$ represents a distribution with point mass at zero. Additionally, $0 < \lambda_1 < \lambda_2$. 

Generated datasets were analysed using the MCMC and VB inference approaches.  Later on, we fit models described in Sections \ref{sec:mixnormal} and \ref{sec:zimp} to the generated data described in (\ref{simdatanorm}) and (\ref{simdatazimp}), respectively. For the MCMC we considered 15,000 iterations with a burn-in of 10,000 and we collected samples every 100th step. Posterior mean parameters are considered as estimates in the MCMC method, while in the VB method posterior mean is replaced by the expectation under the variational distribution with the fitted parameters in place of the posterior.

To compare the MCMC and VB methods within each class of model, we computed the mean squared error
$$\mbox{MSE} = \frac{1}{M n} \sum_{M=1}^{100} \sum_{i=1}^{n} \left[ \frac{1}{L} \sum_{l=1}^{L}(p_{il} - \hat{p}_{il})^2 \right]$$
with $\hat{p}_{il}$ being the estimate of $p_{il}$ under MCMC or VB approach. 

To assess the ability of classification of the mixture normal model, we consider the misclassification rate  at the $p$-percentile 
$$ \mbox{MR}_p = \frac{1}{M} \sum_{M=1}^{100} \frac{1}{n}\sum_{i=1}^{n}  \left[I(\gamma_{i2}=1,\hat{\gamma}_{ip}=0) + I(\gamma_{i2}=0,\hat{\gamma}_{ip}=1)\right],$$
where $\hat{\gamma}_{ip} = I(\hat{\pi}_{i2} > p)$, $p=0.5, 0.75, 0.9$. For the ZIMP model, {we weighted the misclassification between the pure zero class and class 3 (largest mean value) twice as large as the misclassification between the pure zero class and class 2). Let
$\hat{\gamma_{il}} = I( \hat{\pi}_{il} = \max_{\ell} \hat{\pi}_{i\ell})$ and define $L_i = \sum_{\ell=0}^{2} \ell I(\gamma_{i\ell} = 1)$ and $\hat{L}_i = \sum_{\ell=0}^{2} \ell I(\hat{\gamma}_{i\ell} = 1)$ as the true and estimated class labels for subject $i$. The misclassification rate is computed as }
{$$ \mbox{MR} = \frac{1}{M} \sum_{M=1}^{100} \frac{1}{n}\sum_{i=1}^{n} |L_i - \hat{L}_i|.$$}

The secondary goal of the simulation study was to assess the time performance of the MCMC and VB methods considered in Study 1 and Study 2, described in Sections \ref{sec:sce1} and \ref{sec:sce2}, respectively. Simulations were run in laptop using a Intel(R) Core(TM) processor with 8192MB RAM memory.

\subsection{Study 1: segregating functional covariates} \label{sec:sce1}

In this study, the goal is to assess the ability of cluster discrimination when using segregating functional covariates. We generated 100 datasets $\{{\cal D}_{r}^{500}, r=1, \ldots, 100\}$ each consisting of $n=500$ subjects. From this baseline data, set we constructed subsets of 100 and 300 subjects $\{({\cal D}_{r}^{100}, {\cal D}_r^{300}), r=1, \ldots, 100\}$, sampled by random, in such way that ${\cal D}_r^{100} \subset {\cal D}_r^{300} \subset {\cal D}_r^{500}$, for $r = 1, \ldots, 100$.

\begin{figure}[!htb]
\begin{minipage}{0.4\textwidth}
\centering
\includegraphics[height=4cm,width=\textwidth]{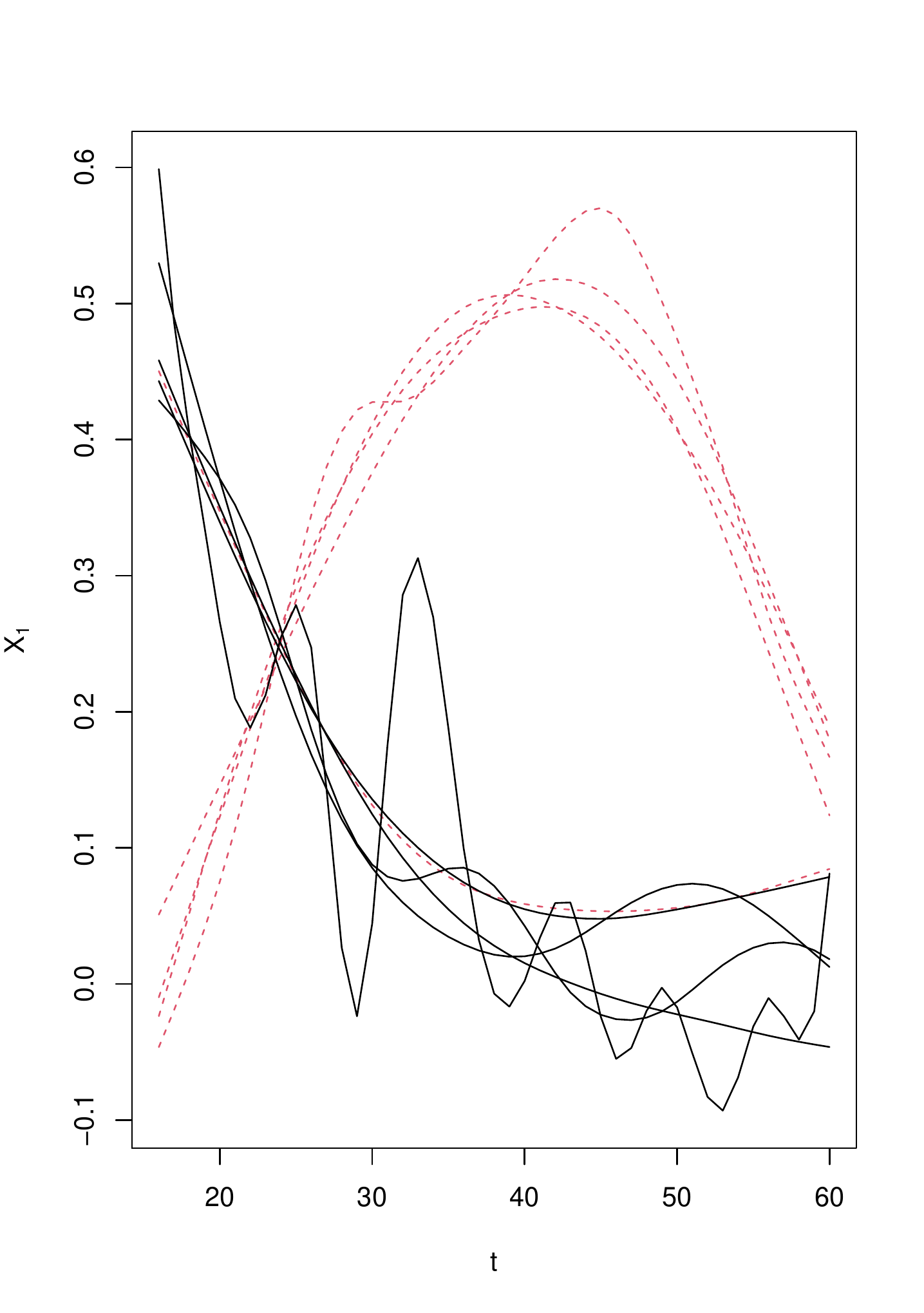}
\end{minipage} 
\begin{minipage}{0.4\textwidth}
\centering
\includegraphics[height=4cm,width=\textwidth]{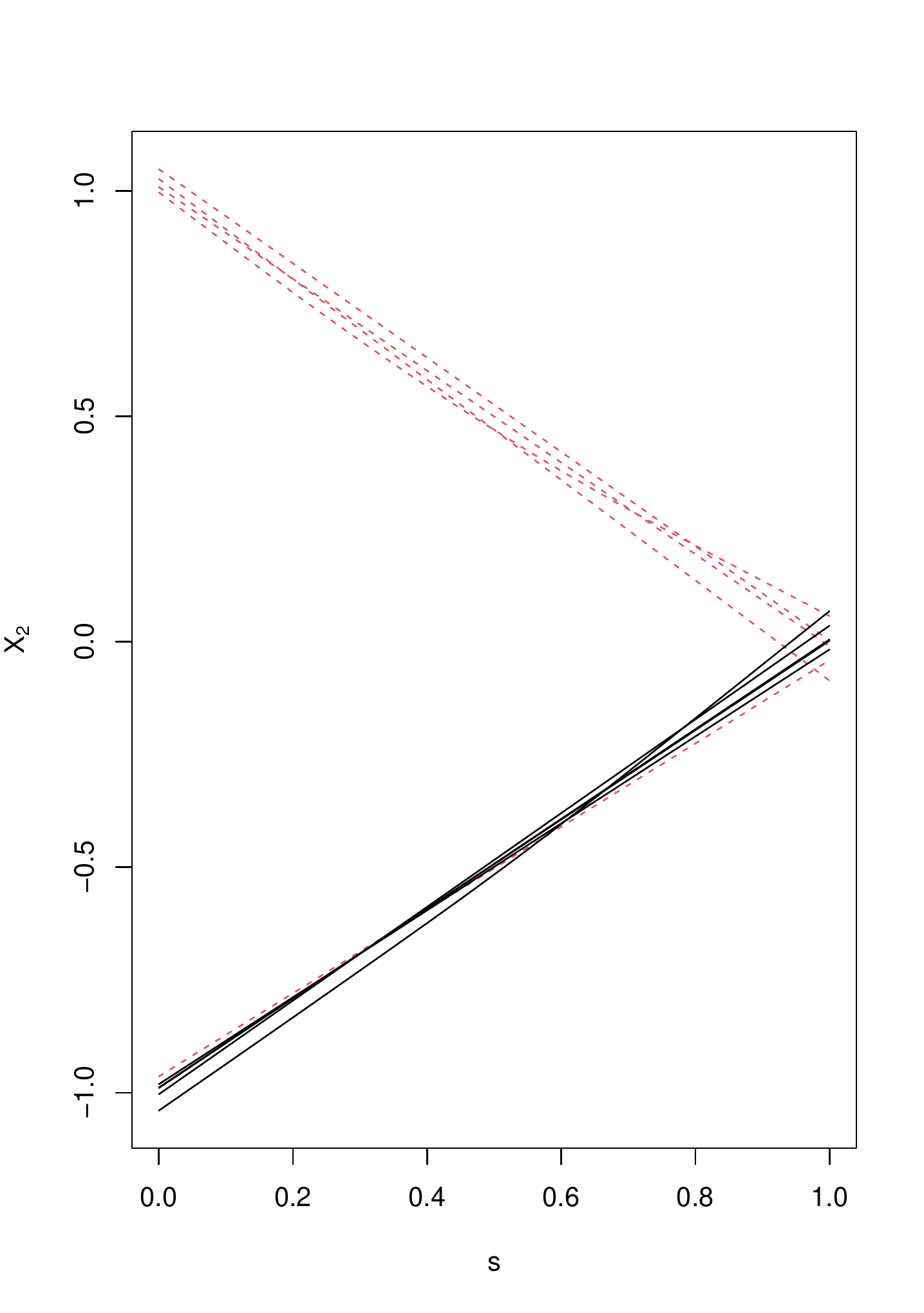}
\end{minipage} 
\begin{minipage}{0.4\textwidth}
\centering
\includegraphics[height=4cm,width=\textwidth]{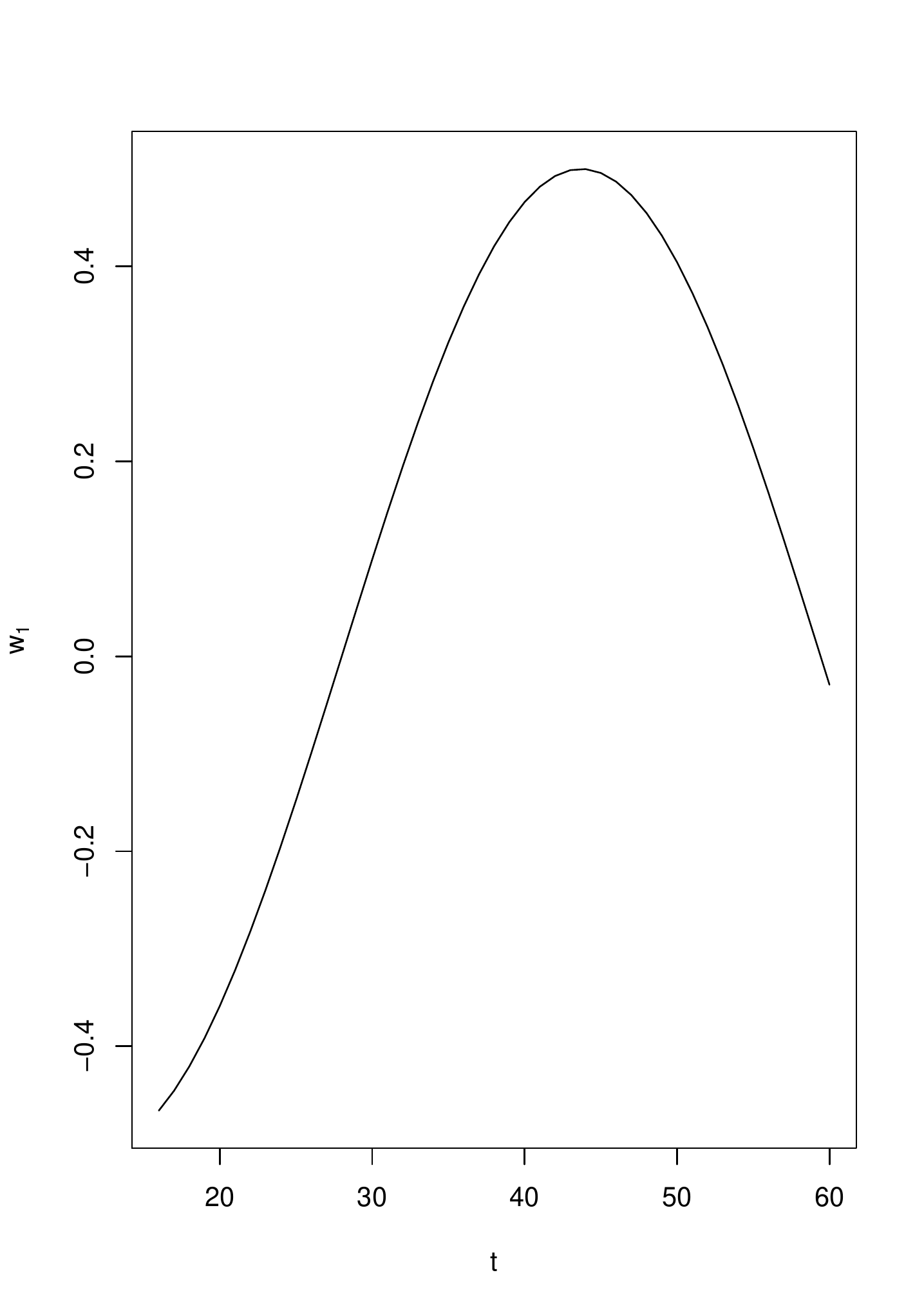}
\end{minipage} 
\begin{minipage}{0.4\textwidth}
\centering
\includegraphics[height=4cm,width=\textwidth]{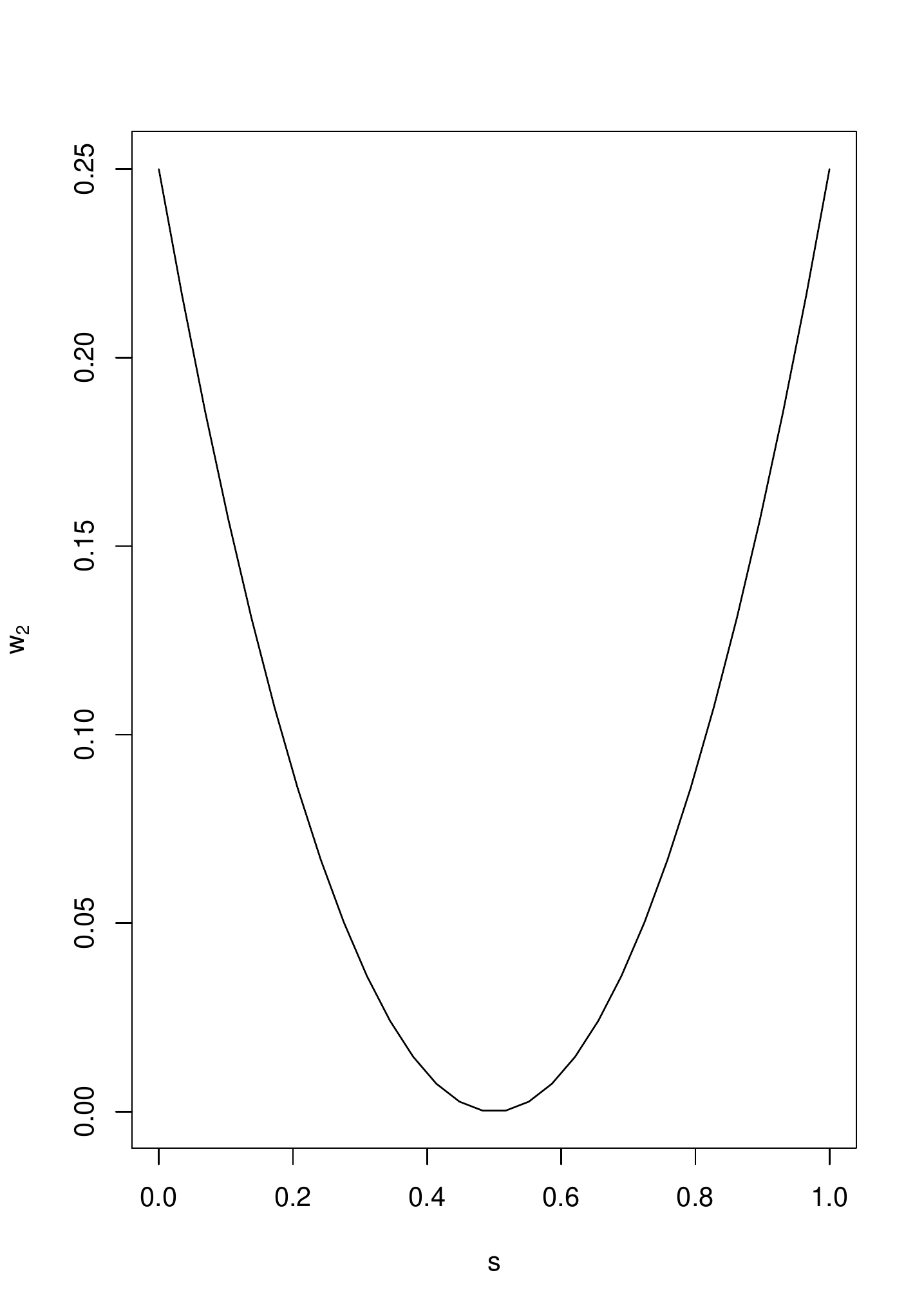}
\end{minipage} 
\caption{Functional covariates $X_{i,1}(\cdot)$ (top left panel) and $X_{i,2}$ (top right panel) for random sample of 10 subjects, $i=1, \ldots, 10$, and weight functions $w_{1}(\cdot)$ {(bottom left panel)} and $w_{2}(\cdot)$ {(bottom right panel)}, for Study 1. Red (traced) and black (solid) lines correspond to {subjects belonging to class 1 and 0, for the normal case, and Class 1 and Class 2 for the ZIMP case, respectively.}} \label{fig:simula1}
\end{figure}

For the baseline dataset, for each subject $i = 1,\ldots, 500$, we simulated the functional covariates $X_{i,1}$ and $X_{i,2}$ in a manner such that subjects cluster segregation was mainly due to the segregating characteristic of the curves, as it can be seen on the top panels of Figure \ref{fig:simula1}.
Notice that the shape of the ten sampled subject's functional covariates discriminates subjects between the two classes. It may happens a subject is eventually misclassified, as it is the case of the red curves among the black curves on the top panels of Figure \ref{fig:simula1}, but it was taken care to be rare event in this Study 1. 

We considered $T=[16, 60]$ and $S=[0,1]$ and the time points $\{t_j \in [16, 60],j=1, 2,\ldots, 45\}$ and $\{s_j \in [0, 1],j=1, 2,\ldots, 30\}$. 

The functional predictors $X_{i,1}$ and $X_{i,2}$  along with functional weights $w_{1}(t), w_{1}^{'}(t) = 0.5 w_{1}(t), w_{2}(t)$ and $w_{2}^{'}(s) = -0.25 w_{2}(s)$ were used to model mixture probabilities as described in (\ref{eq:p12}) and (\ref{eq:p13}). We simulated from the normal mixture model considering (\ref{simdatanorm}) with $\epsilon_i \sim$ iid $N(0,18)$, leading to a mixture of a normal distribution with mean $\mu_0=0$ and another with mean $\mu_1=9$, both having variance $\sigma^2=18$. Further, we simulated from the ZIMP model considering (\ref{simdatazimp}) with $\lambda_1 = 2$ and $\lambda_2 = 10$. 

For all datasets, we ran the MCMC for 15,000 iterations with a burn-in of 10,000, sampled every 100th step. Using MCMC samples of normal mixed model location parameters, we calculated the average of lower and upper bound of 95\% high density posterior (HPD) intervals across the 100 simulated datasets and results are given in Table \ref{table:ICs} along with the 2.5\% and 95\% quantiles calculated from the VB results across the same 100 data, for the same parameters. Results are given according to sample size $n$. Overall, the intervals indicate a good performance as they include their respective true parameter value. Exceptions are the intervals of $\mu_0$ and $\mu_1$ for $n=300$ and $n=500$, as the respective MCMC's intervals does not include the true parameter value and neither that of $n=500$ for the VB method.  

\begin{table}[h!]
\caption{Interval for normal mixed model and ZIMP location parameters, logit link, $n = 100, 300, 500$, and MCMC and VB inference methods in Study 1.} \label{table:ICs}
\begin{tabular}{l|ccccc} \hline 
 &   & \multicolumn{2}{c}{Normal Mixed model} & \multicolumn{2}{c}{ZIMP model}\\ \hline
$n$ &  Method & $\mu_0$ & $\mu_1$ & $\lambda_1$ & $\lambda_2$\\ \hline
\multirow{2}{*}{100} & MCMC & (-2.34;0.23) & (4.30;9.71) & (1.33;3.09) & (8.81;10.88) \\ 
& VB & (-1.82;0.60)  & (6.95;11.15)  &  (1.44;3.11) & (9.00;11.00)  \\ \hline
\multirow{2}{*}{300} & MCMC & (-1.83;-0.40) & (6.35;8.72) & (1.91;2.80) & (9.12;10.26) \\
 & VB & (-1.2;0.08) & (7.52;10.32) & (1.70;2.66) & (9.46;10.50) \\ \hline
\multirow{2}{*}{500} & MCMC & (-1.44;-0.27) & (7.01;8.69) & (1.93;2.61) & (9.33;10.22)   \\
& VB & (-1.06;-0.11) & (7.84,10.14) & (1.81;2.42) & (9.53;10.33) \\ \hline
\end{tabular}
\end{table}

The MSE results in Table \ref{table:1} show a slight advantage of VB over the MCMC method as the MSE is at most one point smaller in the hundredths place. The MR$_{p}$ for $p=0.75$ and $p=0.9$ point to this direction also. These results follow for $n=100, 300$ and $500$. For MR$_{.5}$, VB show greater improvement over MCMC  for $n=100, 300, 500$, with MCMC results being about 1.5 larger than those of VB method. The MSE and MR$_{p}$ results in Table \ref{table:2} show similar results for all sample sizes and methods, with small variation due to method and sample size, with a slight advantage of VB over the MCMC method as the MSE is at most one point smaller in the hundredths place. Elapsed time for obtaining results from each method is given in Table \ref{table:3}. In Study 1, the computational time of the VB method was substantially smaller than the MCMC method for fitting normal mixture and ZIMP models.

\begin{table}[h!]
\caption{MSE and misclassification rate at $p$-percentile (MR$_{p}$) for $p=0.5, 0.75,$ and $0.9$  (MR$_{p}$) for logit  link, normal mixed model, $n = 100, 300, 500$, VB and MCMC inference methods in Study 1.} \label{table:1}
\begin{tabular}{l|ccccc} \hline 
$n$ & Method & MR$_{.5}$ & MR$_{.75}$ & MR$_{.9}$& MSE \\ \hline
\multirow{2}{*}{100} & MCMC & 0.14 & 0.09 & 0.08 & 0.11\\
                       & VB & 0.10 & 0.09 & 0.08 & 0.10\\ \hline
\multirow{2}{*}{300} & MCMC & 0.13 & 0.09 & 0.08& 0.11\\
                       & VB & 0.09 & 0.07 & 0.07& 0.09\\ \hline
\multirow{2}{*}{500} & MCMC & 0.12 & 0.08 & 0.08 & 0.10\\
                       & VB & 0.08 & 0.07 & 0.07 & 0.09 \\ \hline
\end{tabular}
\end{table}

\begin{table}[h!]
\caption{MSE and misclassification rate (MR) for logit link, ZIMP model, $n = 100, 300, 500$, VB and MCMC inference methods in Study 1.} \label{table:2}
\begin{tabular}{l|ccc} \hline 
$n$ &  Method & {MR} & MSE \\ \hline
\multirow{2}{*}{100} & MCMC & 0.03 & 0.16 \\ 
                     & VB & 0.03  & 0.16\\ \hline
\multirow{2}{*}{300} & MCMC & 0.04 & 0.16 \\
                       & VB & 0.03  & 0.15\\ \hline
\multirow{2}{*}{500} & MCMC & 0.04 & 0.16 \\
                       & VB & 0.03 & 0.15 \\ \hline
\end{tabular}
\end{table}

\begin{table}[h!]
\caption{Mean and standard deviation of elapsed time (in minutes) resulted from normal mixture and ZIMP model for fitting data in Study 1 using MCMC and VB methods.} \label{table:3}
\begin{tabular}{l|ccc} \hline 
$n$ & Method & Normal mixture & ZIMP \\ \hline
\multirow{2}{*}{100} & MCMC & 1.43 (0.10)  & 6.03 (1.43)\\
                     & VB   & 0.29 (0.14) & 1.04 (0.23)\\ \hline
\multirow{2}{*}{300} & MCMC & 2.00 (0.20) & 8.43 (1.81)\\
                     & VB   & 0.41 (0.41) & 2.93 (0.69) \\ \hline
\multirow{2}{*}{500} & MCMC & 2.58 (0.18) & 10.51 (2.44)\\
                     & VB   & 0.04 (0.09) & 5.45 (0.66)\\ \hline
\end{tabular}
\end{table}

\subsection{Study 2: non segregating functional covariates} \label{sec:sce2}

In this study the goal is to assess the ability of cluster discrimination when using non segregating functional covariates. Inspired by the example in \cite{mousavi2018functional}, our second scenario was constructed as follows: for the first step, we considered $T=[0, 10]$ and $S=[0,1]$ and generated  150 functional predictors $X_{i1}$ and $X_{i2}$, $i=1, \ldots, 150$ using basis expansions on the form
\begin{eqnarray*}
X_{i1}(t) &=& \sum_{k=1}^{13} C_{1ik} B^{(1)}_k(t) \\
X_{i2}(s) &=& \sum_{k=1}^{13} C_{2ik} B^{(2)}_k(s)
\end{eqnarray*}
for $i = 1, \ldots 150$, where $B^{(1)}$ and $B^{(2)}$ are cubic B-splines corresponding to nine equally spaced knots over the intervals $[0,10]$ and $[0,1]$ respectively.  
The coefficients $C_{1ik}$ and $C_{2ik}$ are the elements of the matrices $C_1$ and $C_2$ which are $13 \times 13$ matrices simulated as
$$C_1 = Z_1 U_1 \quad \mbox{and} \quad C_2 = Z_2 U_2$$ where $Z_1$ is a $150 \times 13$ matrix formed by iid $N(0.1,1)$ random variables, $Z_2$ is a $150 \times 13$ matrix of iid $N(0,1)$ random variables and $U_1$ and $U_2$ are $13 \times 13$ matrices of iid [0,1]-uniform random variables. 
For each subject, the functional covariates were sampled at 256 equally spaced time points $\{t_j \in T,j=1, 2,\ldots, 256\}$ and $\{s_j \in S,j=1, 2,\ldots, 256\}$.

\begin{figure}[!htb]
\begin{minipage}{0.4\textwidth}
\centering
\includegraphics[height=4cm,width=\textwidth]{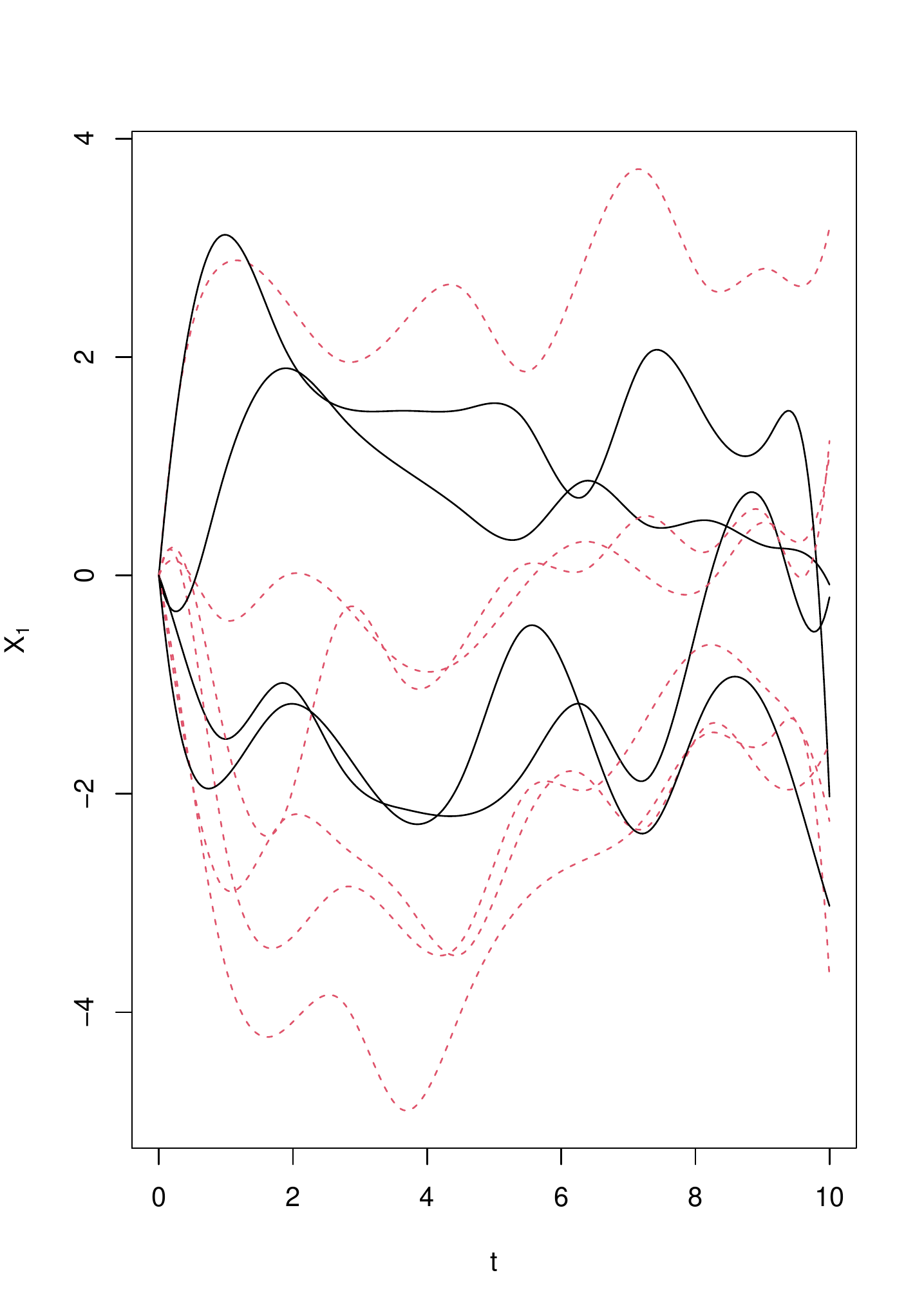}
\end{minipage} 
\begin{minipage}{0.4\textwidth}
\centering
\includegraphics[height=4cm,width=\textwidth]{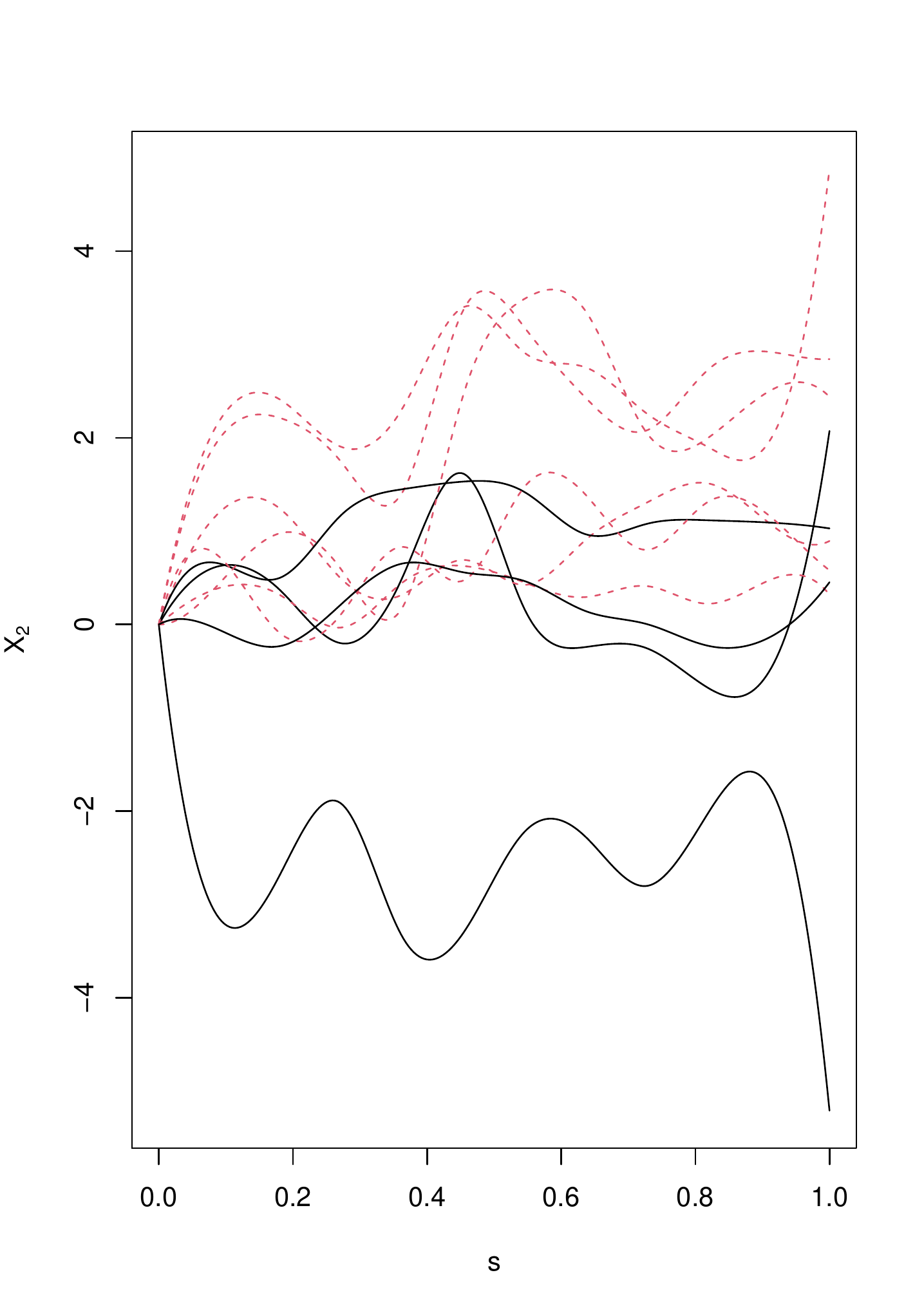}
\end{minipage} 
\begin{minipage}{0.4\textwidth}
\centering
\includegraphics[height=4cm,width=\textwidth]{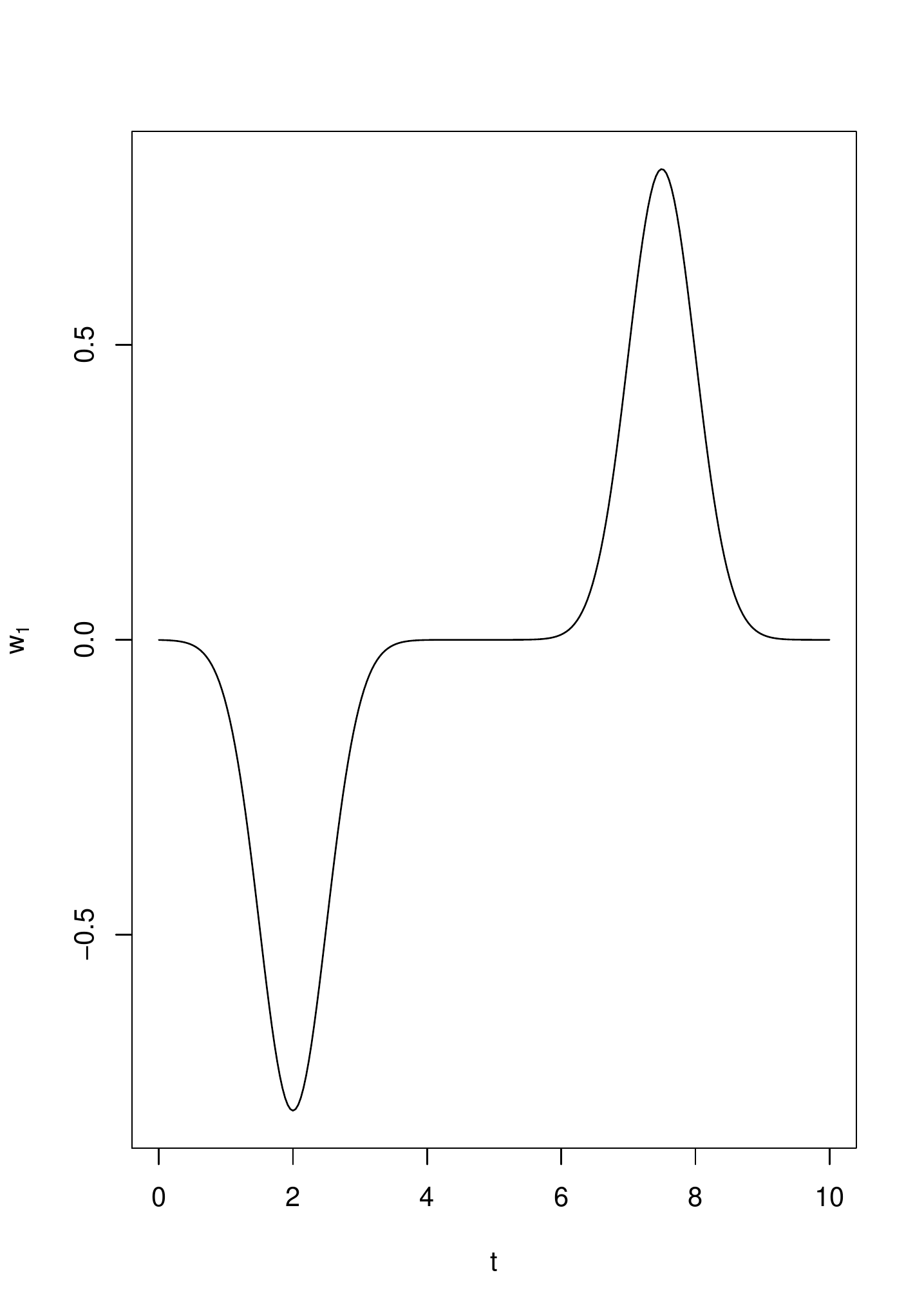}
\end{minipage} 
\begin{minipage}{0.4\textwidth}
\centering
\includegraphics[height=4cm,width=\textwidth]{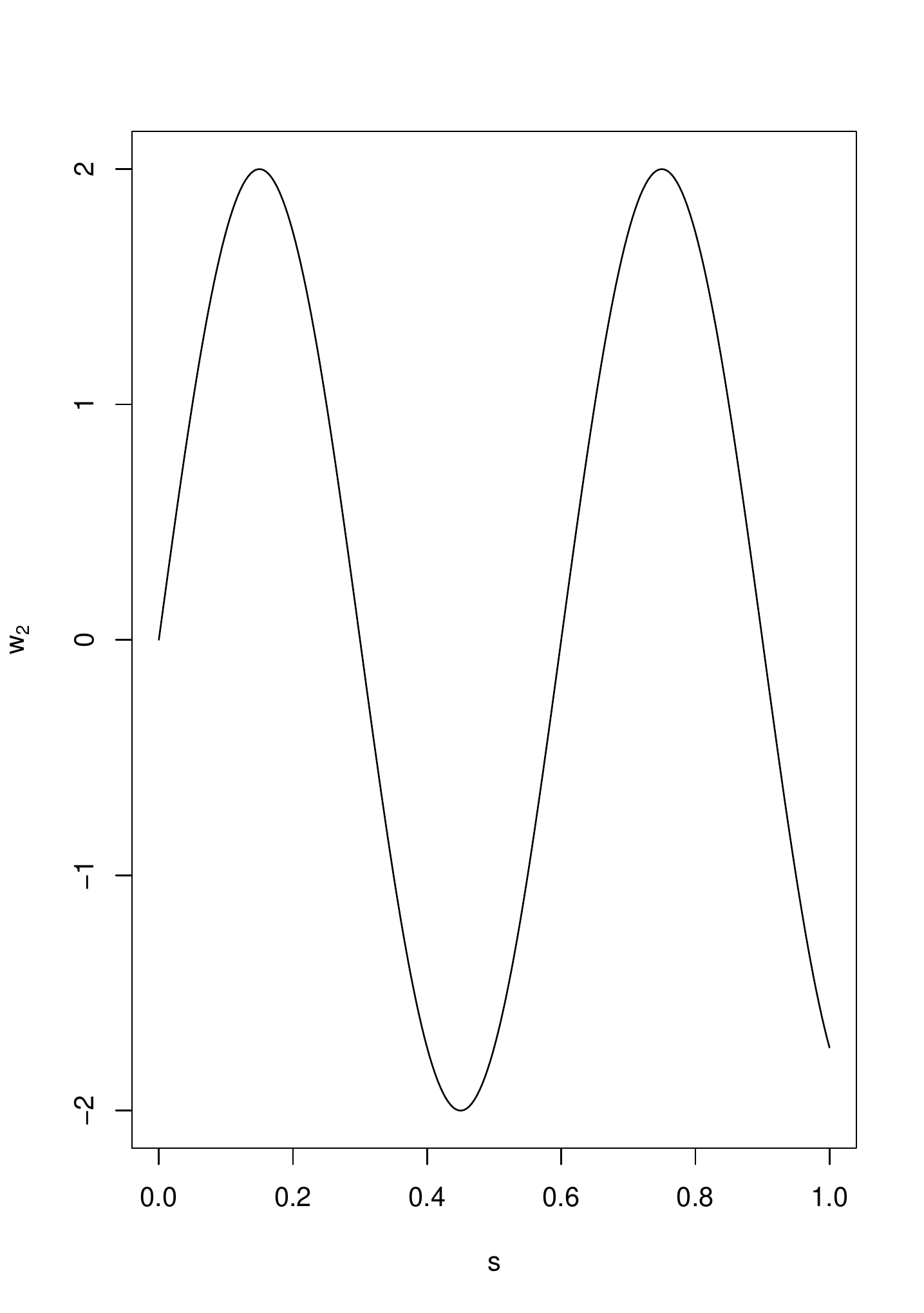}
\end{minipage} 
\caption{Functional covariates $X_{i,1}$ and $X_{i,2}$, $i=1, \ldots, 10$ and weight functions $w_{1}(\cdot)$ and $w_{2}(\cdot)$, for Scenario 2. Red (traced) lines correspond to $\gamma_i=1$, black (solid) lines correspond to $\gamma_i=0$.} \label{fig:scenario2}
\end{figure}

We show, in the top panels of Figure \ref{fig:scenario2}, a sample of 10 curves for $X_{i1}$ and $X_{i2}$. In Study 1, the separation of covariates was clear, but in this Study 2, discriminating the sub-populations is not so clear,  see plot on curves $X_{i1}$ and $X_{i2}$ on the top panels of Figure \ref{fig:scenario2}.

Further, we generated 50 datasets with weight functions given by
\begin{eqnarray*}
w_1(t) & = & -\phi(t;2,0.5)+\phi(t,7.5,.5) \\
w_2(s) & = & 2\sin(10s\pi/3)
\end{eqnarray*}
where $\phi(\cdot;\mu,\sigma)$ is the normal density with mean $\mu$ and standard deviation $\sigma$, shown in the bottom panels of Figure \ref{fig:scenario2}. To calculate the mixture probabilities in (\ref{eq:p12}) and (\ref{eq:p13}) we considered the functional covariates $X_{i1}(t)$ and $X_{i2}(s)$ along with their respective functional weights $w_1(t)$ and $w_2(s)$. We then sampled data from mixture normal model in (\ref{simdatanorm}) with $\epsilon_i \sim$ iid $N(0,18)$, leading to a mixture of a normal distribution with mean $\mu_0=0$ and another with mean $\mu_1=9$, both variances equal to $\sigma^2=18$, and the ZIMP model in (\ref{simdatazimp}) with $\lambda_1 = 2$ and $\lambda_2 = 8.5$. 

For all datasets, we ran the MCMC for 15,000 iterations with a burn-in of 10,000, sampled every 100th step. Using MCMC samples of normal mixed model location parameters, the average of lower and upper bound of 95\% high density posterior (HPD) intervals across the 100 simulated datasets are given by $(-1.35; 1.28)$ and $(7.76; 9.81)$ for $\mu_0$ and $\mu_1$, respectively. Considering the VB method for the same datasets we obtain the 2.5\% and 95\% quantiles across the 100 data sets for $\mu_0$ and $\mu_1$ as $(-0.94;3.36)$ and $(5.31; 9.66)$, respectively. Results show the true location parameter $\mu_0 = 0$ and $\mu_1=8.5$ lie within the lower and upper limits of the respective intervals. 

Although the MSE for the normal mixture model fitted with MCMC is smaller than the VB method as shown, in Table \ref{table:5}, misclassification rate for $p=0.5, 0.75$ and $0.9$ are smaller under VB method.

\begin{table}[h!]
\caption{MSE and misclassification rate for logit link, normal mixed model, $n = 150$, VB and MCMC inference methods for Study 2.} \label{table:5}
\begin{tabular}{l|cccc} \hline 
 &  \multicolumn{3}{c}{Misclassification rate (MR$_{p}$)} & MSE\\ \hline
 Method & 0.5 & 0.75 & 0.9 &  \\ \hline
 MCMC & 0.09 & 0.09 & 0.13 & 0.02\\
 VB & 0.06 & 0.07 & 0.10 & 0.05\\ \hline
\end{tabular}
\end{table}

For the ZIMP model fitted with the MCMC method, the mean of lower and upper bound of 95\% high density posterior (HPD) intervals for $\lambda_1$ and $\lambda_2$ across the 100 datasets are given by $(1.62;2.65)$ and $(7.32;9.87)$. Using the VB method for the same datasets we obtain the 2.5\% and 95\% quantiles for $\lambda_1$ and $\lambda_2$ as $(1.67;2.41)$ and $(7.5;9.36)$, respectively. Results show the true location parameter $\lambda_1 = 2$ and $\lambda_2 = 8.5$ lie within the lower and upper limits of the respective intervals. 

 MCMC and the VB methods produced close MSE and misclassification rate for the ZIMP model - one point of advantage in the hundredths place to MCMC - as shown in Table \ref{table:6}.

\begin{table}[h!]
\caption{MSE and misclassification rate (MR) for logit link, ZIMP model, $n = 150$, MCMC and VB inference methods in Study 2.} \label{table:6}
\begin{tabular}{l|cc} \hline 
Method & {Misclassification rate (MR)} & MSE \\ \hline
MCMC & 0.06 & 0.12 \\ 
VB & 0.07  & 0.13\\ \hline
\end{tabular}
\end{table}

As shown in Table \ref{table:7}, the computational time consuming of the VB method is substantially smaller than the MCMC method for normal mixture and ZIMP models.

\begin{table}[h!]
\caption{Mean (standard deviation) of elapsed time (in minutes) of MCMC and VB methods in Study 2.} \label{table:7}
\begin{tabular}{l|ccc} \hline 
$n$ & Method  & Normal mixture & ZIMP \\ \hline
\multirow{2}{*}{150} & MCMC & 4.02 (0.61) & 14.62 (1.80)\\
                     & VB   & 1.95 (1.79) & 4.90 (0.44)\\ \hline
\end{tabular}
\end{table}

\section{Applications} \label{sec:realdata}

\subsection{Identification of early responders using EEG data.} \label{sec:early}

Placebo responders  are those patients whose response is termed ``non-specific'', e.g., in a drug trial, an improvement in symptoms that is not due to the effect of the active chemicals in the drug. There is an intense debate about how to identify placebo-responders  in clinical trials of medications, in particular, for major depressive disorder (MDD) \citep{walsh2002placebo} since, there could be placebo responders among either the control or the treatment group. Furthermore, it is known that there is a high rate of placebo responders among patients in MDD treatment trials and in some experiments with selective serotonin reuptake inhibitors (SSRIs) it was found that some patients can have a better response using placebo \citep{gueorguieva2011trajectories}. Identifying such patients using covariates would be an important tool in  clinical research. Scalar covariates such as sex and  disease severity are typically included in the modeling. On the other hand, there are several studies relating differences in neural processing between placebo and active treatments (see for example, \cite{leuchter2002changes}, \cite{watson2007placebo}, \cite{zhang2009transferable}, \cite{wager2015neuroscience},  \cite{ciarleglio2018constructing}, and references therein). One way of measuring neural processing is through Electroencephalography (EEG). It is a fast,  inexpensive and non-invasive procedure that has been used for decades for recording brain activity.

 One disease for which this is a particularly crucial problem is major depressive disorder (MDD). Recent studies have suggested that less than 40\% of MDD patients achieve remission after completing a lengthy course of first-line treatment (McGrath et al., 2013). Such a low remission rate may be greatly improved if clinicians are better able to identify patient characteristics that define subgroups of patients who will benefit mostly from a given treatment. Furthermore, placebo response rates can be high in MDD treatment trials and analyses of results from previous trials that have compared placebo to active medications, including a class of commonly used antidepressants know as selective serotonin reuptake inhibitors (SSRIs), have found that some subjects worsen with an antidepressant, i.e., would fare better on placebo (Gueorguieva et al., 2011). 

\cite{jiang2017latent} analysed data from a randomized placebo controlled depression clinical trial of sertraline in order to identify early responders to treatment (which is indicative of a placebo response since it is believed that response to the active treatment is not immediate). The dataset consists of 96 MDD patients, randomized to either a drug or placebo treatment.  For each subject, several scalar and categorical covariates are available, as well as their resting state electroencephalography (EEG) under a closed eyes condition. This EEG data contains the current source density amplitude spectrum values (V/m2)
\citep{nunez2006electric} at a total of 14  electrodes ($P_9$, $P_{10}$, $P_7$, $P_8$, $P_5$, $P_6$, $PO_7$, $PO_8$, $PO_3$, $PO_4$, $O_1$, $O_2$, $PO_Z$ and
$O_Z$) located in occipital and
parietal brain regions.  Each electrode is measured at 45 frequencies at a 0.25 Hz  resolution
within the theta (4 -- 7 Hz) and alpha (7 -- 15 Hz) frequency bands. The response variable for each subject is the Hamilton Depression Rating Scale (HAM-D), measured before the treatment (baseline) and after one week into
the study. It is believed that the active drug treatment can  only have an effect on 
symptoms after two  weeks. Therefore, any improvement observed after one week is likely to
be due to placebo effect (or spontaneous improvement). For more details, we refer to  \cite{jiang2017latent} and
references therein. For EEG location maps see Figure 7 in \cite{rupasov2012time}.

\begin{figure}
    \centering
    \begin{minipage}{0.4\textwidth}
        \includegraphics[width=7cm]{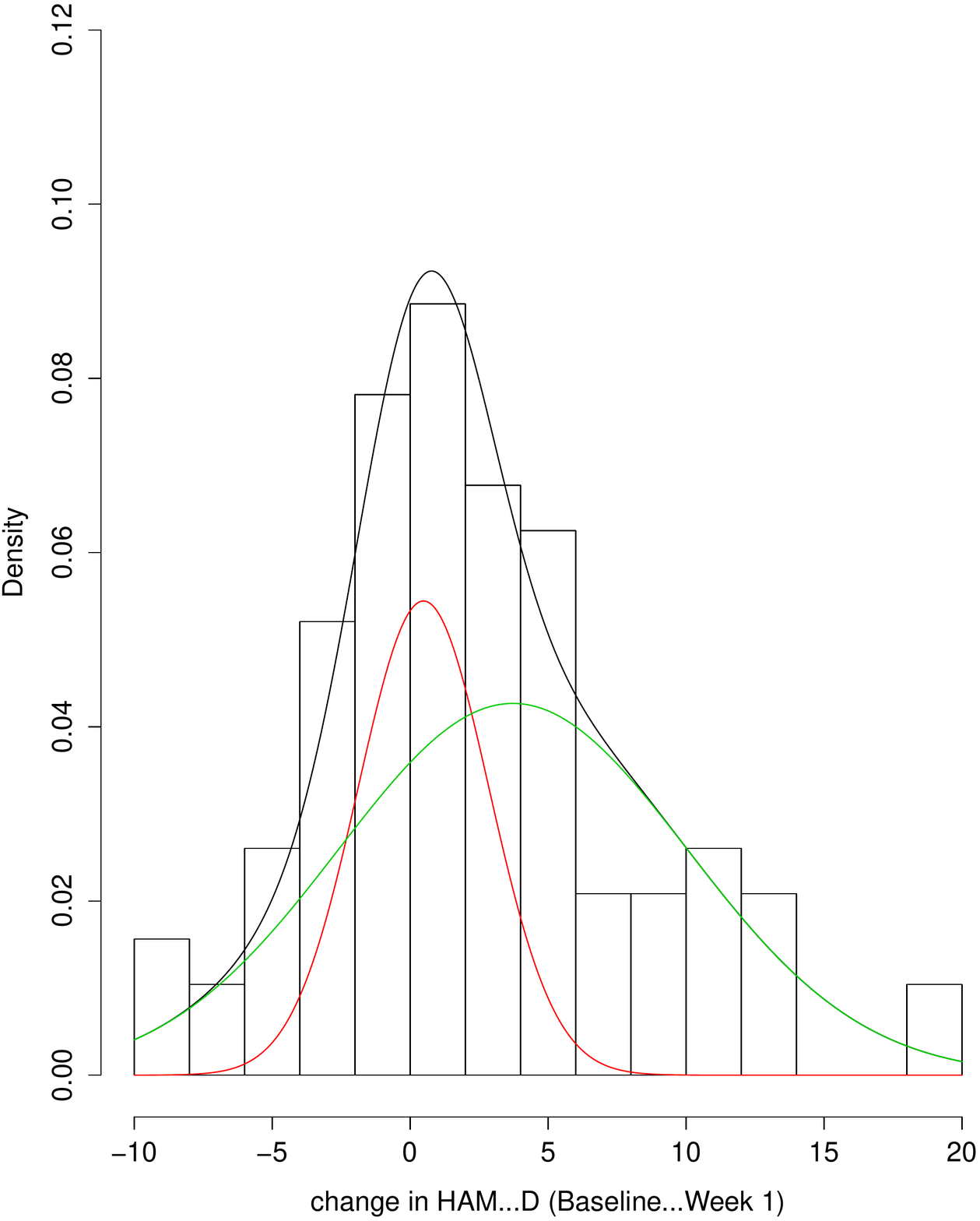}
       \end{minipage}
    \begin{minipage}{0.4\textwidth}
        \includegraphics[width=7cm]{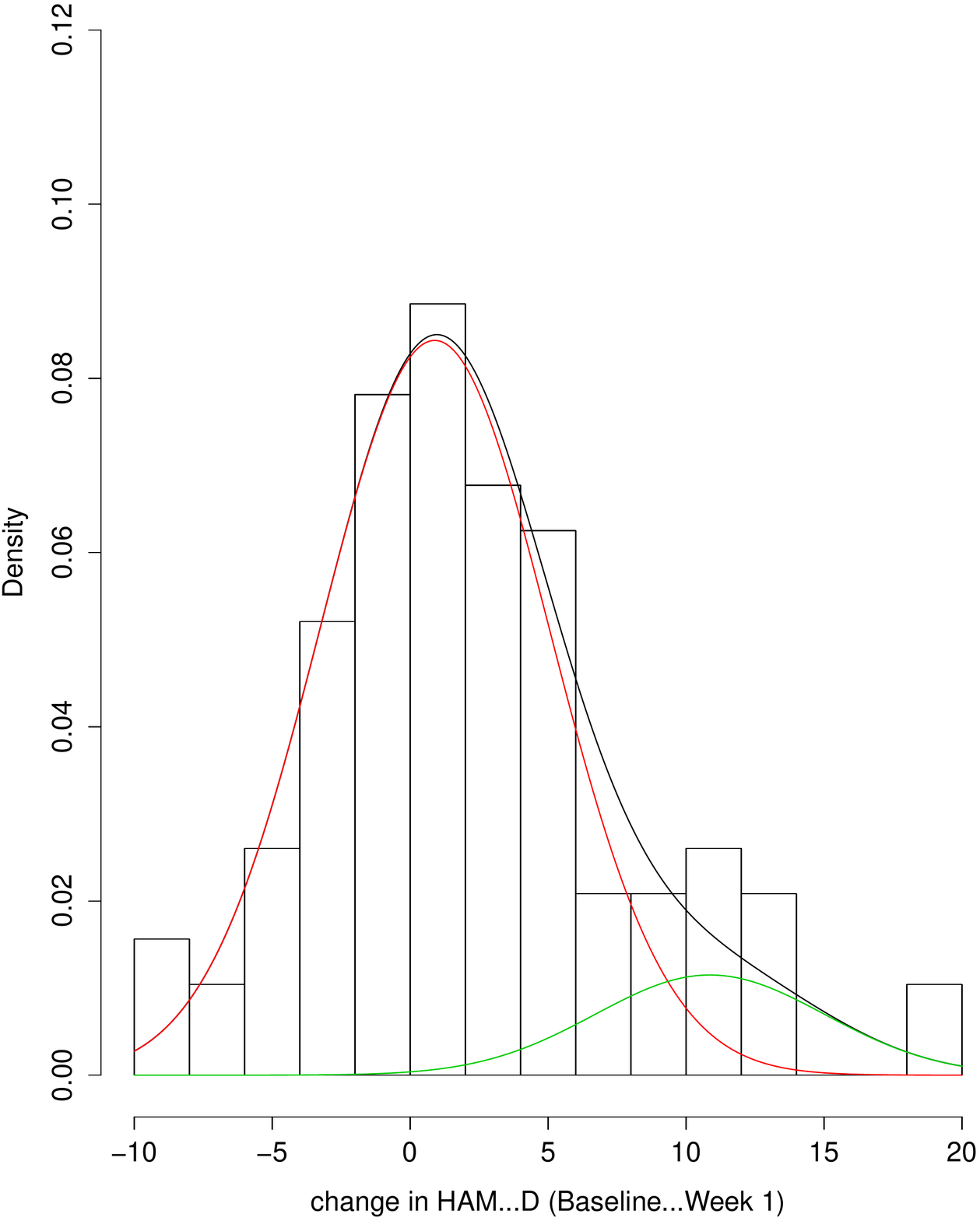}
     \end{minipage}
  \caption{Histogram of the change in HAM-D (baseline - week 1) showing the amount of improvement in depression symptoms after 1 week for both drug and placebo treated patients. The left panel presents the normal mixture with two components given by the EM algorithm without covariates.  The right panel, shows the fit using  mixture of two normal distributions $0.84 \times {\cal N}(0.82,17.08) + 0.16 \times {\cal N}(10.72,17.08)$ chosen by the linear model with Normal prior and logit link.  Red and green curves are the mixture components.} \label{fig:1}
\end{figure}

Let $y_{i}$ denote the change in the HAM-D (baseline - week 1) for subject $i$, $i=1, \ldots, 96$, where a positive change indicates diminished depression symptom severity. In order
to compare our results with \cite{jiang2017latent}, we will focus on the same scalar covariates,
sex and chronicity, and functional covariates given by data taken from 14 EEG electrodes.  Figure \ref{fig:1} shows histograms of the change in HAM-D (baseline - week 1) showing the amount of improvement in depression symptoms after 1 week, a positive change indicates improvement in symptoms. Notice that there is a strong indication  of a mixture of two distributions. To explore the data set, we fit a parametric model using the EM algorithm for a mixture of Gaussian distributions (different means and different variances) with no covariates, and the two fitted Gaussian curves are shown in the left panel of Figure \ref{fig:1}, with more than 40\% of the subjects are classified as early responders (green curve). That is, this model has low power to discriminate the subjects into two classes. 

\cite{jiang2017latent} analyzed this dataset using the same hierarchical model given by \eqref{eq:normal}. The unobserved binary subgroup indicators are modeled via a hierarchical probit model as a function of the baseline EEG measurements and other scalar covariates of interest.  In their work, instead of the regression component given by \eqref{eq:pi}, they propose to use the EEG data in the form of a (14 x 45) matrix-valued covariate. However, instead of focusing on estimating the coefficients for the entire matrix, they assume a low-dimensional structure through CP decomposition \citep{kolda2009tensor}, reducing the matrix dimension to $9 \times 4$.  One disadvantage of this
approach is that it does not take  advantage of the functional nature of the data and it lacks direct interpretability for the estimated parameters. Also, the analysis uses a tensor product and so the results will likely depend on how the electrodes are ordered in the matrix. 

\begin{figure}[!htb]
\hspace{-1cm}\includegraphics[width=10cm]{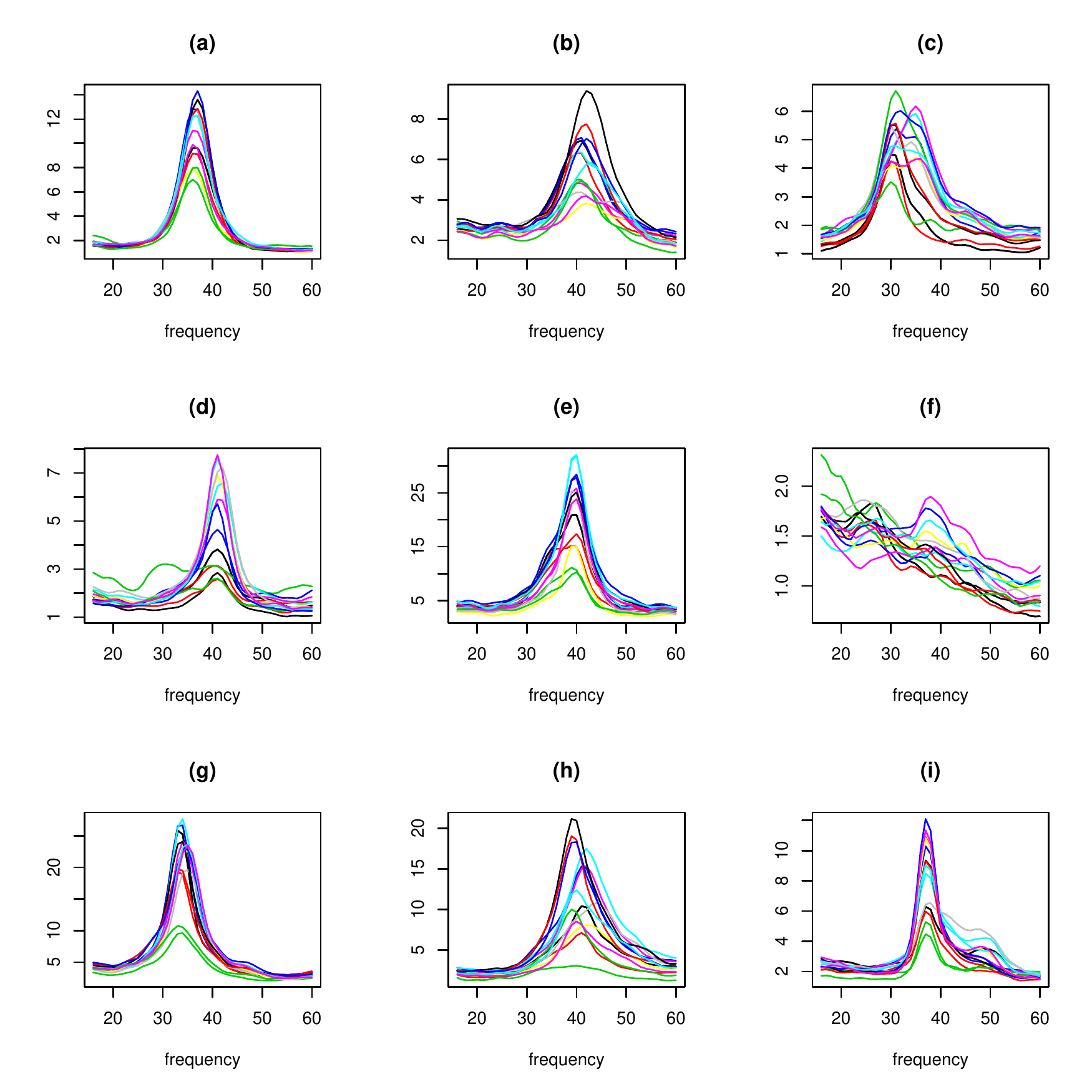}
\caption{EEG for 9 subjects} \label{fig:2}
\end{figure}

Just to get some idea on the behavior of the functional covariates, Figure  \ref{fig:2}  presents data taken from 14 EEG electrodes for 9 subjects. Each panel represents a subject and each curve is the EEG for different electrodes. The colors are consistent across all plots. As  can be seen, there is a large variability in the range of values for each subject. Therefore, we used as functional covariates the  EEG signal for each subject that has been standardized to have zero mean and unit standard deviation. This standardizing will allow us to use the same number of basis functions and same knot location for all functions $B^{(X,j)}$ in \eqref{eq:weight.1}. 

To sample from the posterior distribution, we use the Gibbs sampler scheme considering the posterior calculation described in Section \ref{sec:posteriors}. Table \ref{table:9} presents the posterior estimates of the scalar parameters using logit link function and $t$, Cauchy and normal priors for the parameters in $(\btheta, \bphi_1, \ldots, \bphi_J)$ and linear vs.\ non-linear functional models. The value $\hat{p}$ in each model represents 
$$ \hat{p} = \frac{1}{96} \sum_{i=1}^{96} I(\hat{p}_{i2} > 1/2).$$

\begin{table}[h!]
\caption{Posterior mean (standard deviation) of $\mu_0, \mu_1, \sigma^{2}$, $\btheta$ and $p$ for different prior distributions and linear and nonlinear functional model.} \label{table:9}
{\scriptsize
\begin{tabular}{lcccccc} \hline 
& \multicolumn{6}{c}{logit link} \\ \hline
& \multicolumn{3}{c}{Nonlinear model} & \multicolumn{3}{c}{Linear model} \\
Prior & {Student t} & {Normal} & {Cauchy} & {Student t} & {Normal} & {Cauchy} \\ \hline  
$\mu_0$ &  0.88 (0.48) & 0.87 (0.47) &  0.89 (0.48) & 0.86 (0.61) & 0.82 (0.61) & 0.89 (0.60)\\ 
$\mu_1$ & 11.52 (1.18) & 11.57 (1.18) & 11.44 (1.16) & 10.81 (1.57) & 10.72 (1.52) & 10.86 (1.55)\\
Intercept & -1.07 (0.34) & -0.008 (0.005) & -1.24 (0.44) & -1.07 (0.34) & -1.07 (0.40) & -1.26 (0.79)\\
Sex & -1.85 (0.48) & -0.34 (0.21) & -1.62 (0.42) & -1.85 (0.50) & -1.85 (0.44) & -1.79 (0.54) \\
Chronicity & -1.76 (0.45) & -0.44 (0.19) & -1.64 (0.41) & -1.76 (0.44) & -1.79 (0.44)  & -1.70 (0.54) \\
$\sigma^2$ & 15.82 (2.30) & 15.81 (2.34)  & 15.95 (2.48) & 17.21 (3.00) & 17.08 (2.85) & 17.32 (2.93)\\
$\hat{p}$ & 0.18 & 0.16 & 0.18 & 0.15 & 0.16 & 0.15 \\ \hline
\end{tabular}}
\end{table}

Notice that all results in Table \ref{table:9}  are very similar. To choose the best model, we used the posterior predictive checks (see Table \ref{table:10} for the proportion of the sample above a threshold $t$),
$$h(\mathbf{y}) = (1/96) \sum_{i=1}^{96} I(y_i > t),$$
for $t=-5, 0, 5$ and $10$. 

\begin{table}[h!]
\caption{Posterior predictive checks as a function of the threshold ($t$) for different prior distributions and linear and nonlinear functional model.} \label{table:10}
\begin{tabular}{r|rrrrrr} \hline 
 & \multicolumn{6}{c}{logit link} \\ \hline
$t$  & \multicolumn{3}{c}{Nonlinear model} & \multicolumn{3}{c}{Linear model} \\
 & {Student t} & {Normal} & {Cauchy} & {Student t} & {Normal} & {Cauchy} \\ \hline
$-5$ & 0.51 & 0.51 & 0.53 & 0.44 & 0.45 & 0.46 \\ 
$0$ & 0.61 & 0.61 & 0.63 & 0.62 & 0.62 & 0.63 \\
$5$ & 0.34 & 0.33 & 0.36 & 0.47 & 0.48 & 0.48 \\
$10$ & 0.44 & 0.47 & 0.47 & 0.39 & 0.39 & 0.40 \\ \hline
\end{tabular}
\end{table}

\begin{figure}[!htb]
\hspace{-1cm}\includegraphics[height=4cm,width=6cm]{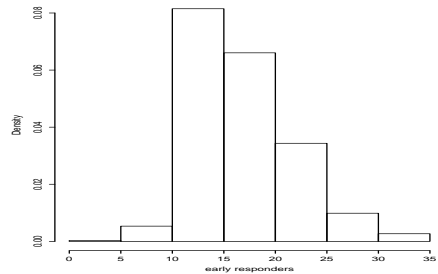}
\caption{Posterior distribution for the total number of early responders using linear model with logit link function and Normal prior} \label{fig:3}
\end{figure}

Although the non-linear model is slightly better than the linear model for $t=-5$ and $t=10$ as shown in Table \ref{table:11} independent of the prior distribution, by the parsimonious criterion we chose the linear model fitted with the Normal prior distribution and consider this model to classify subjects as early responders based on the maximum posterior estimate of $p(\gamma_{i}|{\mathbf{y}})$, see Table \ref{table:11}.  Specifically, 15 subjects (15.6\%) were
classified to the early responder subgroup (posterior mean probability is $> 1/2$), with the change in HAM-D (baseline -
week1) centering at 10.72 (high density posterior (HPD) 95\% credible interval: [7.80; 13.55]), while the others
were assigned to the other subgroup, with the change in HAM-D (baseline - week1)
centering at 0.82 (HPD 95\% Credible interval: [-0.26;2.08]). The effect of sex and
chronicity are significant since the intercept (male and low chronicity) has mean value
-1.07 (HPD 95\% Credible interval: [-1.81;-0.48]), the added effect of being female has
mean -1.85 (HPD 95\% Credible interval: [-2.73;-1.04]) whereas the added effect of high
chronicity has mean -1.79 (HPD 95\% Credible interval: [-2.68;-1.11]). The right panel of Figure \ref{fig:1} shows the fitted distribution using the mean values of the posterior parameters estimated by the  linear model with Normal prior and logit link. Comparing left and right panels of Figure \ref{fig:1} and looking at Table \ref{table:11} we can see that the fitted model has a much better ability to discriminate between the two groups. The most likely number of early responders is between 10 {(10.4\%)} and 15 {(15.6\%)} as can be seen from the posterior distribution for $\sum_{i=1}^{96} \gamma_{i}$, the total number of early responders, shown in Figure \ref{fig:3}.

\begin{table}[h!]
\caption{Frequency table for the posterior mean for the probability of being and early responder.} \label{table:11}
\begin{tabular}{l|c} \hline 
Probability & Number of subjects \\ \hline
0.00 - 0.10 & 65 \\ \hline
0.11 - 0.20  & 6 \\ \hline
0.21 - 0.30  & 4 \\ \hline
0.31 - 0.40  & 4 \\ \hline
0.41 - 0.50  & 2 \\ \hline
0.51 - 0.60  & 2 \\ \hline
0.61 - 0.70  & 1 \\ \hline
0.71 - 0.80  & 1 \\ \hline
0.81 - 0.90  & 1 \\ \hline
0.91 - 1.00  & 10 \\ \hline
Total & 96 \\ \hline
\end{tabular}
\end{table}

\begin{table}[]
    \centering
    \begin{tabular}{c|ccccc} \hline
     Quantile   &   2.5\% & 25\%  & 50\% &  75\% &  97.5\% \\ \hline 
        $\mu_0$ &  -0.40 &  0.40 &  0.84 &  1.22 &  2.00 \\
        $\mu_1$ &   8.01 & 9.65 & 10.64 & 11.74 & 13.91 \\
        $\sigma^2$ & 12.20 & 15.23 & 16.74 & 18.82 & 23.17 \\
        Intercept & -1.87 & -1.27 & -1.02 & -0.84 & -0.51 \\
        Sex & -2.73 & -2.13 & -1.83 & -1.57 & -1.04 \\
        Chronicity & -2.67 & -2.13 & -1.78 & -1.43 & -1.08 \\ \hline 
    \end{tabular}
    \caption{Posterior quantiles of normal mixture parameters using MCMC sample for logit link and Normal prior.}
    \label{table:quantiles}
\end{table}

For comparison between MCMC and Variational Bayes results,  Table \ref{table:1vb} shows the
expected value estimates for the parameters of the VB distributions for the linear/nonlinear model, logit/probit link and  Normal prior ({\it cf.} Table \ref{table:1}). We present only the results for the Normal prior since there was very little effect of the prior on the estimate of VB parameters.

\begin{table}[h!]
\caption{Posterior VB parameters $m_0$, $m_1$, $s_0^2$, $s_1^2$, expectation under the variational distribution with VB fitted parameters and $\hat{p} = (1/96) \sum_{i=1}^{96} I(E_{q^{*}}(\alpha_i)>1/2)$.} \label{table:1vb}
\begin{tabular}{lrrrr} \hline  
& \multicolumn{2}{c}{Logit link} &  \multicolumn{2}{c}{Probit link} \\ \hline
Model & Nonlinear & Linear & Nonlinear & Linear \\ \hline
$E_{q^*}[\mu_0]$ &  1.40  & 1.13 &  1.15  & 0.32  \\
$E_{q^*}[\mu_1]$ &  13.34  & 12.38 & 12.61  & 9.89 \\
$E_{q^*}(\sigma^2)$ &  18.14 & 16.68 &  16.51 & 14.64 \\
Intercept ($E_{q^*}[\beta_0]$) & -0.01  & -0.30 & -0.01  & -0.33  \\
Female ($E_{q^*}[\beta_1])$ & -0.14 & -0.93 & -0.01 & -0.89 \\
Chronicity ($E_{q^*}[\beta_2]$) & -0.13  & -1.39 & -0.17 & -1.44 \\
$\hat{p}$ & 0.10 & 0.14  & 0.14 & 0.24  \\  \hline 
\end{tabular}
\end{table}

\begin{figure}[!htb]
\hspace{-1cm}\includegraphics[height=5cm,width=15cm]{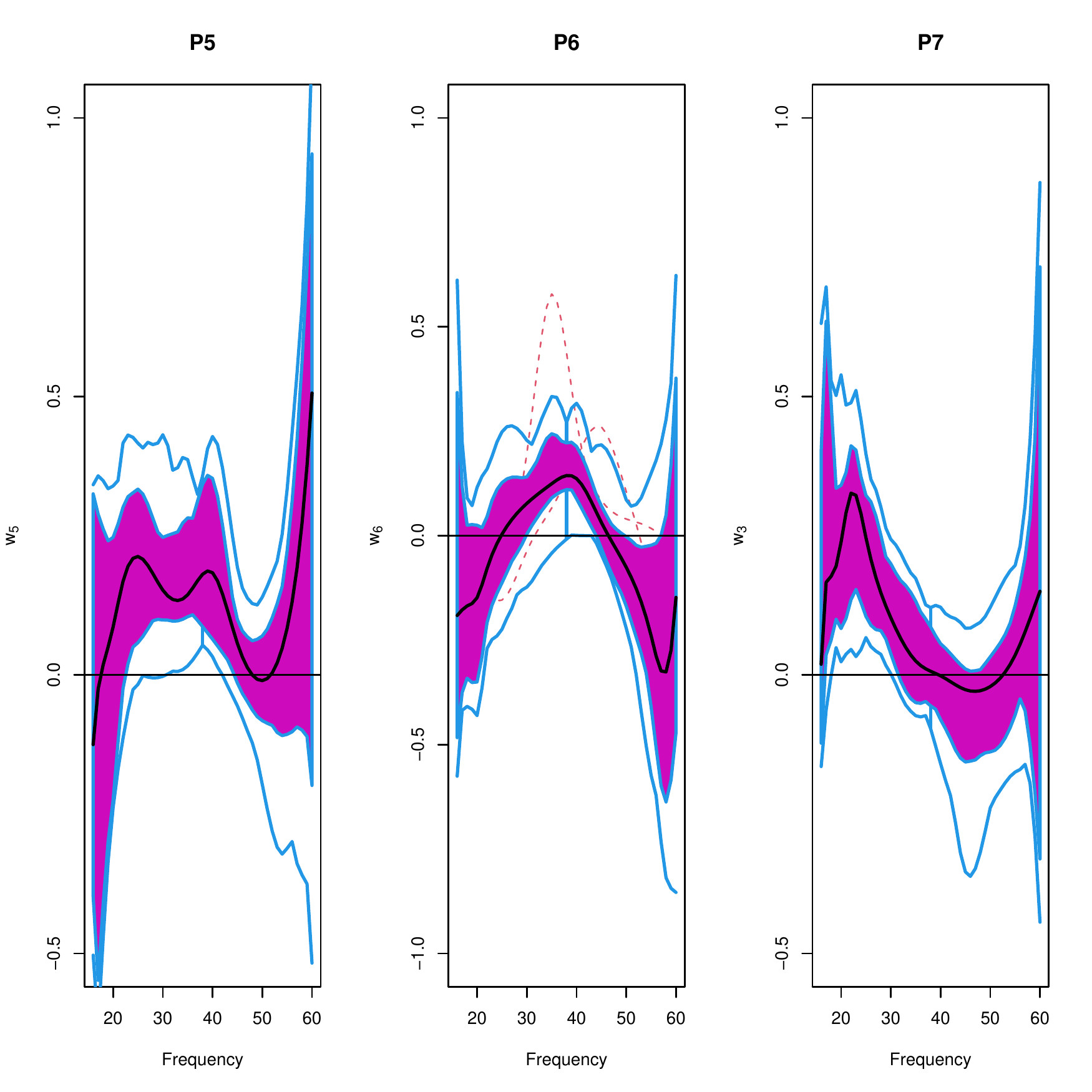}
\caption{Functional boxplots for the significant weight functions  using linear model with logit link function and Normal prior. The magenta area is the 50\% central region, red dashed curves are outlier candidates.} \label{fig:4}
\end{figure}

For the linear model, we estimate the weight functions $w_j$, $j=1, \ldots, 14$, which gives the effect
to what extent each resting state EEG alpha and theta power in the posterior region
of brain under a closed eyes condition could help identify a potential early responders sub-group (which is believed to consist of subjects susceptible to non-specific placebo effects). Figure \ref{fig:4} presents the functional boxplots for the  weight functions $\{w_j(t); j = 3, 5 \;\mbox{and}\; 6\}$ which were the only ones significant to predict early respondents using the linear model with the logit link and Normal prior and which are related to the EEGs located at $P_5$, $P_6$ and $P_7$. The functional boxplots are the equivalent to usual boxplots. They are a graphical method to display  five  descriptive statistics: the median, the first and third quartiles, and the non-outlying minimum and maximum observations. For a nice review on the subject see \cite{sun2011functional}.

\subsection{Predicting illness for milking cows based on functional covariates measured 30 days before lactation} 
Understanding factors that affect productivity of dairy cattle is essential for optimizing profitability and sustainability of dairy farms. The transition from late pregnancy into early lactation in cows is one of the most important stages of the lactation cycle of these animals as cows are at greatest risks of experiencing health disorders during this period. The occurrence of health disorders affects the productivity of cows, and thus is a determining factor of the profitability of dairy herds.

A potential strategy to identify cows with health disorders in early lactation for treatments and other interventions is through
the use of automated monitoring of cow behavioral, physiological, and performance parameters with automated health monitoring systems based on sensor data \citep{stangaferro2016usea,stangaferro2016useb,stangaferro2016usec}. This is because it has been demonstrated that multiple sensor parameters, such as for example rumination time, physical activity, resting time, body temperature, milk volume and component yield, are useful for monitoring cow health as they are dramatically altered during episodes of health disorders \citep{stangaferro2016usea,stangaferro2016useb,stangaferro2016usec} and thus, can be used to predict the health status of cows. Moreover, data from these sensors systems can also be combined with non-sensor data to increase the accuracy of alerts used to identify cows with health disorders. Our dataset for this application consists of data collected in order to train, validate, and test machine-learning algorithms (MLA) models created through a combination of sensor data from Automated Health Monitoring System (AHMS) and non-sensor data available at commercial dairy farms. This project funded by the USDA-NIFA was conducted by the Dairy Cattle Biology and Management Laboratory at Cornell University. We have data on daily clinical examination of cows during the first 40 Days after calving (days in milk) from 258 cows. Information on whether a cow had a health disorder (1) or not (0) was collected by the research team on a daily basis. The data consists of 41 rows (0-40) per cow with sensor and non-sensor data from the previous lactation (i.e., before calving) and after calving. We focused on four sensor parameters related to physical activity for the 30 days prior to calving. The selected parameters were: Activity  (Total number of steps in a given day divided by 24), Number of Resting Bouts, Average Rest Time and Total Rest Time per day. Moreover, in this analysis, in addition to the 4 functional covariates as described above, we considered  4 scalar variables related to previous lactation period known to be associated with health outcomes after calving: $Z_1=$ Age of first calving, $Z_2=$ Previous lactation days in milk, $Z_3=$ Previous lactation health event (0 or 1), and $Z_4=$ Previous lactation number of previous health events.

The objective of this application example was to model the number of days a cow was sick in the first 40 days after calving. Our main goal was to classify cows into 2 or 3 classes of status by using the scalar and functional
covariates related to activity and resting times. There was high variability in the dataset as seen in Table \ref{tab:vacas}. During the observation period, 175 cows were not diagnosed with a health disorder whereas 83 cows with at least one health disorder event with a mean value  of 1.60 sick days and a variance of 10.30. If we just consider the values not equal to zero we get a mean value of 4.96 sick days and variance 15.95. As previously noted, we have 68\% of zeros in the sample (175/258) and probably a zero inflated distribution to accommodate overdispersion caused by zeros. On the other hand, if we consider only the non-zero observations, we still have a large variance as compared with the mean. Therefore, we fitted a zero inflated mixture of two Poisson distributions.

\begin{figure}
    \centering
    \includegraphics[height=3cm,width=8cm]{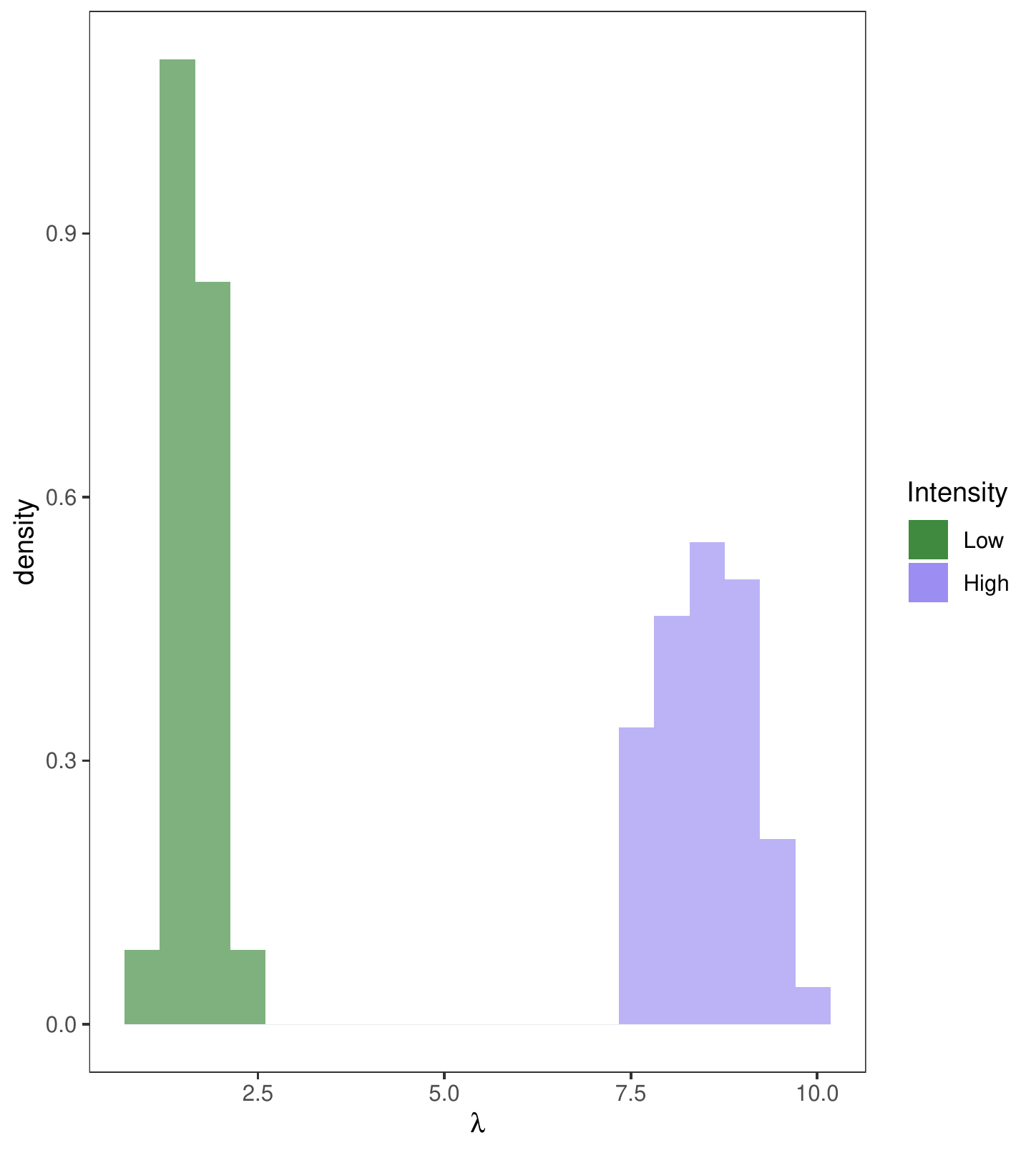}
    \caption{Histogram for posterior distribution of $\lambda_1$ and $\lambda_2$ for the ZIMP model.}
    \label{fig:ZIP_lambda}
\end{figure}

\begin{figure}
    \begin{minipage}{.3\linewidth}
    \includegraphics[height=4cm,width=5cm]{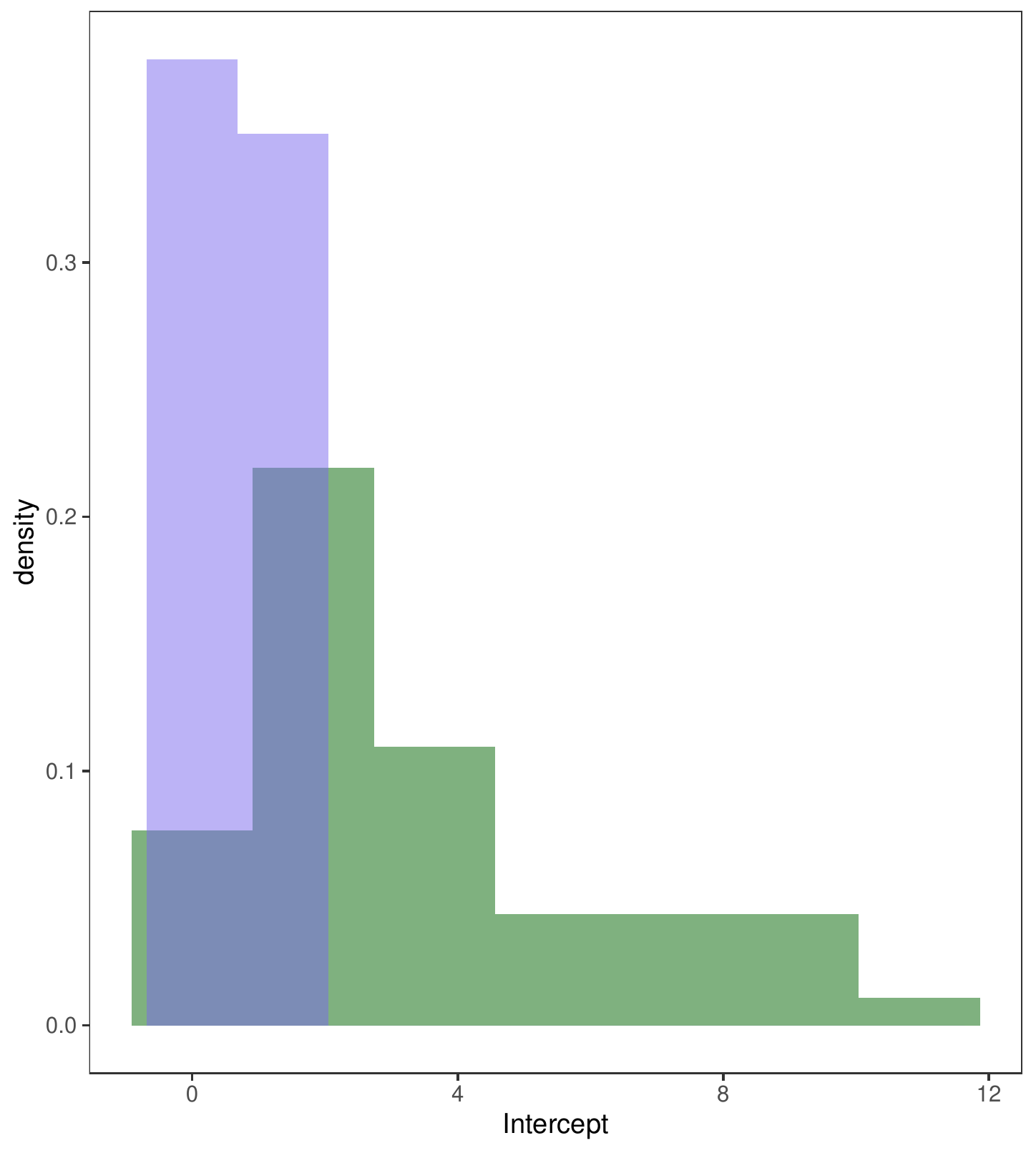}
    \end{minipage}
    \begin{minipage}{.3\linewidth}
    \includegraphics[height=4cm,width=5cm]{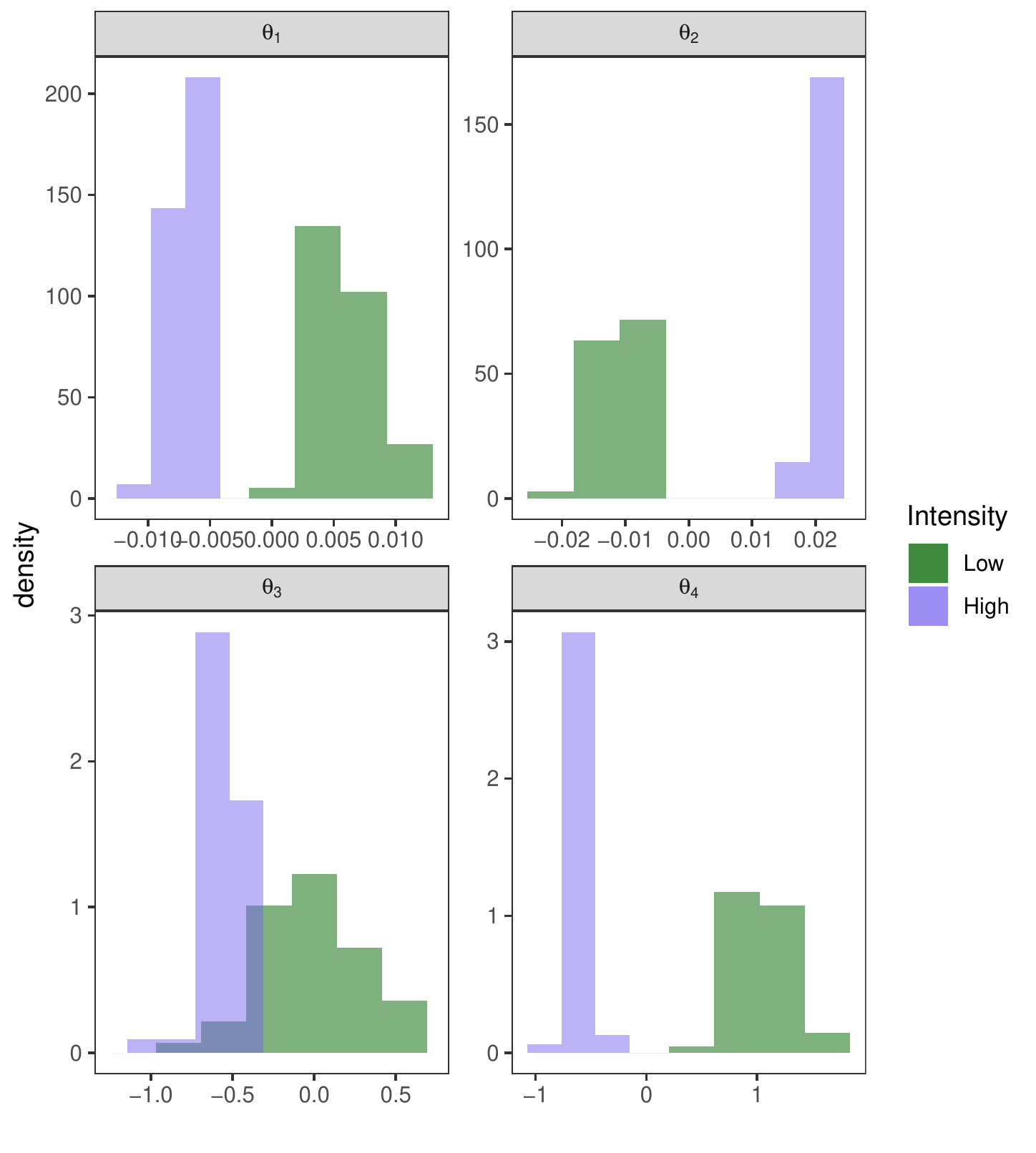}
    \end{minipage}
    \caption{Histogram for posterior distribution of Intercepts and coefficients related to the scalar variables ($\theta_1$, $\theta_2$, $\theta_3$ and $\theta_4$) for the ZIP model.}\label{fig:ZIP_coeff}
\end{figure}

In this case, we fitted the model to 208 cows randomly selected as a training set and used the remaining 50 cows to evaluate the prediction using the model. After fitting the model, we obtained the following classification using the posterior mean for the gamma variable. Notice, from Table \ref{tab:zip_class} that the classification according to the maximum value of the estimated probability is the same using MCMC and VB estimates. From the 138 cows in the training set that did not get sick during lactation, 136 were classified as belonging to the pure-zero class with high probability while the other 2 cows were classified as Low Incidence. All 43 cows that were sick between 1 and 4 days were classified as Low Incidence whereas only one of the cows diagnosed with a health disorder for 5 days was classified as Low Incidence and the other one was classified as High Incidence. All cows with 6 or more days with a health disorder were classified as High incidence.
The posterior mean and standard deviation and quantiles of the posterior distribution as well as the mean for the variational Bayes mean value for these classes are shown in Table  \ref{tab:ZIP_quantiles_cows} and Figure \ref{fig:ZIP_lambda}. Notice that the estimates of the mean and standard deviation of the posterior distribution for $\lambda_1$ and $\lambda_2$ are very similar using VB and MCMC. On the other hand, from Table \ref{tab:ZIP-coef} and Figure \ref{fig:ZIP_coeff}, we can see that the VB estimates for the mean of the intercept and scalar coefficients for variables are not so good for the coefficients that are not significant whereas they are similar for the ones which are significant.
Direct comparison for the coefficients of the expansion for the weights of the functional covariates are not meaningful
and we constructed the functional boxplot for the posterior estimates of the weight functions as shown in Figures  \ref{fig:ZIP_weights_class1} and  \ref{fig:ZIP_weights_class2}. Notice that apparently, none of the functional covariates are significant for the regression term for Low Incidence
Class whereas only the covariates Activity and, to a lesser extent, Number of Rest Times and Average Rest
Times are significant to determine the probabilities of the High Incidence latent class.

\begin{table}[]
    \centering
    \begin{tabular}{c|ccccccccccccccc|c} \hline
        Sick days &  0 &  1 &  2 &  3 &  4 &  5 &  6 &  7 &  8 &  9 & 10 & 11 & 12 &  13 & 18  & Total \\ \hline
        Freq & 175  & 22 &  10 &   7 &   7 &   4 &   5  &  4 &   7 &   6  &  2 &  3 &   3  &  2 &   1 & 258 \\ \hline
    \end{tabular}
    \caption{Frequency of sick days for lactating cows (training + testing set)}
    \label{tab:vacas}
\end{table}

As we noted before, we have 68\% of zeros in the sample (175/258), {indicative of} a zero inflated distribution. On the other hand, if we consider only the non-zero observations, we still have a large variance as compared with the mean. Therefore, we fitted a zero inflated mixture of two Poisson distributions. In this case, we fitted the model to 208 cows and used the remaining 50 cows to check the prediction using the model. After fitting the model, we classified cows health disease state in "Pure zero", "Low Incidence" and "High Incidence" classes using the posterior mean for the gamma variable. Results are given in Table \ref{tab:zip_class}. Notice, that the classification according to the maximum value of the estimated probability is the same using MCMC and VB estimates. From the 138 cows that did not   get sick during the lactating period, 136 are classified as belonging to the ``pure zero'' class with high probability, whereas only 2 cows were classified as Low Incidence. All 43 cows that were sick between 1 and 4 days were classified as Low Incidence whereas only one of the cows who got sick for 5 days was classified as Low Incidence where the other one was classified as High Incidence. All cows with 6 or more sick days were classified as High incidence.

\begin{table}[]
    \centering
    \begin{tabular}{c|cccccccccccccc} \hline
      Class    & \multicolumn{13}{c}{Number of sick days} \\ 
         & 0  & 1 &  2 &  3 &  4 &  5 &  6  & 7 &  8 &  9 & 11 & 12 & 13 & 18 \\ \hline
     Pure Zero  & 136 &  0  & 0 &  0  & 0 &  0 &  0  & 0  & 0  & 0 &  0  & 0 &  0 & 0 \\
     Low Incidence  & 2  & 20 &  10  & 6 &  5 &  1 &  0 &  0 &  0 &  0 &  0   & 0  & 0 & 0 \\
     High Incidence   & 0  &  0   & 0  &  0  &  0 &   1  &  5   &  4  &  6   & 5  &  2   & 3  &  1 & 1 \\ \hline 
    \end{tabular}
    \caption{Classification for the training set according to the class presenting the maximum mean value of the posterior distribution and VB estimation of the latent variable $\gamma$}
    \label{tab:zip_class}
\end{table}

The posterior mean and standard deviation and quantiles of the posterior distribution as well as the mean for the variational Bayes mean value for these classes are shown in Table \ref{tab:ZIP_quantiles_cows} and Figure \ref{fig:ZIP_lambda}.
Notice that the estimates of the mean and standard deviation of the posterior distribution for $\lambda_1$ and $\lambda_2$ are very similar using VB and MCMC. On the other hand, from Table \ref{tab:ZIP-coef} and Figure \ref{fig:ZIP_coeff}, we can see that the VB estimates for the mean of the Intercept and scalar coefficients for variables are not so good for the coefficients that are not significant whereas they are similar for the ones which are significant. 

Direct comparison for the coefficients of the expansion for the weights of the functional covariates are not meaningful and we constructed the functional boxplot for the posterior estimates of the weight functions as shown in Figures \ref{fig:ZIP_weights_class1} and \ref{fig:ZIP_weights_class2}.  Notice that apparently, none of the functional covariates are significant for the regression term for Low Incidence Class whereas only the covariates ``Activity'' and, to a lesser extent, ``Number of Rest Times" and ``Average Rest Times" are  significant to determine the probabilities of the High Incidence latent class.   

\begin{table}[]
    \centering
    \begin{tabular}{c|c|c|ccccc} \hline
  &  & & \multicolumn{5}{c}{Quantiles - MCMC} \\  
   Parameter & VB-Mean (SD) & MCMC Mean & 2.5\% & 25\%  & 50\% &  75\% &  97.5\% \\ \hline  
        $\lambda_1$  & 1.60 (.18) & 1.62 (.26) & 1.18  & 1.40 &  1.65 &  1.80 &  2.16 \\
        $\lambda_2$  &  8.45 (.53) & 8.52 (.60)  & 7.58 & 8.06 & 8.46 & 8.99 & 9.39 \\ \hline
  \end{tabular}
        \caption{VB Mean and Posterior mean and quantiles for the parameters according to MCMC sample for logit link and Normal prior for Zero Inflated mixture of two Poisson distributions.}
    \label{tab:ZIP_quantiles_cows}
\end{table}

\begin{table}[] 
    \centering
    {\small
    \begin{tabular}{c|ccccc} \hline
         &  Intercept & $\theta_1$ & $\theta_2$ & $\theta_3$ & $\theta_4$ \\ \hline
         Low Intensity & \multicolumn{5}{c}{} \\ \hline
MCMC  & 3.40 (2.74) & 0.006 (.002) & -0.011 (.004) & -0.025 (.294) & 1.030 (.245) \\
VB  & 0.310 (.988) &   0.007 (.003) & -0.010 (.005) &  0.163 (.532) &  0.886 (.265) \\ \hline
High Intensity & \multicolumn{5}{c}{} \\ \hline
MCMC & 0.727 (.232) &  -0.007 (.001) &  0.020 (.001) & -0.545 (.132) & -0.641 (.100)  \\
VB &  0.178 (.989) & -0.005 (.004) &  0.022 (.005) & -0.535 (.620) & -0.553 (.354) \\ \hline 
    \end{tabular}
    \caption{Mean and standard deviation for the estimated coefficients for the Intercept and scalar variables for MCMC and VB algorithms.} \label{tab:ZIP-coef}}
\end{table}

\begin{figure}
    \begin{minipage}{.6\linewidth}
    \includegraphics[height=4cm,width=10cm]{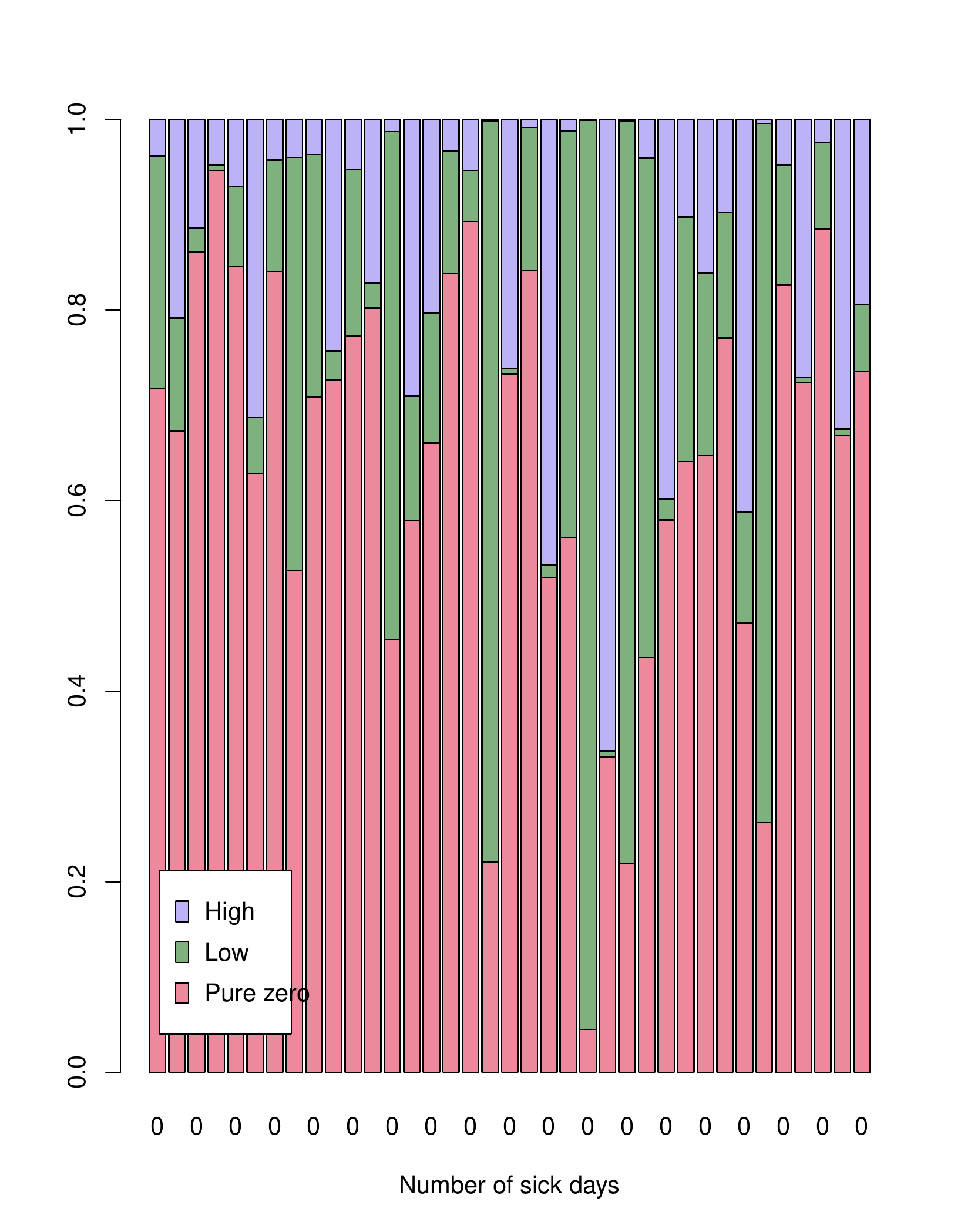}
    \end{minipage}
    \begin{minipage}{.35\linewidth}
    \includegraphics[height=4cm,width=6cm]{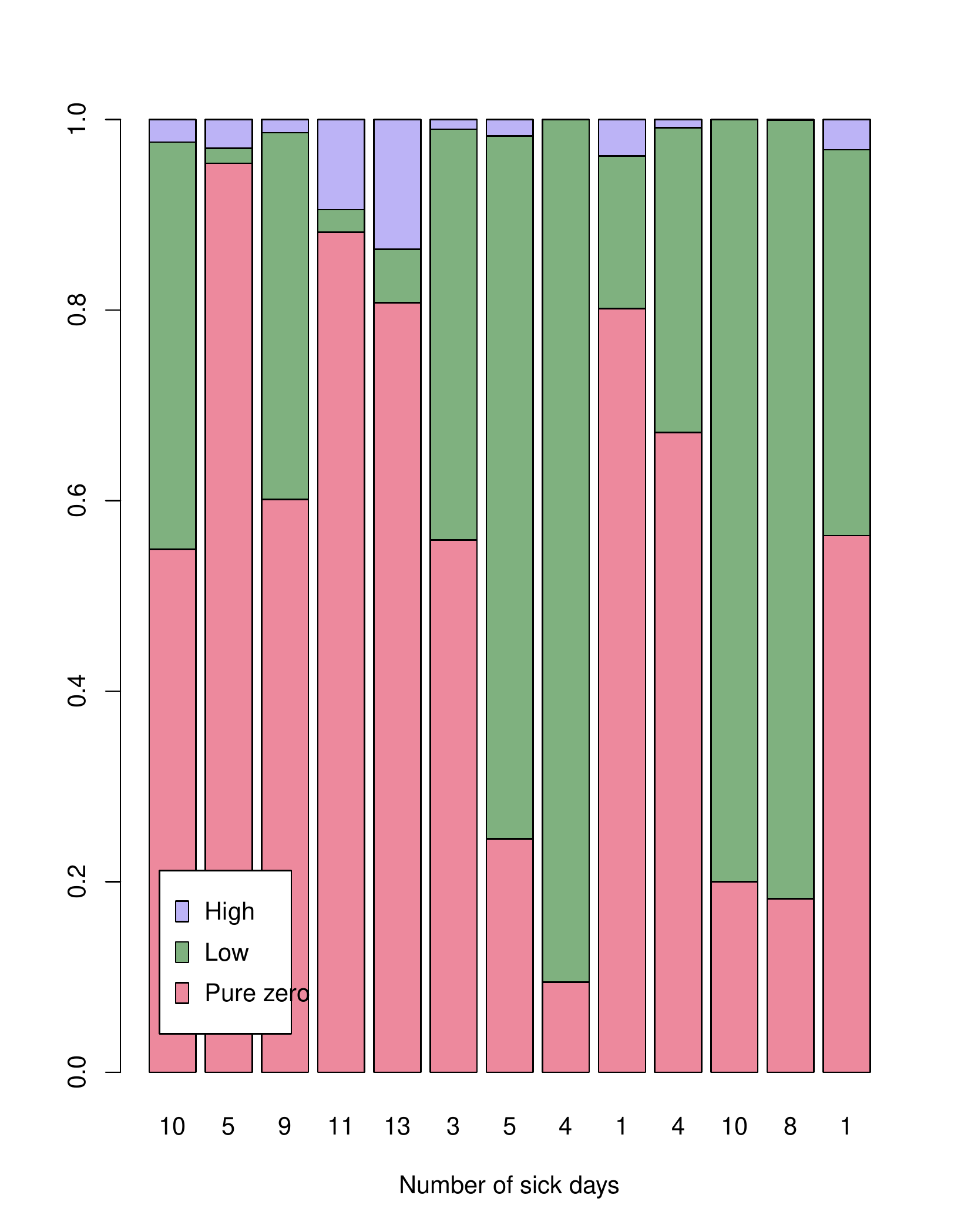}
    \end{minipage}
    \caption{Predicted probability of Classes for test set using the ZIP model. The numbers under the bars the number of sick days.}
    \label{fig:ZIP_gamma-pred}
\end{figure}

\begin{table}[h!]
\caption{Frequency table f estimated probabilities $\hat{p}_{0i}$, $\hat{p}_{1i}$ and $\hat{p}_{2i}$ for the Zero Inflated mixture of Poisson using MCMC posterior mean and VB.} \label{table:cows}
\begin{tabular}{l|ccc|ccc} \hline 
& \multicolumn{3}{c}{MCMC} & \multicolumn{3}{c}{VB} \\
Probability & ``Pure zero" & Class 1 & Class 2 & `Pure zero" & Class 1 & Class 2 \\ \hline
0.00 - 0.10 & 70 & 132 & 176 & 70 & 146 & 176 \\ \hline
0.11 - 0.20  & 0 & 20 & 4 & 0 & 9  & 0\\ \hline
0.21 - 0.30  & 0 & 7 & 0  & 1 & 5 & 1\\ \hline
0.31 - 0.40  & 1 & 4 & 0 & 0 & 2 & 1 \\ \hline
0.41 - 0.50  & 1 & 1 & 0 & 1 & 2 & 2\\ \hline
0.51 - 0.60  & 0 & 1 & 0 & 2 & 3 & 0\\ \hline
0.61 - 0.70  & 3 & 1 & 1 & 2 & 1 & 0\\ \hline
0.71 - 0.80  & 4 & 0 & 0 & 4 & 2 & 1\\ \hline
0.81 - 0.90  & 20 & 3 & 0 & 9 & 0 & 0\\ \hline
0.91 - 1.00  & 109 & 39 & 27 & 119 & 38 & 27\\ \hline
Total & 208 & 208 & 208 & 208  & 208 & 208 \\ \hline
\end{tabular}
\end{table}

 \begin{figure}
    \centering
    \includegraphics[height=4cm,width=10cm]{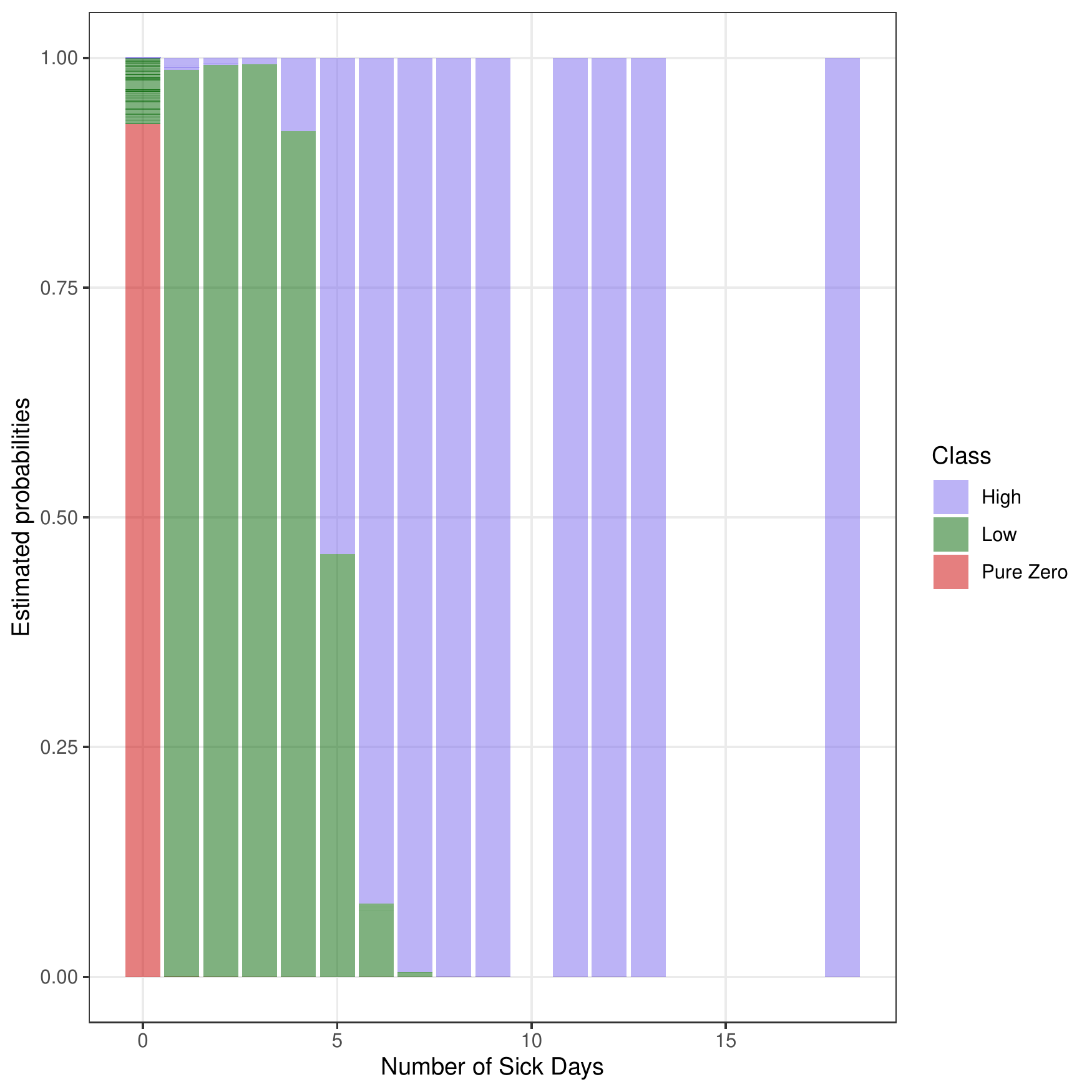}
    \caption{Barplot for the estimated probability of classes Pure Zero, Low Intensity and High Intensity according to the the number of sick days for the ZIP model.}
    \label{fig:ZIP_gamma_est}
\end{figure}

\begin{figure}
    \centering
    \includegraphics[height=6cm,width=10cm]{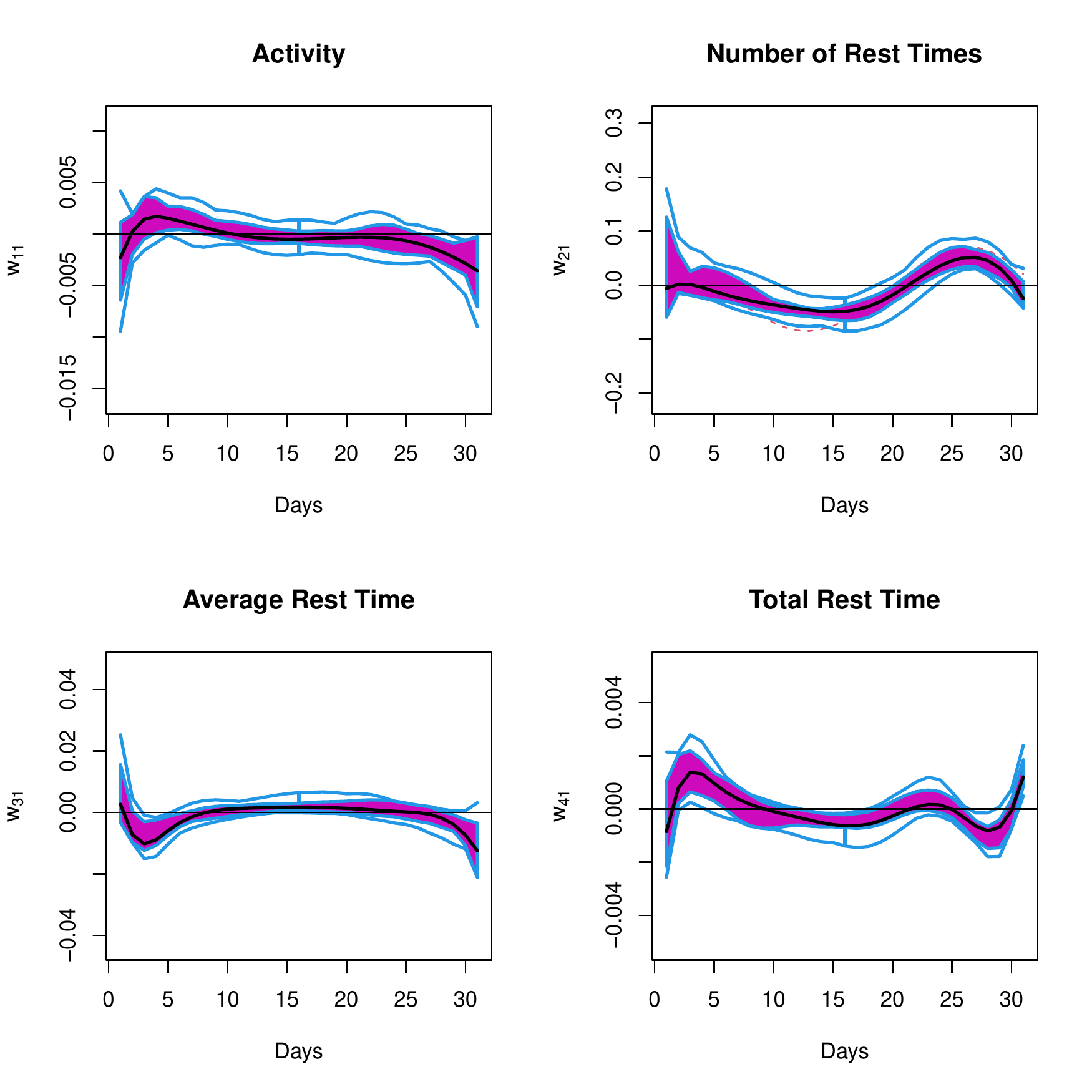}
    \caption{Functional boxplots for the weight functions corresponding Class 1 (Low Intensity) and  covariates ``Activity'', ``Number of Rest Times", ``Average Rest Time'' and ``Total Rest Time'' respectively, under linear model with logit link function and Normal prior.} 
    \label{fig:ZIP_weights_class1}
\end{figure}

\begin{figure}
    \centering
    \includegraphics[height=6cm,width=10cm]{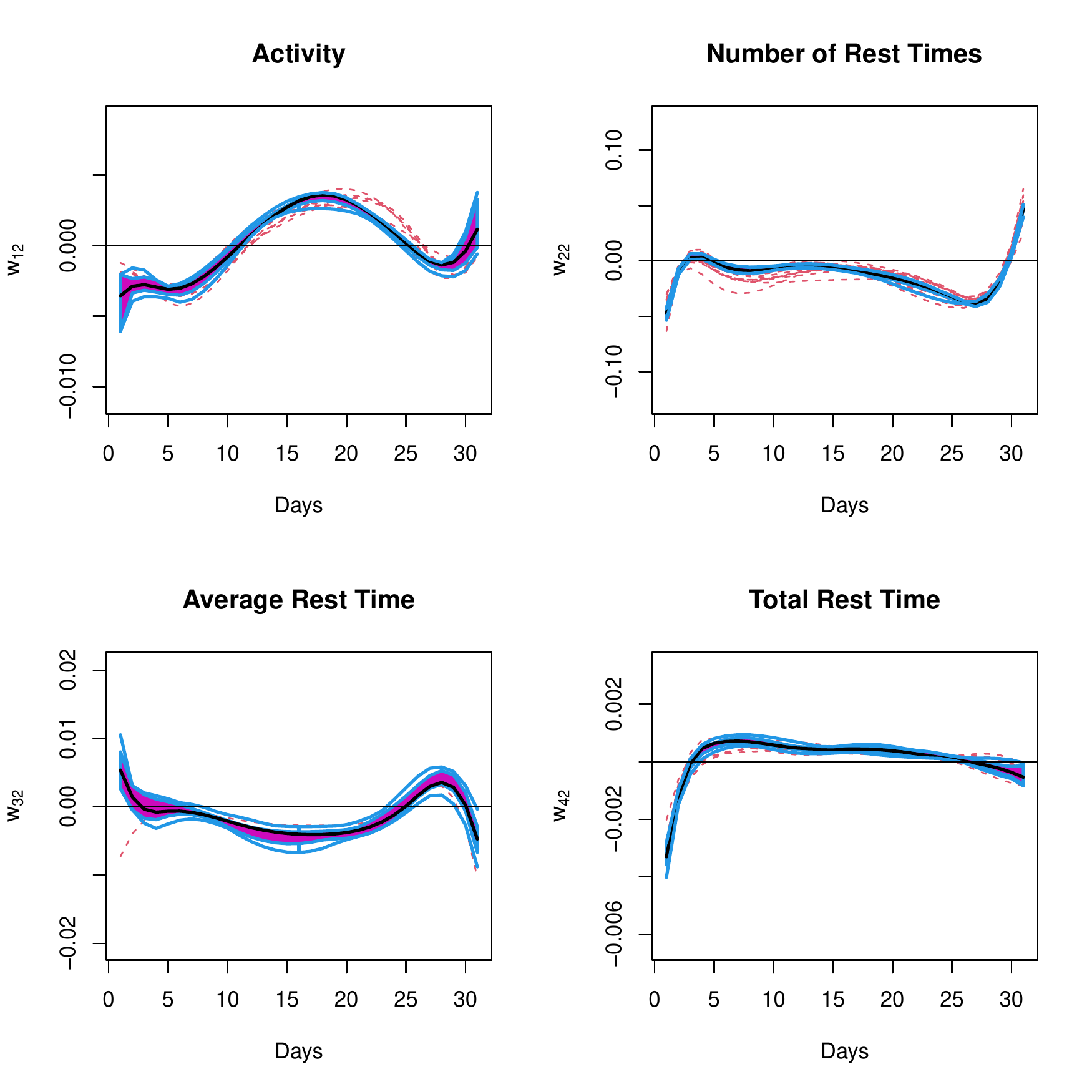}
    \caption{Functional boxplots for the weight functions corresponding Class 2 (High Intensity) and  covariates ``Activity'', ``Number of Rest Times", ``Average Rest Time'' and ``Total Rest Time'' respectively, under linear model with logit link function and Normal prior.} 
    \label{fig:ZIP_weights_class2}
\end{figure}

\section{Discussion}

In this paper, we have considered a mixture model driven by latent variables. We used
a semi-parametric regression model incorporating functional covariates as predictors for
the latent group membership. The main features of our methodology are:
\begin{enumerate}
\item the non-parametric approach of expanding the unknown functions $w_j$ and $F_j$ into B-splines basis reduces the dimension of the problem;
\item linear and non-linear regressions can be implemented; 
\item it can be used with any link function $g$;
\item it incorporates Student $t$ prior information for the regression coefficients applying an efficient approximate EM algorithm already implemented in R; 
\item it can be used with any distributions; 
\item the functional covariates do not need to be observed concurrently and they can even be different for each subject;
\item in the case of the linear model, it has the added advantage of interpretability of
the weight functions which might naturally incorporate prior information that is available to experts in the field.
\end{enumerate}

To show the strength of our method, and also to compare the performance of the MCMC with Variational Bayes, we  ran several simulation scenarios with different link functions and prior distributions. Also, we analyzed two datasets, one comes from a placebo controlled clinical trial to investigate whether the EEG alpha and theta powers can be used to identify an early placebo responder, and the other dataset comes from the Animal Health Monitoring System (AHMS) - USDA project from the Dairy Cattle Biology and Management Laboratory at Cornell University to study the factors that affect the health of lactating cows.

\paragraph{Acknowledgments} This work was partially financed by NIMH grant 5 R01 MH099003, USDA National Institute of Food and Agriculture Animal Health program award 2017-67015-26772 to Julio Giordano, FAPESP grants 2017/15306-9, 2018/06811-4 and  2019/10800-0, CNPq grants 302598/2014-6, 442012/2014-4 and 304148/2020-2. A special thanks to Alberto Saa for helping to solve a computational bottleneck and Guilherme J.M. Rosa for fruitful discussions. Any opinions, findings, conclusions, or recommendations expressed in this publication are those of the author(s) and do not necessarily reflect the view of the National Institute of Food and Agriculture (NIFA) or the United States Department of Agriculture(USDA).

\bibliography{early-methodological.bib}{}


\appendix
\section{Mixture of normal distributions}

\subsection{Full Conditional posterior distributions}

\subsubsection{Full Conditional posterior distribution of $\gamma_i$} 
Let $\bgamma_{-i}$ be the vector $\bgamma = (\gamma_1,\ldots,\gamma_{i-1},\gamma_{i+1},\ldots,\gamma_n)$. The full conditional posterior distribution of $\gamma_i$ is given by
\begin{eqnarray*}
\lefteqn{P(\gamma_i=1|\bTheta,\mathbf{y},\bgamma_{-i})  = \frac{P(\gamma_i = 1,\bTheta,\mathbf{y},\bgamma_{-i})}{P(\bTheta,\mathbf{y},\bgamma_{-i})} } \\ \nonumber
&=& \frac{P(\gamma_i = 1,\bTheta,\mathbf{y},\bgamma_{-i})}{P(\gamma_i=1,\bTheta,\mathbf{y},\bgamma_{-i})+P(\gamma_i=0,\bTheta,\mathbf{y},\bgamma_{-i})} \\ \nonumber
&=& \frac{\phi\left(\frac{y_i-\mu_1}{\sigma}\right) g^{-1}\left(\mathbf{z}_{i} \btheta + \sum_{j=1}^{J} \bphi_{j} \mathbf{R}_{ij}\right)}{\phi\left(\frac{y_i-\mu_1}{\sigma}\right) g^{-1}\left(\mathbf{z}_{i} \btheta + \sum_{j=1}^{J} \bphi_{j} \mathbf{R}_{ij}\right)+\phi\left(\frac{y_i-\mu_0}{\sigma}\right) \left(1-g^{-1}\left(\mathbf{z}_{i} \btheta + \sum_{j=1}^{J} \bphi_{j} \mathbf{R}_{ij}\right)\right)}
\end{eqnarray*}
since $P(\gamma_i = x,\bTheta,\mathbf{y},\bgamma_{-i}) = \phi\left(\frac{y_i-\mu_1 x - \mu_0 (1-x)}{\sigma}\right) g^{-1}\left(\mathbf{z}_{i} \btheta + \sum_{j=1}^{J} \bphi_{j} \mathbf{R}_{ij}\right)$. Here, $g(\cdot)$ is link function.

 \subsubsection{Full Conditional posterior distribution of $\mu_0$ and $\mu_1$} \label{ap:fulleta}
 We update $\mu_0$ using a normal distribution with mean $$\left(1/\sigma_{0}^{2}+\sum_{i=1}^{n} I(\gamma_{i}=0)/\sigma^{2}\right)^{-1} \sum_{i=1}^{n} y_i I(\gamma_{i}=0)$$ and variance $$\left(1/\sigma_{0}^{2}+\sum_{i=1}^{n} I(\gamma_{i}=0)/\sigma^{2}\right)^{-1}.$$

 On the other hand, conditionally on $\mu_0$, $\mu_1$ is updated with a truncated normal distribution  with mean $$\left(1/\sigma_{0}^{2}+\sum_{i=1}^{n} I(\gamma_{i}=1)/\sigma^{2}\right)^{-1} \sum_{i=1}^{n} y_i I(\gamma_{i}=1)$$ and variance $$\left(1/\sigma_{0}^{2}+\sum_{i=1}^{n} I(\gamma_{i}=1)/\sigma^{2}\right)^{-1}$$
 restricted to $(\mu_0,\infty)$.

 \subsubsection{Full Conditional posterior distribution of $\sigma^2$} \label{ap:fullsigma}

 We update $\sigma^{2}$ using an inverse-gamma distribution with parameters $$a_0 \quad \mbox{and} \quad  b_0+\sum_{i=1}^{n} (\eta_0+\eta_1 \, \gamma_{i})^{2},$$ with $a_0,b_0,c_0$ and $d_0$ are the parameters of the prior distributions.

\subsection{Full conditional distribution for $\bbeta$}

\noindent 
Update $[\bdelta_{ij}|.,v_{i},x_{ij}] \sim N(\mathbf{M}^{-1} (\bbeta_{j}[\mu_{v}(-j)-v_i+\mu_{x}(-j)-x_{ij}]+\bDelta^{-1} \bmu_{\delta}),\mathbf{M}^{-1})$, with 
\begin{eqnarray*}
\mathbf{M} &=& \bbeta_{j} \, \bbeta_{j}^{\top} \left( \frac{\sigma^{2}_{x}+1}{\sigma^{2}_{x}}\right) + \bDelta^{-1}  \nonumber \\
\mu_{v}(-j) &=& \mathbf{z}_{i}^{\top} \btheta_{i} + \sum_{j^{'} \ne j} \bbeta_{j^{'}}^{\top} \delta_{i,j^{'}} \nonumber \\
\mu_{x}(-j) &=& \sum_{j^{'} \ne j} \bbeta_{j^{'}}^{\top} \delta_{i,j^{'}} \nonumber \\
\end{eqnarray*}

In detail:
\begin{eqnarray*}
[\bdelta_{ij}|.,v_{i},x_{ij}] &\propto& \phi(v_i;\mu_{v},1) \times \phi(x_{ij};\mu_{x},\sigma_{x}^{2}) \times \phi(\delta_{ij};\bmu_{v\delta},\bDelta) \nonumber \\
&=& exp\left\{ -\frac{1}{2} \left[ \bdelta_{ij}^{\top} \bbeta_{j} \bbeta_{j}^{\top} \bdelta_{ij} - 2 \bdelta_{ij}^{\top} \bbeta_{j} (\mu_{v}(-j) - v_i) \right. \right. \nonumber \\
&&\;\; + \bdelta_{ij}^{\top} \left[ \frac{(\bbeta_{j} \bbeta_{j}^{\top})}{\sigma^{2}_{x}} \right] \bdelta_{ij} - 2 \bdelta_{ij}^{\top} \bbeta_{j} \left[ \frac{(\mu_{v}(-j) - v_i)}{\sigma^{2}_{x}} \right]  \nonumber \\ 
&&\;\;+ \left. \left.  \bdelta_{ij}^{\top} \bDelta^{-1} \bdelta_{ij} - 2 \bdelta_{ij}^{\top} \bDelta^{-1} \bmu_{\delta} \right]  \right\} \nonumber
\end{eqnarray*}

\subsection{Variational distributions}  \label{ap:vb_normal}

\subsubsection{Computing $q^*(\gamma_i)$} 

 ({\it cf.} Equations (22), (23) and (24) from \cite{blei2017variational}) 
\begin{eqnarray*}
\lefteqn{\log q^*(\gamma_i) =} \\
&=& C_{1i} + \{ E_{\bnu(-\gamma_i)}[\log p(\gamma_i|\btheta,\bphi) + \log p(y_i|\gamma_i,\mu)] \} \nonumber \\
&=& C_{1i} + E_{\bnu(-\gamma_i)} \left\{ \gamma_i \log \left[g^{-1}\left(\mathbf{z}_{i} \btheta +  \sum_{j=1}^J \bphi_j \mathbf{R}_{ij}\right)\right] + (1-\gamma_i) \log \left[1-g^{-1}\left(\mathbf{z}_{i} \btheta +  \sum_{j=1}^J \bphi_j \mathbf{R}_{ij}\right)\right] \right. \nonumber \\
& & \hspace{3cm}\left. - \gamma_i \frac{1}{2\sigma^{2}} (y_i-\mu_1)^{2} - (1-\gamma_i) \frac{1}{2 \sigma^{2}} (y_i-\mu_0)^{2} \right\} \nonumber \\
&=& C_{1i} + \gamma_i E_{\bnu(-\gamma_i)} \left\{ \log \left[g^{-1}\left(\mathbf{z}_{i} \btheta +  \sum_{j=1}^J \bphi_j \mathbf{R}_{ij}\right)\right] - \frac{1}{2\sigma^{2}} (y_i-\mu_1)^{2} \right\} \nonumber \\
&& \hspace{.8cm}+(1-\gamma_i) E_{\bnu(-\gamma_i)} \left\{ \log \left[1-g^{-1}\left(\mathbf{z}_{i} \btheta +  \sum_{j=1}^J \bphi_j \mathbf{R}_{ij}\right)\right] -\frac{1}{2\sigma^{2}}(y_i-\mu_0)^{2} \right\} \nonumber \\
&=& C_{1i}+ \gamma_i \left\{ E_{q^*(\btheta,\bphi)} \log \left[g^{-1}\left(\mathbf{z}_{i} \btheta +  \sum_{j=1}^J \bphi_j \mathbf{R}_{ij}\right)\right] - E_{q^*(\sigma^{2})} \left( \frac{1}{2\sigma^{2}} \right) E_{q^*(\mu_1)}(y_i-\mu_1)^{2} \right\} \nonumber \\
&& \hspace{.8cm} + (1-\gamma_i)  \left\{ E_{q^*(\btheta,\bphi)} \log \left[1-g^{-1}\left(\mathbf{z}_{i} \btheta +  \sum_{j=1}^J \bphi_j \mathbf{R}_{ij}\right)\right] - E_{q^*(\sigma^{2})} \left( \frac{1}{2\sigma^{2}} \right) E_{q^*(\mu_0)}(y_i-\mu_0)^{2} \right\} \nonumber \\
&=& C_{1i} +\gamma_i \left\{ E_{q^*(\btheta,\bphi)} \log \left[g^{-1}\left(\mathbf{z}_{i} \btheta +  \sum_{j=1}^J \bphi_j \mathbf{R}_{ij}\right)\right] - E_{q^*(\sigma^{2})} \left( \frac{1}{2\sigma^{2}} \right) [(y_i-m_1)^{2} + s^{2}_{1}] \right\} \nonumber \\
&& \hspace{.8cm} +(1-\gamma_i)  \left\{ E_{q^*(\btheta,\bphi)} \log \left[1-g^{-1}\left(\mathbf{z}_{i} \btheta +  \sum_{j=1}^J \bphi_j \mathbf{R}_{ij}\right)\right] - E_{q^*(\sigma^{2})} \left( \frac{1}{2\sigma^{2}} \right) [(y_i-m_0)^{2} + s^{2}_{0}] \right\} \nonumber 
\end{eqnarray*}
where $C_{1i}=C(\bnu(-\gamma_i), y_i,\bkappa)$. 

\subsubsection{Computing $q^*(\mu_0|\bkappa)$ and $q^*(\mu_1|\bkappa)$}

For $k=0$, we have
\begin{eqnarray*}
\log q^*(\mu_0) &=&  C_{2,k,0} -\frac{\mu_{0}^{2}}{2 \tau^{2}_{0}} - \sum_{i=1}^{n} E_{q^*(\sigma^{2}, \gamma_i)} \left[\frac{1}{2 \sigma^2}(1 - \gamma_i) (y_i - \mu_0)^2\right] \nonumber \\
&=& C'_{2,k,0} -\frac{\mu_{0}^{2}}{2 \tau^{2}_{0}} - \sum_{i=1}^{n} E_{q^*(\sigma^{2})}  \left[\frac{1}{2 \sigma^2}\right] (1 - \alpha_i) (y_i^2 - 2y_i\mu_0 + \mu^2_0). \label{eq:qmu0} 
\end{eqnarray*}

Similarly, for $k=1$,
\begin{eqnarray}
\log q^*(\mu_1) &=&  C_{2,k,1} -\frac{\mu_{1}^{2}}{2 \tau^{2}_{1}} - \sum_{i=1}^{n} E_{q^*(\sigma^{2}, \gamma_i)} \left[\frac{1}{2 \sigma^2} \gamma_i (y_i - \mu_1)^2\right] \nonumber \\
&=& C'_{2,k,1} -\frac{\mu_{1}^{2}}{2 \tau^{2}_{1}} - \sum_{i=1}^{n} E_{q^*(\sigma^{2})} \left[ \frac{1}{2 \sigma^2}\right] \alpha_i (y_i - \mu_0)^2 \nonumber \\
&=& C_{2,k,1} -\frac{\mu_{1}^{2}}{2 \tau^{2}_{1}} - \sum_{i=1}^{n} E_{q^*(\sigma^{2})} \left[ \frac{1}{2 \sigma^2}\right] \alpha_i (y^2_i - 2y_i\mu_1+\mu^2_1)^2 
\label{eq:qmu1*} 
\end{eqnarray}

This calculation reveals that the coordinate-optimal variational density of $\mu_k$, $k=0,1$, are in the exponential family with natural parameters and sufficient statistics given in Table \ref{table:npmuk}.

\begin{table}[htb]
    \centering
    \begin{tabular}{c|c|c}
        & Natural parameters & Sufficient statistics \\ \hline 
  $k=0$       & $\mu_0$ & $E_{q^*(\sigma^{2})} [1/\sigma^2] \sum_{i=1}^n (1-\alpha_i) y_i$ \\
  & $\mu_0^2$ &  $  (1/2) (E_{q^*(\sigma^{2})} [1/\sigma^2] \sum_{i=1}^n (1-\alpha_i) + 1/\tau_0^2)$ \\ \hline
   $k=1$      & $\mu_1$ & $E_{q^*(\sigma^{2})} [1/\sigma^2] \sum_{i=1}^n \alpha_i y_i$ \\
   & $\mu_1^2$ &  $  (1/2) (E_{q^*(\sigma^{2})} [1/\sigma^2] \sum_{i=1}^n \alpha_i + 1/\tau_1^2)$  \\ \hline
    \end{tabular}
    \caption{Natural parameters and sufficient statistics for distributions $q^*(\mu_0)$ and $q^*(\mu_1)$}
    \label{table:npmuk}
\end{table}

 That is, the distribution is Gaussian, expressed in terms of variational mean and variance, the updates for $q^*(\mu_k)$ are

\begin{equation*}
     m_0 =  \frac{E_{q^*(\sigma^{2})} [1/\sigma^2] \sum_{i=1}^n (1-\alpha_i)y_i}{E_{q^*(\sigma^{2})} [1/\sigma^2] \sum_{i=1}^n (1-\alpha_i) +  1/\tau_0^2}, \quad   s_0^2 =  \frac{1}{E_{q^*(\sigma^{2})} [1/\sigma^2] \sum_{i=1}^n (1-\alpha_i) +  1/\tau_0^2}
\end{equation*}
and 
\begin{equation*}
   m_1 =  \frac{E_{q^*(\sigma^{2})} [1/\sigma^2] \sum_{i=1}^n \alpha_iy_i}{E_{q^*(\sigma^{2})} [1/\sigma^2] \sum_{i=1}^n \alpha_i +  1/\tau_1^2}, \quad 
     s_1^2 =  \frac{1}{E_{\sigma^{2}} [1/\sigma^2] \sum_{i=1}^n \alpha_i +  1/\tau_1^2} 
\end{equation*}
where $\tau_0^2$ and $\tau_1^2$ are the parameters from the prior distribution.

\subsubsection{Computing $q^*(\sigma^{2}|\bkappa)$}

 The variational density of $q(\sigma^{2})$ considers the likelihood and the prior distribution of $\sigma^{2} \sim IG(a_0,b_0)$. Therefore,
 
\begin{eqnarray}
\log q(\sigma^{2}) &= & C_{3} + E_{(\mu_0,\mu_1,\bgamma)} \left[\sum_{i=1}^n \gamma_i \log \phi(y_i; \mu_1, \sigma^2) + (1-\gamma_i) \log \phi(y_i; \mu_0, \sigma^2)  \right] \nonumber \\
& & \quad + (a_0 + 1) \log (1/\sigma^2) - \frac{b_0}{\sigma^2} \nonumber \\
& = & C_4 + \frac{n}{2} \log (1/\sigma^2) - \sum_{i=1}^n E_{(\mu_1,\bgamma)} \left[ \frac{\gamma_i}{2 \sigma^2} (y_i - \mu_1)^2 \right] - \sum_{i=1}^n E_{(\mu_1,\bgamma)} \left[ \frac{(1-\gamma_i)}{2 \sigma^2} (y_i - \mu_0)^2 \right] \nonumber \\
& & \quad + (a_0 + 1) \log (1/\sigma^2) - \frac{b_0}{\sigma^2} \nonumber \\
& = & C_4 + \frac{n}{2} \log (1/\sigma^2) - \sum_{i=1}^n  \frac{\alpha_i}{2 \sigma^2} E_{\mu_1}(y_i - \mu_1)^2  - \sum_{i=1}^n  \frac{(1-\alpha_i)}{2 \sigma^2} E_{\mu_0}(y_i - \mu_0)^2  \nonumber \\
& & \quad + (a_0 + 1) \log (1/\sigma^2) - \frac{b_0}{\sigma^2} \nonumber \\
& = & C_4 + \frac{n}{2} \log (1/\sigma^2) - \sum_{i=1}^n  \frac{\alpha_i}{2 \sigma^2} [(y_i - m_1)^2 + s_1^2]  - \sum_{i=1}^n  \frac{(1-\alpha_i)}{2 \sigma^2} [(y_i - m_0)^2 + s_0^2]  \nonumber \\
& & \quad + (a_0 + 1) \log (1/\sigma^2) - \frac{b_0}{\sigma^2} \nonumber \\
\end{eqnarray}

which is an inverse gamma with parameters $A_0 = a_0+n/2$
and $ B_0 = b_0 + B_{0,q^*}$ 
where
$$ B_{0,q^*} =    \sum_{i=1}^n  \frac{\alpha_i}{2} [(y_i - m_1)^2 + s_1^2]  + \sum_{i=1}^n  \frac{(1-\alpha_i)}{2} [(y_i - m_0)^2 + s_0^2].$$

\subsection{Calculating the ELBO} \label{ap:elbo_normal}

The ELBO is given by
\begin{eqnarray*}
\mbox{ELBO}(\bkappa) &=&  
\sum_{i=1}^{n} (E_{q^*}[\log p(y_i|\gamma_i,\mu_0,\mu_1,\sigma^{2}, \bbeta)] + E_{q^*}[\log p(\gamma_i)] \nonumber \\
&+& \sum_{k=0}^{1} E_{q^*}[\log p(\mu_k)] + E_{q^*}[\log p(\sigma^{2})] + E_{q^*}[\log p(\bbeta)] \nonumber \\
&-& \sum_{i=1}^{n} (E_{q^*}[\log q^*(\gamma_i)] - \sum_{k=0}^{1} E_{q^*}[\log q^*(\mu_k)] \nonumber \\ 
&-& E_{q^*}[\log q^*(\sigma^{2})] - E_{q^*}[\log q^*(\bbeta)] \nonumber \\
& = & E_0 + E_1 + E_2 + E_3 + E_4 + E_5  - F_1 - F_2 - F_3 - F_4 - F_5
\end{eqnarray*}
with the expectation is taken with respect to $q^*(\mu_0,\mu_1,\sigma^{2},\bgamma,\bbeta|\bkappa)$, that is $E_{q^*} := E_{q^*(\mu_0,\mu_1,\sigma^{2},\bgamma,\bbeta|\bkappa)}$. Recall that $q^*(\mu_0,\mu_1,\sigma^{2},\bgamma,\bbeta) = q^*(\mu_0)q^*(\mu_1)q^*(\sigma^{2})q^*(\bgamma)q^*(\bbeta)$. Let $\Psi(\cdot)$ denote the digamma function. \\

Therefore, to compute the ELBO we need to compute the following pieces:

\begin{eqnarray*}
E_0 & = &  E_{q^*}(\log p(\mathbf{y}|\mu_0,\mu_1,\sigma^{2}, \bgamma))  \nonumber \\
& = & \sum_{i=1}^{n} E_{q^*}[\gamma_i \log \phi(y_i;\mu_1,\sigma^{2}) + (1-\gamma_i) \log \phi(y_i;\mu_0,\sigma^{2})] \nonumber \\
&=& -\frac{n}{2} \log 2\pi - \frac{1}{2} E_{q^*}(1/\sigma^{2}) \sum_{i=1}^{n} \{\alpha_i [(y_i-m_1)^{2} + s_{1}^{2}] + (1-\alpha_i) [(y_i-m_0)^{2} + s_{0}^{2}]  \} \nonumber \\
& & \quad - (n/2) E_{q^*}(\log \sigma^{2}) \nonumber \\
&=& -\frac{n}{2} \log 2\pi -\frac{1}{2} \frac{A_0}{B_0} \sum_{i=1}^{n} \{\alpha_i [(y_i-m_1)^{2} + s_{1}^{2}] + (1-\alpha_i) [(y_i-m_0)^{2} + s_{0}^{2}]  \} \nonumber \\
& & \quad + \frac{n}{2} \left(\log(B_0) - \Psi(A_0)\right).
\end{eqnarray*}

Since, 
\begin{eqnarray*}
E_{q^*}(\log p(\gamma_i)) &=& E_{q^*}[\gamma_i \log [g^{-1}(\mathbf{x}_{i} \bbeta)] + (1-\gamma_i) \log [1-g^{-1}(\mathbf{x}_{i} \bbeta)] \nonumber \\
&=& \alpha_i \int \log g^{-1}(\mathbf{x}_{i} \bbeta) q^*(\bbeta) \, d\bbeta] + (1-\alpha_i) \int \log [1-g^{-1}(\mathbf{x}_{i} \bbeta)] q^*(\bbeta) \, d\bbeta, 
\end{eqnarray*}
we have
\begin{equation}
\label{eq:e1}
    E_1 = \sum_{i=1}^n \left[\alpha_i \int \log g^{-1}(\mathbf{x}_{i} \bbeta) q^*(\bbeta) \, d\bbeta] + (1-\alpha_i) \int \log [1-g^{-1}(\mathbf{x}_{i} \bbeta)] q^*(\bbeta) \, d\bbeta \right]
\end{equation}
with $q^*(\bbeta)$ is computed in Section \ref{sec:VBbeta}. The high dimensional integral in \eqref{eq:e1} can be computed efficiently transforming it into a one-dimensional integral as described in Appendix \ref{ap:fast}. 

Also, for $k=0,1$, we have 
\begin{eqnarray}
E_{q^*}(\log p(\mu_k)) &=& E_{q^*}(\log \phi(\mu_k;0,\sigma^{2}_{\mu})) \nonumber \\
&=& -\frac{1}{2} \log (2\pi \tau^{2}_{k})  -(1/2\tau^{2}_{k}) E_{q^*}(\mu_k^{2}) \nonumber \\
&=& -\frac{1}{2} \log (2\pi  \tau^{2}_{k})  -\frac{1}{2\tau^{2}_{k}} (m_{k}^{2} + s^{2}_{k}).  \nonumber 
\end{eqnarray}

Then, 
\begin{equation*}
E_2  =  -\frac{1}{2} \log (2\pi  \tau^{2}_{0})  -\frac{1}{2\tau^{2}_{0}} (m_{0}^{2} + s^{2}_{0})
\end{equation*}
and 
\begin{equation*}
E_3  =  -\frac{1}{2} \log (2\pi  \tau^{2}_{1})  -\frac{1}{2\tau^{2}_{1}} (m_{1}^{2} + s^{2}_{1}). 
\end{equation*}

Moreover,
\begin{eqnarray*}
E_4 & = & E_{q^*}(\log p(\sigma^{2})) \,=\, E_{q^*}  \left[\log \left( \frac{b_0^{a_0}}{\Gamma(a_0)} (1/\sigma^2)^{a_0 + 1} \exp \left( \frac{-b_0}{\sigma^2} \right) \right) \right] \nonumber  \\
& = & a_0 \log b_0 - \log(\Gamma(a_0)) + (a_0 + 1)  \log\left(B_0 - \Psi(a_0 + n/2) \right)  - b_0 A_0 B_0^{-1}.
\end{eqnarray*}

Finally,
\begin{eqnarray*}
E_5 & = & E_{q^*}(\log p(\bbeta))  \,=\, E_{q^*}[\log \phi(\bbeta;\bmu_{\bbeta},\Sigma_{\bbeta})] \nonumber \\
&=& - \frac{R}{2} \log 2 \pi - \frac{1}{2} \log |\Sigma_{\bbeta}| - (1/2) E_{q^*}\left[(\bbeta - \bmu_{\bbeta})^{\top} \Sigma_{\bbeta}^{-1} (\bbeta - \bmu_{\bbeta})\right] \nonumber \\
&=& - \frac{R}{2} \log 2 \pi - \frac{1}{2} \log |\Sigma_{\bbeta}|  - \frac{1}{2} tr(\Sigma_{\bbeta}^{-1} \Sigma_{q^*(\bbeta)}) - \frac{1}{2} [(\bmu^* - \bmu_{\bbeta})^{\top} \Sigma_{\bbeta}^{-1} (\bmu^* - \bmu_{\bbeta})]
\end{eqnarray*}
where $R$ is the dimension of the $\bbeta$ vector. \\

On the other hand,
\begin{eqnarray*}
F_1 & = & \sum_{i=1}^n E_{q^*}(\log q^*(\gamma_i)) \nonumber \\ 
 &=&  \sum_{i=1}^n [\alpha_i \log \alpha_i + (1-\alpha_i) \log (1 - \alpha_i)].
\end{eqnarray*}

Also,
\begin{eqnarray*}
F_2 & = & E_{q^*}(\log q^*(\mu_0))  \,=\, E_{q^*}(\log \phi(\mu_0;m_0,s^{2}_{0})) \nonumber \\
&=& -\frac{1}{2} \log 2\pi -\frac{1}{2} \log s^2_{0} - \frac{1}{2s^{2}_{0}} E_{q^*} (\mu_0 - m_0)^{2} \nonumber \\
&=& -\frac{1}{2} \log 2\pi - \frac{1}{2} \log s^2_{0}(1/2s^{2}_{0}) - \frac{1}{2}\nonumber \\ 
&=& -\frac{1}{2} \log 2\pi - \frac{1}{2}- \frac{1}{2} \log s^2_{0}(1/2s^{2}_{0}). 
\end{eqnarray*}

Analagously, 
\begin{eqnarray*}
F_3 & = & E_{q^*}(\log q^*(\mu_1)) \nonumber \\ 
 &=& E_{q^*}(\log \phi(\mu_k;m_k,s^{2}_{k})) \nonumber \\
&=& -\frac{1}{2} \log 2\pi -\frac{1}{2} \log s^2_{k} - \frac{1}{2s^{2}_{k}} E_{q^*} (\mu_k - m_k)^{2} \nonumber \\
&=& -\frac{1}{2} \log 2\pi - \frac{1}{2} \log s^2_{k}(1/2s^{2}_{k}) - \frac{1}{2}\nonumber \\ 
&=& -\frac{1}{2} \log 2\pi - \frac{1}{2}- \frac{1}{2} \log s^2_{1}(1/2s^{2}_{1}) 
\end{eqnarray*}

Moreover, 
\begin{eqnarray*}
F_4 & = & E_{q^*}(\log q^*(\sigma^{2})) \nonumber \\
 &=& E_{q^*}  \left[\log \left( \frac{B_0^{A_0}}{\Gamma(A_0)} (1/\sigma^2)^{A_0+ 1} \exp \left( \frac{-B_0}{\sigma^2} \right) \right) \right] \nonumber  \\
& = &  E_{q^*}  \left[ A_0 \log B_0 - \log \Gamma(A_0) + (A_0 + 1) \log (1/\sigma^2) - \frac{B_0}{\sigma^2} \right] \nonumber \\
& = &  A_0 \log B_0 - \log \Gamma(A_0) + (A_0 + 1) E_{q^*}\left[\log (1/\sigma^2)\right] - B_0 E_{q^*}\left[1/\sigma^2\right]  \nonumber \\
& = & A_0 \log B_0 - \log \Gamma(A_0) + (A_0 + 1) \left(\log(B_0) - \Psi(A_0)\right)  - A_0.
\end{eqnarray*}

Finally,
\begin{eqnarray*}
F_5 & = & E_{q^*}(\log q^*(\bbeta)) \nonumber \\
 &=& E_{q^*}[\log \phi(\bbeta;\bmu_{q^*(\bbeta)},\Sigma_{q^*(\bbeta}))] \nonumber \\
&=& - \frac{R}{2} \log 2 \pi - \frac{1}{2} \log |\Sigma_{q^*(\bbeta)}| - (1/2) E_{q^*(\bbeta)}[(\bbeta - \bmu_{q^*(\bbeta)})^{\top} \Sigma_{q^*(\bbeta)}^{-1} (\bbeta - \bmu_{q^*(\bbeta}))] \nonumber \\
&=& - \frac{R}{2} \log 2 \pi - \frac{1}{2} \log |\Sigma_{q^*(\bbeta)}|  - \frac{1}{2} tr(\Sigma_{q^*(\bbeta)}^{-1} \Sigma_{q^*(\bbeta)}) \nonumber \\
&=& - \frac{R}{2} \log 2 \pi - \frac{1}{2} \log |\Sigma_{q^*(\bbeta)}|  - \frac{R}{2}.
\end{eqnarray*}

\subsection{A fast way to compute $E_{q^*}^*(\bbeta)[\log g^{-1}({\bf x}' \bbeta)]$} \label{ap:fast}

Let $\mathbf{x} \in \mathbb{R}^d$ and $F: \mathbb{R}^d \rightarrow \mathbb{R}^d $, we want to compute
$$E_{q^*}[F({\bf x}' \bbeta)] =  \int_{\mathbb{R}^d} F(x'\bbeta) \frac{1}{(2\pi)^{p/2} \sqrt{\det(\Sigma)}} \exp \left( -\frac{1}{2}  (\bbeta - \bmu)' \Sigma^{-1} (\bbeta - \bmu) \right) d\bbeta.$$

Construct an orthonormal matrix $S$ such as the first column of $S$ is $x/\|x\|$ and let $S \bzeta = (\bbeta - \mu)$. Since $S$ is orthonormal, the Jacobian of this transformation is 1, $S'= S^{-1}$, $(S'\Sigma^{-1} S)^{-1} = (S'\Sigma S)$ and $\det(S'\Sigma S) = \det\Sigma$.  Therefore,
\begin{eqnarray}
\lefteqn{E_{q^*}[F({\bf x}' \bbeta)] =} \nonumber \\
&=& \int_{\mathbb{R}^d} F(x'S \bzeta+ x'\bmu) \frac{1}{(2\pi)^{p/2} \sqrt{\det(\Sigma)}} \exp \left( -\frac{1}{2}  \bzeta' S'\Sigma^{-1} S \bzeta \right) d\bzeta \nonumber \\
& = & \int_{\mathbb{R}} F(\|x\| \zeta_1+ x'\bmu) \int_{\mathbb{R}^{p-1}} \frac{1}{(2\pi)^{p/2} \sqrt{\det(\Sigma)}} \exp \left( -\frac{1}{2}  \bzeta' S'\Sigma^{-1} S \bzeta \right) d\zeta_p \ldots d\zeta_2 d\zeta_1  \nonumber \\
& = & \int_{\mathbb{R}} F(\|x\| \zeta_1+ x'\bmu) \frac{1}{(2\pi)^{1/2} \sqrt{\Sigma_{(1,1)}}} \exp \left( -\frac{1}{2}  \zeta_1 (S'\Sigma^{-1} S)_{(1,1)} \zeta_1 \right)  d\zeta_1. \label{eq:int1}
\end{eqnarray}

\section{Zero Inflated mixture of Poisson distributions}

\subsection{Full Conditional posterior distribution of $\gamma_i$} 
Let $\bgamma_{-i0}$, $\bgamma_{-i1}$ and $\bgamma_{-i2}$ be the vector $\bgamma_{j} = (\bgamma_{1j},\ldots,\bgamma_{nj})$ without observation $\bgamma_{ij}$, $j=0,1,2$, respectively. The full conditional posterior distribution of $\bgamma_i$ is given by
\begin{eqnarray*}
P(\gamma_{i1}=1|\bTheta,\mathbf{y},\bgamma_{-i1}) &=&
 \frac{e^{-\lambda_{1}} \lambda_{1}^{y_i} \exp({\bf x}_i \bbeta_1)}{I(y_i=0) + e^{-\lambda_{1}} \lambda_{1}^{y_i} \exp({\bf x}_i \bbeta_1) + e^{-\lambda_{2}} \lambda_{2}^{y_i} \exp({\bf x}_i \bbeta_2) }, 
\end{eqnarray*}

\begin{eqnarray*}
P(\gamma_{i2}=1|\bTheta,\mathbf{y},\bgamma_{-i2}) &=&
 \frac{e^{-\lambda_{2}} \lambda_{2}^{y_i} \exp({\bf x}_i \bbeta_2)}{I(y_i=0) + e^{-\lambda_{1}} \lambda_{1}^{y_i} \exp({\bf x}_i \bbeta_1) + e^{-\lambda_{2}} \lambda_{2}^{y_i} \exp({\bf x}_i \bbeta_2) }, 
\end{eqnarray*}
and
\begin{eqnarray*}
P(\gamma_{i0}=1|\bTheta,\mathbf{y},\bgamma_{-i0}) &=&
 \frac{I(y_i=0)}{I(y_i=0) + e^{-\lambda_{1}} \lambda_{1}^{y_i} \exp({\bf x}_i \bbeta_1) + e^{-\lambda_{2}} \lambda_{2}^{y_i} \exp({\bf x}_i \bbeta_2) }.
\end{eqnarray*}

\subsection{Variational Bayes} \label{ap:vb_zip}

Analogously to the normal case, we define the variational densities as \\$q^*(\bnu|\bkappa) = q^*(\lambda_1|\bkappa) q^*(\lambda_2|\bkappa) q(\bbeta_1|\bkappa) q(\bbeta_1|\bkappa) q(\gamma|\bkappa)$ and have to calculate
\begin{enumerate}
\item $q^*(\lambda_1|\bkappa) \propto \exp\left\{ E_{q^*(\bnu(-\lambda_1))} \log \, p(\lambda_1|\bnu(-\lambda_2),\mathbf{y}),\bkappa  \right\}$
\item $q^*(\lambda_2|\bkappa) \propto \exp\left\{ E_{q^*(\bnu(-\lambda_2))} \log \, p(\lambda_2|\bnu(-\lambda_2),\mathbf{y}),\bkappa  \right\}$
\item $q^*(\bbeta_1|\bkappa) \propto \exp\left\{ E_{q^*(\bnu(-\bbeta_1)|\bkappa)} \log \, p(\bbeta_1|\bnu(-\bbeta_1),\mathbf{y},\bkappa)  \right\}$
\item $q^*(\bbeta_2|\bkappa) \propto \exp\left\{ E_{q^*(\bnu(-\bbeta_2)|\bkappa)} \log \, p(\bbeta_2|\bnu(-\bbeta_2),\mathbf{y},\bkappa)  \right\}$
\item $q^*(\bgamma|\bkappa) \propto \exp \left\{ E_{q^*(\bnu(-\gamma),|\bkappa)} \log \, p(\bgamma|\bnu(-\bgamma),\mathbf{y},\bkappa)  \right\}$
\end{enumerate}
where $\bkappa = (\balpha,\psi_1,\zeta_1,\psi_2,\zeta_2,\bmu^*_{\bbeta_1}, \bmu^*_{\bbeta_2}, \mathbf{V}_{\bbeta_1}, \mathbf{V}_{\bbeta_1})$ is the vector of variational parameters. 

For all the cases the Variational densities $q^*(\bbeta_1|\bkappa)$ and $q^*(\bbeta_2|\bkappa)$  will have exactly the same computations as in the normal case, see Section \ref{sec:VBbeta}.

The  vector of unknowns is 
$\bnu = (\lambda_1,\lambda_2,\bbeta_1, \bbeta_2, \bgamma) = (\bTheta,\bgamma),$
where $\bgamma=(\bgamma_1,\ldots,\bgamma_n)$.

\subsubsection{Variational density $q^*(\bgamma_i|\bkappa)$}\label{sec:vb_gamma_p1_a}

 If we consider $q^*(\lambda_1|\bkappa)$ and $q^*(\lambda_2|\bkappa)$ belonging to the gamma family of  distributions with parameters $(\psi_1,\zeta_1)$ and $(\psi_2,\zeta_2)$ respectively,  we get 

\begin{eqnarray*}
\lefteqn{p(\bgamma_i|\bnu(-\bgamma_i),\mathbf{y},\bkappa)=} \\
& \propto & 
\left(\frac{I(y_i=0)}{\log(1 + \exp({\bf x}_{i} \bbeta_1) + \exp({\bf x}_{i} \bbeta_2))}\right)^{\gamma_{i0}} \times  \left(\frac{e^{-\lambda_1} \lambda_1^{y_i} e^{{\bf x}_{i} \bbeta_1}} {\log(1 + \exp({\bf x}_{i} \bbeta_1) + \exp({\bf x}_{i} \bbeta_2))}\right)^{\gamma_{i1}}  \\
&& \hspace{6.4cm}\times \left(\frac{e^{-\lambda_2} \lambda_2^{y_i} e^{{\bf x}_{i} \bbeta_2}} {\log(1 + \exp({\bf x}_{i} \bbeta_1) + \exp({\bf x}_{i} \bbeta_2))}\right)^{\gamma_{i2}} \nonumber \\
& \propto & \left(I(y_i=0)\right)^{\gamma_{i0}} \left(e^{-\lambda_1} \lambda_1^{y_i} e^{{\bf x}_{i} \bbeta_1} \right)^{\gamma_{i1}} \left(e^{-\lambda_2} \lambda_2^{y_i} e^{{\bf x}_{i} \bbeta_2}\right)^{\gamma_{i2}}.  
\end{eqnarray*}

Therefore, we have 
    \begin{eqnarray*}
\lefteqn{E_{q^*(\bnu(-\bgamma_i))} [ \log p(\bgamma_i|\bnu(-\bgamma_i),\mathbf{y},\bkappa)] =} \\ &=& 
 C_{1i} + \gamma_{i0} I[y_i=0] + \gamma_{i1} \left(E_{q^*(\lambda_1)}[-\lambda_1 +  y_i \log(\lambda_1)] +  E_{q^*(\bbeta_1)}[{\bf x}_{i} \bbeta_1] \right) \nonumber \\
 && \quad + \gamma_{i1} \left(E_{q^*(\lambda_2)}[-\lambda_2 +  y_i \log(\lambda_2)] +  {\bf x}_{i} E_{q^*(\bbeta)}[ \bbeta_1] \right)
 \nonumber \\
 & = & C_{1i} + \gamma_{i1} \left[-\frac{\psi_1}{\zeta_1} +  y_i (-\log(\zeta_1) + \Psi(\psi_1))\right] +  E_{q^*(\bbeta_1)}[{\bf x}_{i} \bbeta_1]  \nonumber \\
 && \quad \gamma_{i2} \left[-\frac{\psi_2}{\zeta_2} +  y_i (-\log(\zeta_2) + \Psi(\psi_2))\right] +  E_{q^*(\bbeta)}[{\bf x}_{i} \bbeta_2].
 \end{eqnarray*}

Therefore, $\bgamma_i$ is a multinomial random variable Multinomial$(1,\alpha_{i0},\alpha_{i1},\alpha_{i2})$ where
\begin{eqnarray*}
    \alpha_{i0} &\propto& I(y_i=0),  \\
    \alpha_{i1} &\propto& \exp\left(-\frac{\psi1}{\zeta_1} +  y_i (-\log(\zeta_1) + \Psi(\psi_1)) +  E_{q^*(\bbeta_1)}[{\bf x}_{i} \bbeta_1 \right),   \\
\mbox{and} && \nonumber \\
    \alpha_{i2} &\propto& \exp\left(-\frac{\psi_1}{\zeta_1} +  y_i (-\log(\zeta_2) + \Psi(\psi_2)) +  E_{q^*(\bbeta_2)}[{\bf x}_{i} \bbeta_2 \right)  
\end{eqnarray*}
with  $\alpha_{i0} + \alpha_{i1} + \alpha_{i2} = 1$.

\subsubsection{Variational density $q^*(\lambda_1|\bkappa)$ and $q^*(\lambda_2|\bkappa)$}

For $k = 1,2$ we have the prior distributions
\begin{equation*}
    p(\lambda_k) = \frac{1}{\Gamma(a_k)} b_k^{a_k} (\lambda_k)^{a_k -1} e^{-b_k \lambda_k}.
\end{equation*}

Also, We choose the family of gamma distributions as the variational family for  $q^*(\lambda_1|\bkappa)$ and $q^*(\lambda_2|\bkappa)$. Therefore, 
\begin{eqnarray*}
\log q^*(\lambda_1) &=&  C_{2,0} + (a_1 -1) \log(\lambda_1) -b_1 \lambda_1  + \sum_{i=1}^{n} E_{q^*(\bgamma_i)} \gamma_{i1}  \left[ - \lambda_1  + y_i \log(\lambda_1) \right] \nonumber \\
&=& C_{2,0} + (a_1 -1) \log(\lambda_1) - b_1 \lambda_1  + \sum_{i=1}^{n}  \alpha_{i1}  \left[ - \lambda_1  + y_i \log(\lambda_1) \right] \nonumber \\
&=& C_{2,0} + \left(\sum_{i=1}^n  \alpha_{i1} y_i + a_1 - 1 \right) \log(\lambda_1) - \left(\sum_{i=1}^n   \alpha_{i1}  + b_1 \right) \lambda_1 
\label{eq:qxi01} 
\end{eqnarray*}

Similarly, 
\begin{eqnarray*}
\log q^*(\lambda_2) &=&  C_{2,1} + (a_2 -1) \log(\lambda_2) -b_2 \lambda_2  + \sum_{i=1}^{n} E_{q^*(\bgamma_i)}  \gamma_{i2}  \left[ -  \lambda_2  + y_i \log(\lambda_2) \right] \nonumber \\
&=& C_{2,1} + (a_2 -1) \log(\lambda_2) -b_2 \lambda_2  + \sum_{i=1}^{n} \alpha_{i2}  \left[ - \lambda_2  + y_i \log(\lambda_2) \right] \nonumber \\
&=& C_{2,1} + \left(\sum_{i=1}^n \alpha_{i2} y_i + a_2 - 1 \right)\log(\lambda_2) - \left(\sum_{i=1}^n  \alpha_{i2}  + b_2 \right) \lambda_2   
\label{eq:qxi11} 
\end{eqnarray*}
which characterizes $q^*(\lambda_1)$ and $q^*(\lambda_2)$ as  gamma distribution with parameters 
\begin{equation}
    \label{eq:zip_psi_zeta_1_a}
    \psi_1:= a_1 + \left(\sum_{i=1}^n  \alpha_{i1} y_i \right) \quad \mbox{and} \quad  \zeta_1:= b_1 + \sum_{i=1}^n  \alpha_{i1}
\end{equation}
and
\begin{equation}
    \label{eq:zip_psi_zeta_2_a}
    \psi_2:= a_2 + \left(\sum_{i=1}^n \alpha_{i2} y_i \right) \quad \mbox{and} \quad  \zeta_2:= b_2 + \sum_{i=1}^n    \alpha_{i2},
\end{equation}
respectively.

\subsection{Calculating the ELBO} \label{ap:elbo_zip}

The ELBO is given by
\begin{eqnarray*}
\lefteqn{\mbox{ELBO}(\psi_1,\zeta_1,\psi_2, \zeta_2, \balpha,\mu^{*}_{\bbeta_1},\mathbf{V}_{\bbeta_1},  \mu^{*}_{\bbeta_2},\mathbf{V}_{\bbeta_2})=}\\ 
&=&  
\sum_{i=1}^{n} (E_{q^*}[\log p(y_i|\gamma_i,\lambda_{1},\lambda_{2})] + E_{q^*}[\log p(\bgamma_i)] \nonumber \\
&& + \, E_{q^*}[\log p(\lambda_{1})]  +  E_{q^*}[\log p(\lambda_{2})]  + E_{q^*}[\log p(\bbeta_1)]  + + E_{q^*}[\log p(\bbeta_2)]\nonumber \\
&& - \, \sum_{i=1}^{n} (E_{q^*}[\log q^*(\bgamma_i)] - E_{q^*}[\log q^*(\lambda_{1})] -  E_{q^*}[\log q^*(\lambda_{2})] \nonumber \\
&& -  E_{q^*}[\log q^*(\bbeta_1)] -  E_{q^*}[\log q^*(\bbeta_2)] \nonumber \\
&=& E_0+ E_1 + E_2 + E_3 + E_4 + E_5- F_1 - F_2 - F_3 -F_4 - F_5
\end{eqnarray*}
with the expectation is taken with respect to $q^*(\lambda_1,\lambda_2,\bgamma,\bbeta|\bkappa)$, that is $E_{q^*} := E_{q^*(\lambda_1,\lambda_2,\bgamma,\bbeta|\bkappa)}$. Therefore, we have to compute 

\begin{eqnarray*}
E_0:= \lefteqn{E_{q^*}(\log p(\mathbf{y}|\lambda_1,\lambda_2, \bgamma)) } \nonumber \\
& = & E_{q^*}[\sum_{i=1}^{n} \gamma_{i0}I(y_i=0) + (-\log y_i!)  + \gamma_{i1} (- \lambda_{1} + y_i \log(\lambda_{1}))] +  E_{q^*}[\gamma_{i2} (-  \lambda_{2} + y_i \log(\lambda_{2}))] \nonumber \\
&=& \sum_{i=1}^{n}  \alpha_{i0}I(y_i=0) + \left\{ (-\log y_i!)  + \alpha_{i1} \left(  -  \frac{\psi_{1}}{\zeta_{1}}  + y_i \left( -\log(\zeta_{1}) + \Psi(\psi_{1})\right) \right) \right.\nonumber \\
& & \quad + \left. \alpha_{i2} \left( -\frac{ \psi_{2}}{\zeta_{2}}  + y_i \left(- \log(\zeta_{2}) + \Psi(\psi_{2}) \right) \right) \right\} \nonumber \\
\end{eqnarray*}
where $\psi_{1}, \zeta_{1}, \psi_{2}$ and $\zeta_{2}$ are given by \eqref{eq:zip_psi_zeta_1_a} and \eqref{eq:zip_psi_zeta_2_a}
and $\Psi$ is the digamma function. 

\begin{eqnarray}
\lefteqn{E_1:= \sum_{i=1}^n E_{q^*}(\log p(\gamma_i))=} \nonumber \\ 
&=& \sum_{i=1}^n E_{q^*}\left[\gamma_{i0} I(y_i=0) + \gamma_{i1} {\bf x'}_{i} \bbeta_1) + \gamma_{i2} \exp({\bf x'}_{i} \bbeta_2) - \log[1 + \exp({\bf x'}_{i} \bbeta_1) + \exp({\bf x'}_{i} \bbeta_2)] \right. \nonumber \\
&=& \sum_{i=1}^n  \alpha_{i0} + \alpha_{i1} {\bf x'}_{i} \mu^*_{\bbeta_1} + \alpha_{i2} {\bf x'}_{i} \mu^*_{\bbeta_2} \nonumber \\
&& \quad - \int \log[1 + \exp({\bf x'}_{i} \bbeta_1) + \exp({\bf x'}_{i} \bbeta_2))] q^*(\bbeta_1) q^*(\bbeta_2) \, d\bbeta_1 d\bbeta_2 
\end{eqnarray}
with $q^*(\bbeta_1)$ and $q^*(\bbeta_2)$ are obtained similarly to Section \ref{sec:VBbeta}. 

\begin{eqnarray}
\lefteqn{E_2:= E_{q^*}(\log p(\lambda_{1}))=} \nonumber \\ &=&   - \log(\Gamma(a_1)) + a_1 \log(b_1) + (a_1 -1) E_{q^*}[\log(\lambda_{1})] -b_1 E_{q^*}[\lambda_{1}]  \nonumber \\
&=&  -  \log(\Gamma(a_1)) +  a_1 \log(b_1) + (a_1 -1)  (-\log(\zeta_{1}) +  \Psi(\psi_{1})) -b_1 \frac{\psi_{1}}{\zeta_{1}}  
\end{eqnarray}
and 
\begin{eqnarray}
\lefteqn{E_3:= E_{q^*}(\log p(\lambda_{2}))=} \nonumber \\ &=&   - \log(\Gamma(a_2)) + a_2 \log(b_2) + (a_2 -1) E_{q^*}[\log(\lambda_{2})] -b_2 E_{q^*}[\lambda_{2}]  \nonumber \\
&=&  -  \log(\Gamma(a_2)) +  a_2 \log(b_2) + (a_2 -1)  (-\log(\zeta_{2}) +  \Psi(\psi_{2})) -b_2 \frac{\psi_{2}}{\zeta_{2}}.  
\end{eqnarray}
Also,
\begin{eqnarray}
\lefteqn{E_4:= E_{q^*}(\log p(\bbeta_1)) =} \nonumber \\
&=& E_{q^*}[\log \phi(\bbeta_1;\bmu_{\bbeta_1},\Sigma_{\bbeta_1})] \nonumber \\
&=& - \frac{n}{2} \log 2 \pi - \frac{1}{2} \log |\Sigma_{\bbeta_1}| - (1/2) E_{q^*}\left[(\bbeta_1 - \bmu_{\bbeta_1})^{\top} \Sigma_{\bbeta_1}^{-1} (\bbeta_1 - \bmu_{\bbeta_1})\right] \nonumber \\
&=& - \frac{R}{2} \log 2 \pi - \frac{1}{2} \log |\Sigma_{\bbeta_1}|  - (1/2) [(\mu^{*}_{\bbeta_1} - \bmu_{\bbeta_1})^{\top} \Sigma_{\bbeta_1}^{-1} (\mu^{*}_{\bbeta_1} - \bmu_{\bbeta_1})]
\end{eqnarray}
where $R$ is the dimension of the $\bbeta_1$ vector.  Similarly,
\begin{eqnarray}
\lefteqn{E_5:= E_{q^*}(\log p(\bbeta_2))=} \nonumber \\
&=& E_{q^*}[\log \phi(\bbeta;\bmu_{\bbeta},\Sigma_{\bbeta})] \nonumber \\
&=& - \frac{R}{2} \log 2 \pi - \frac{1}{2} \log |\Sigma_{\bbeta_2}|  - (1/2) [(\mu^{*}_{\bbeta_2} - \bmu_{\bbeta_2})^{\top} \Sigma_{\bbeta_2}^{-1} (\mu^{*}_{\bbeta_2} - \bmu_{\bbeta_2})]
\end{eqnarray}

On the other hand,
\begin{eqnarray}
\lefteqn{F_1:=E_{q^*}(\log q^*(\bgamma)) =} \nonumber \\
&=& \sum_{i=1}^n \alpha_{i0} \log \alpha_{i0} I(y_i=0) + \alpha_{i1} \log \alpha_{i1} + \alpha_{i2} \log  \alpha_{i2}, 
\end{eqnarray}

\begin{eqnarray}
\lefteqn{F_2:= E_{q^*}(\log q^*(\lambda_1))=} \nonumber \\
&=& \left\{ - \log\Gamma(\psi_{1}) + \psi_{1} \log(\zeta_{1}) + (\psi_{1} - 1) E_{q^*}[\log(\lambda_1)] - (\zeta_{1}) E_{q^*}[\lambda_1] \right\} \nonumber \\
&=& \left\{ -\log \Gamma(\psi_{1}) + \psi_{1} \log(\zeta_{1}) + (\psi_{1} - 1)(-\log(\zeta_{1}) +\Psi(\psi_{1}))  - \psi_{1}/\zeta_1 \right\},
\end{eqnarray}
and 
\begin{eqnarray}
\lefteqn{F_3:= E_{q^*}(\log q^*(\lambda_2))=} \nonumber \\
&=&  \left\{ - \log\Gamma(\psi_{2}) + \psi_{2} \log(\zeta_{2}) + (\psi_{2} - 1) E_{q^*}[\log(\lambda_{2})] - (\zeta_{2}) E_{q^*}[\lambda_{2}] \right\} \nonumber \\
&=&  \left\{ - \log\Gamma(\psi_{2}) + \psi_{2} \log(\zeta_{2}) + (\psi_{2} - 1)(-\log(\zeta_{2}) +\Psi(\psi_{2}))  - \psi_{2}/\zeta_2 \right\}.
\end{eqnarray}

Finally,
\begin{eqnarray}
F_4:= & = & E_{q^*}(\log q^*(\bbeta_1)) \nonumber \\
 &=& E_{q^*}[\log \phi(\bbeta_1;\bmu_{q^*(\bbeta_1)},\Sigma_{q^*(\bbeta_1}))] \nonumber \\
&=& - \frac{R}{2} \log 2 \pi - \frac{1}{2} \log |\Sigma_{q^*(\bbeta_1)}| - (1/2) E_{q^*(\bbeta_1)}[(\bbeta_1 - \bmu_{q^*(\bbeta_1)})^{\top} \Sigma_{q^*(\bbeta_1)}^{-1} (\bbeta_1 - \bmu_{q^*(\bbeta_1}))] \nonumber \\
&=& - \frac{R}{2} \log 2 \pi - \frac{1}{2} \log |\Sigma_{q^*(\bbeta_1)}|  - \frac{1}{2} tr(\Sigma_{q^*(\bbeta_1)}^{-1} \Sigma_{q^*(\bbeta_1)}) \nonumber \\
&=& - \frac{R}{2} \log 2 \pi - \frac{1}{2} \log |\Sigma_{q^*(\bbeta_1)}|  - \frac{R}{2}.
\end{eqnarray}
and 
\begin{eqnarray}
F_5:= & = & E_{q^*}(\log q^*(\bbeta_2)) \nonumber \\
 &=& E_{q^*}[\log \phi(\bbeta_2;\bmu_{q^*(\bbeta_2)},\Sigma_{q^*(\bbeta_2}))] \nonumber \\
&=& - \frac{R}{2} \log 2 \pi - \frac{1}{2} \log |\Sigma_{q^*(\bbeta_2)}| - (1/2) E_{q^*(\bbeta_2)}[(\bbeta_2 - \bmu_{q^*(\bbeta_2)})^{\top} \Sigma_{q^*(\bbeta_2)}^{-1} (\bbeta_2 - \bmu_{q^*(\bbeta_2}))] \nonumber \\
&=& - \frac{R}{2} \log 2 \pi - \frac{1}{2} \log |\Sigma_{q^*(\bbeta_2)}|  - \frac{1}{2} tr(\Sigma_{q^*(\bbeta_2)}^{-1} \Sigma_{q^*(\bbeta_2)}) \nonumber \\
&=& - \frac{R}{2} \log 2 \pi - \frac{1}{2} \log |\Sigma_{q^*(\bbeta_2)}|  - \frac{R}{2}.
\end{eqnarray}

\subsection{A fast way to compute $E_{q^*(\bbeta)}[\log (1 + \exp({\bf x}' \bbeta_1) + \exp({\bf x}' \bbeta_1)]$} \label{ap:fast2}

Similarly to Section \ref{ap:fast} we can construct an orthonormal matrix $S$ such as the first column of $S$ is $x/\|x\|$. We have that under the variational densities, $\\beta_1$ and $\bbeta_2$ are independent $d$-variate normal random vectors with mean vectors $\mu^*_{\bbeta_1}$ and $\mu^*_{\bbeta_2}$ and covariance matrices $\Sigma_{1}$ and $\Sigma_2$, respectively. Letting $S \beeta = (\bbeta_1 - \mu_{\bbeta_1})$ and $S \bzeta = (\bbeta_2 - \mu_{\bbeta_2})$ we can write
\begin{eqnarray}
\lefteqn{E_{q^*}[\log(1+\exp({\bf x}' \bbeta_1)+\exp({\bf x}' \bbeta_2))] =} \nonumber \\&=& \int_{\mathbb{R}^{2d}} [\log(1+\exp({\bf x}' \bbeta_1)+\exp({\bf x}' \bbeta_2))] q^*(\bbeta_1) q^*(\bbeta_2) \,d\bbeta_1 \, d\bbeta_2 \nonumber \\
& = & \int_{\mathbb{R}^{d}}  F_1(x'\bbeta_1) q^*(\bbeta_1) d\bbeta_1 \label{eq:int2}
\end{eqnarray}
where
\begin{equation}
    \label{eq:F1}
   F_1(\bbeta_1) \,=\, \left( \int_{\mathbb{R}^{d}} [\log(1+\exp({\bf x}' \bbeta_1)+\exp({\bf x}' \bbeta_2))]  q^*(\bbeta_2) \,d\bbeta_2 \right).
\end{equation}

Plugging \eqref{eq:int1} into \eqref{eq:F1} we get
\begin{eqnarray}
    \label{eq:F1.1}
F_1(\bbeta_1) &=&\int_{\mathbb{R}} \log(1+ \exp({\bf x}' \bbeta_1) + \exp(\|x\| \zeta_1 + x'\bmu_{\bbeta_2})) \nonumber \\
&& \quad \times \frac{1}{(2\pi)^{1/2} \sqrt{\Sigma_{2,(1,1)}}} \exp \left( -\frac{1}{2}  \eta_1 (S'\Sigma_{2}^{-1} S)_{(1,1)} \zeta_1 \right)   d\zeta_1. 
\end{eqnarray}

Now, we can apply \eqref{eq:int1} and \eqref{eq:F1.1} into \eqref{eq:int2} and we get
\begin{eqnarray*}
\lefteqn{E_{q^*}[\log(1+\exp({\bf x}' \bbeta_1)+\exp({\bf x}' \bbeta_2))] =} \nonumber \\
&=&  \int_{\mathbb{R}}  \int_{\mathbb{R}} \log(1+ \exp(\|x\| \eta_1 + x'\bmu_{\bbeta_1}) + \exp(\|x\| \zeta_1 + x'\bmu_{\bbeta_2})) \nonumber \\
&& \quad \quad \quad \times \frac{1}{(2\pi)^{1/2} \sqrt{\Sigma_{2,(1,1)}}} \exp \left( -\frac{1}{2}  \zeta_1 (S'\Sigma_{2}^{-1} S)_{(1,1)} \zeta_1 \right)  d\zeta_1 \nonumber \\
&& \quad \quad \quad \times \frac{1}{(2\pi)^{1/2} \sqrt{\Sigma_{1,(1,1)}}} \exp \left( -\frac{1}{2}  \eta_1 (S'\Sigma_{1}^{-1} S)_{(1,1)} \right)  \, d\eta_1. 
\end{eqnarray*}

\end{document}